\documentclass{jfm}

\usepackage{graphicx}
\usepackage{natbib}
\usepackage{multirow}
\usepackage{booktabs}
\usepackage{amsmath}
\usepackage{amssymb}
\usepackage{color}
\usepackage{moreverb}
\usepackage[geometry]{ifsym}
\usepackage{MnSymbol}

\ifCUPmtlplainloaded \else
  \checkfont{eurm10}
  \iffontfound
    \IfFileExists{upmath.sty}
      {\typeout{^^JFound AMS Euler Roman fonts on the system,
                   using the 'upmath' package.^^J}%
       \usepackage{upmath}}
      {\typeout{^^JFound AMS Euler Roman fonts on the system, but you
                   dont seem to have the}%
       \typeout{'upmath' package installed. JFM.cls can take advantage
                 of these fonts,^^Jif you use 'upmath' package.^^J}%
      }
  \else
  \fi
\fi

\ifCUPmtlplainloaded \else
  \checkfont{msam10}
  \iffontfound
    \IfFileExists{amssymb.sty}
      {\typeout{^^JFound AMS Symbol fonts on the system, using the
                'amssymb' package.^^J}%
       \usepackage{amssymb}%
         
       \let\ge=\geqslant  \let\geq=\geqslant
      }{}
  \fi
\fi

\ifCUPmtlplainloaded \else
  \IfFileExists{amsbsy.sty}
    {\typeout{^^JFound the 'amsbsy' package on the system, using it.^^J}%
     \usepackage{amsbsy}}
    {\providecommand\boldsymbol[1]{\mbox{\boldmath $##1$}}}
\fi

\newsavebox{\astrutbox}
\sbox{\astrutbox}{\rule[-5pt]{0pt}{20pt}}

\newcommand\eg{e.g.\ }
\def\tm{\leavevmode\hbox{$\rm {}^{TM}$}}

\title[Decay of multi-scale grid turbulence]{The decay of 
  turbulence generated by a class of multi-scale grids}

\author[P. C. Valente and J. C. Vassilicos]%
{P.\ns C.\ns V\ls A\ls L\ls E\ls N\ls T\ls E \break
J.\ns C.\ns V\ls A\ls S\ls S\ls I\ls L\ls I\ls C\ls O\ls S
 }

\affiliation{Department of Aeronautics, Imperial College London,
London SW7 2AZ, United Kingdom}

\pubyear{2011}
\volume{??}
\pagerange{??}
\date{21 August 2011}

\begin{document}

\maketitle

\begin{abstract}
A new experimental investigation of decaying turbulence generated by a low-blockage space-filling fractal square grid is presented. We find agreement with previous works by Seoud \& Vassilicos [``Dissipation and decay of fractal-generated turbulence", \textit{Phys. Fluids} \textbf{19}, 035103 (2007)] and Mazellier \& Vassilicos [``Turbulence without the Richardson-Kolmogorov cascade", \textit{Phys. Fluids} \textbf{22}, 075101 (2010)] but also extend the length of the assessed decay region and consolidate the results by repeating the experiments with different probes of increased spatial resolution. It is confirmed that this moderately high Reynolds number $Re_{\lambda}$ turbulence (up to $Re_\lambda \simeq 350$ here) does not follow the classical high Reynolds number scaling of the dissipation rate $\varepsilon \sim u'^{3}/L$ and does not obey the equivalent proportionality between the Taylor-based Reynolds number $Re_{\lambda}$ and the ratio of integral scale $L$ to Taylor micro-scale $\lambda$. Instead we observe an approximate proportionality between $L$ and $\lambda$ during decay. This non-classical behaviour is investigated by studying how the energy spectra evolve during decay and examining how well they can be described by self-preserving single-length scale forms. A detailed study of homogeneity and isotropy is also presented which
reveals the presence of transverse energy transport and pressure transport in the part of the turbulence decay region where we take data (even though previous studies found mean flow and turbulence intensity profiles to be approximately homogeneous in much of the decay region). The exceptionally fast turbulence decay observed in the part of the decay region where we take data is consistent with the non-classical behaviour of the dissipation rate. 
Measurements with a regular square mesh grid as well as comparisons with active grid experiments by Mydlarski \& Warhaft [``On the onset of high-Reynolds-number grid-generated wind tunnel turbulence", \textit{J. Fluid Mech.} vol. \textbf{320} (1996)] and Kang, Chester \& Meveneau [``Decaying turbulence in an active-grid-generated flow and comparisons with large-eddy simulation", \textit{J. Fluid Mech.} vol. \textbf{480} (2003)] are also presented to highlight the similarities and differences between these turbulent flows and the turbulence generated by our fractal square grid.

\end{abstract}


\section{Introduction} \label{sec:introduction}

At high enough Reynolds numbers, the local viscous dissipation rate
$\varepsilon$ of the local average turbulent kinetic energy $K$ scales
with $K$ and a local correlation length scale $L$, i.e. $\varepsilon
\sim K^{3/2}/L$. At least, this is what one reads in turbulence
textbooks \cite[see, for example,][]{Batchelor:book,
  TennekesLumley:book, Lumley92, Townsend:book, Frisch:book, Lesieur,
  MathieuScott,Pope,Cambon}.  \cite{TennekesLumley:book} introduce
this scaling in their very first chapter with the words ``it is one of
the cornerstone assumptions of turbulence
theory''. \cite{Townsend:book} uses it explicitly in his treatment of
free turbulent shear flows \cite[see page 197 in][]{Townsend:book}
which includes wakes, jets, shear layers, etc. Since G.I Taylor
introduced it in 1935 \cite[]{Taylor1935}, this scaling is also
customarily used in theories of decaying homogeneous isotropic
turbulence \cite[see][]{ Batchelor:book, Frisch:book, Rotta:book} and
in analyses of wind tunnel simulations of such turbulence
\cite[e.g.][]{batchelor1948decay, CC66} in the form
\begin{align}
\varepsilon=C_{\varepsilon}\frac{u'^{3}}{L}
\label{Eq:DissipationCoeff}
\end{align}
where $u'$ is the r.m.s. velocity fluctuation, $L$ is an integral
length scale and $C_{\varepsilon}$ is a constant independent of time,
space and Reynolds number when the Reynolds number is large enough.
However, as \cite{Taylor1935} was careful to note, the constant
$C_{\varepsilon}$ does not need to be the same irrespective of the
boundaries (initial conditions) where the turbulence is produced
 \cite*[see][]{Burattini2005,M&V2008,G&V2009}.
 
In high Reynolds number self-preserving free turbulent shear flows,
the cornerstone scaling $\varepsilon \sim K^{3/2}/L$ determines the
entire dependence of $\varepsilon$ on the streamwise coordinate and
ascertains its independence on Reynolds number
\cite[see][]{Townsend:book}. This cornerstone scaling is also
effectively used in turbulence models such as $K-\varepsilon$
\cite[see][]{Pope} and in Large Eddy Simulations
\cite[see][]{Lesieur,Pope}. The assumption that $\varepsilon$ is
independent of Reynolds number when the Reynolds number is large
enough is an inseparable part of the Richardson-Kolmogorov cascade
\cite[][]{TennekesLumley:book,Frisch:book}. This is the celebrated
nonlinear dissipation mechanism of the turbulence whereby, within a
finite time $L/\sqrt{K}$ (the same time scale for all high enough Reynolds
numbers), smaller and smaller ``eddies'' are generated till eddies so
small are formed which can very quickly lose their kinetic energy by
linear viscous dissipation. The higher the Reynolds number, the
smaller the size of these necessary dissipative eddies but the time scale
$L/\sqrt{K}$ for energy to cascade to them from the large eddies remains the
same. The dissipation rate $\varepsilon$ is proportional to $K$
divided by this time, and therefore $\varepsilon \sim K^{3/2}/L$.

In various high Reynolds number self-preserving free turbulent shear
flows as in wind tunnel grid-generated turbulence, $K$ and $L$ vary
with streamwise downstream distance $x-x_{0}$ (where $x_0$ is an
effective/virtual origin) as power laws. Specifically, $K\sim
U_{\infty}^{2} ({x-x_{0}\over L_{B}})^{-n}$ and $L\sim L_{B}
({x-x_{0}\over L_{B}})^{m}$ where $L_B$ is a length-scale
characterising the inlet and $U_{\infty}$ is the appropriate inlet
velocity scale. In table \ref{Table:OtherFlows} we recall the generally accepted values
taken by the exponents $n$ and $m$ in plane wakes, axisymmetric wakes,
self-propelled plane wakes, self-propelled axisymmetric wakes, mixing
layers, plane jets, axisymmetric jets and wind-tunnel grid-generated
turbulence \cite[from][]{TennekesLumley:book, CC66}.
Estimating a Taylor microscale $\lambda$ from $\varepsilon \sim
\nu K /\lambda^{2}$ where $\nu$ is the kinematic viscosity of the
fluid, and then applying the cornerstone assumption $\varepsilon \sim
K^{3/2}/L$ to all these flows yields the following two relations:
\begin{align}
L/\lambda \sim Re_{0}^{1/2} ({x-x_{0}\over L_{B}})^{{m\over 2} -{n\over 4}}
\label{Eq:?}
\end{align}
and 
\begin{align}
\sqrt{K}\lambda/\nu \sim Re_{0}^{1/2} ({x-x_{0}\over L_{B}})^{{m\over 2} -{n\over 4}}
  \label{Eq:???}
\end{align}
where $Re_{0} \equiv U_{\infty}L_{B}/\nu$ is the inlet Reynolds number
and $\sqrt{K}\lambda/\nu$ is a local Taylor microscale-based Reynolds
number. The different values of ${m\over 2} -{n\over 4}$ are given in
table \ref{Table:OtherFlows}. Remarkably, $\varepsilon \sim K^{3/2}/L$
implies that $L/\lambda \sim \sqrt{K}\lambda/\nu$ in all these flows
whatever the values of $n$ and $m$, meaning that $L/\lambda \sim
\sqrt{K}\lambda/\nu$ collapses the $Re_0$ and the $x$ dependencies in
the same way for all these flows. We stress that this collapse is the
immediate consequence of $\varepsilon \sim K^{3/2}/L$. The relation
$L/\lambda \sim \sqrt{K}\lambda/\nu$ simply reflects the
Richardson-Kolmogorov cascade: the higher the Reynolds number, the
smaller the size of the dissipative eddies, i.e. the greater the range
of excited scales and the greater $L/\lambda$.

As noted by \cite{Lumley92}, by 1992 there had not been too much
detailed and comprehensive questioning of data to establish the
validity of $\varepsilon \sim K^{3/2}/L$ but he wrote: ``I hardly
think the matter is really much in question''. He cited the data
compilations of \cite{Sreeni84} which suggested that $C_{\varepsilon}$
does become constant at $Re_{\lambda} = u'\lambda/\nu$ larger than
about 50 for wind tunnel turbulence generated by various biplane
square-mesh grids, but there seemed to be little else at the
time. Since then, direct numerical simulations (DNS) of high Reynolds
number statistically stationary homogeneous isotropic turbulence have
significantly strengthened support for the constancy of
$C_{\varepsilon}$ at $Re_{\lambda}$ greater than about 150 (see
compilation of data in \cite{Burattini2005}, see also
\cite{Sreeni98}). Other turbulent flows have also been tried in the
past fifteen years or so such as various turbulent wakes and jets and
wind tunnel turbulence generated by active grids
\cite[see][]{Burattini2005,M&V2008} with some, perhaps less clear,
support of the constancy of $C_{\varepsilon}$ at large enough
$Re_{\lambda}$ (perhaps larger than about 200 if $L$ is defined
appropriately, see \cite{Burattini2005}) and also some clear
indications that the high Reynolds number constant value of
$C_{\varepsilon}$ is not universal, as indeed cautioned by
\cite{Taylor1935}.

\begin{table}
\caption{Powers law exponents characteristic of the downstream evolutions of $K$, $L$, $L/\lambda$}
\centering
\begin{tabular*}{0.9\textwidth}{@{\extracolsep{\fill}}cccc}
 & $K$ & $L$ &  $L/\lambda$  \\
\midrule
Plane wake & 1 & 1/2 & 0 \\
Axisymmetric wake & 4/3 & 1/3 & -1/6 \\
Self-propelled plane wake & 3/2 & 1/4 & -1/4 \\
Self-propelled axisymmetric wake & 8/5 & 1/5 & -3/10 \\
Mixing layer & 0 & 1 & 1/2 \\
Plane jet & 1 & 1 & 1/4 \\
Axisymmetric jet & 2 & 1 & 0 \\
Regular grid turbulence & 1.25 & 0.35 & -0.14 \\
\end{tabular*}
\label{Table:OtherFlows}
\end{table}

A decade ago, \cite{Vassilicos&QueirosConde} took the opposite
approach and asked whether it might be possible to break $\varepsilon
\sim K^{3/2}/L$ in some fundamental way in some flows, and so they
proposed generating turbulence with fractal/multiscale
objects/stirrers/inlet conditions. Some years later, \cite{H&V2007}
published an exploratory study of wind tunnel grid-generated
turbulence where they tried twenty one different planar grids from
three different families of passive fractal/multiscale grids: fractal
cross grids, fractal I grids and fractal square grids. They
ascertained that the fractal dimension $D_f$ of these grids needs to
take the maximal value $D_f = 2$ for least downstream turbulence
inhomogeneity. They also identified some important grid-defining
parameters (such as the thickness ratio $t_r$, see figure
\ref{Fig:Grid} and table \ref{Table:GridDetails}) and some of their
effects on the flow, in particular on the Reynolds number
$Re_{\lambda}$ which they showed can reach high values with some of
these grids in small and conventional sized wind tunnels, comparable
to values of $Re_{\lambda}$ achieved with active grids in similar wind
tunnels and wind speeds. Their most interesting, and in fact
intriguing, results were for their space-filling ($D_f = 2$)
low-blockage (25\%) fractal square grids (see figure
\ref{Fig:Grid}). Fractal square grids have therefore been the
multiscale grids of choice in most subsequent works on
multiscale/fractal-generated turbulence \cite*[]{S&V2007,nagata2008dns,
nagata2008direct,SPSV2010,M&V2010,suzuki,Sylvain2011}.
For the case of space-filling low-blockage fractal square grids, \cite{H&V2007}
 found a protracted region between the grid and a
distance $x_{peak}$ downstream of the grid where the turbulence
progressively builds up; and a decay region region at $x>x_{peak}$
where the turbulence continuously decays downstream. They reported a
very fast turbulence decay which they fitted with an exponential and
also reported very slow downstream growths of the longitudinal and
lateral integral length-scales and of the Taylor microscale.
(Very recently, \cite{K&D2011} studied the decay behind
multiscale cross grids and found conventional  decay rates. Note that
for multiscale cross grids our prior  publications did not claim fast,
unconventional, decay rates \cite[]{H&V2007}. This may serve
as further justification for focusing attention on fractal square grids
in the present paper. Even so, multiscale cross grids have been used
successfully in some recent studies for enhancing the Reynolds number,
see \cite*{3DPTV} and \cite*{GGL2010}.)

\cite{S&V2007} concentrated their attention on the decay
region of turbulence generated by space-filling low-blockage fractal
square grids and confirmed the results of \cite{H&V2007}.
In particular, they showed that $L/\lambda$ remains
approximately constant whilst $Re_{\lambda}$ decays with downstream
distance $x$ and they noted that this behaviour implies a fundamental
break from (\ref{Eq:DissipationCoeff}) where $C_{\varepsilon}$ is
constant. They also found that one-dimensional longitudinal energy
spectra at different downstream centreline locations $x$ can be made
to collapse with $u'$ and a single length-scale, as opposed to the two
length-scales ($L$ and Kolmogorov microscale) required by
Richardson-Kolmogorov phenomenology. Finally, they also carried out
homogeneity assessments in terms of various profiles (mean flow,
turbulence intensity, turbulence production rate) as well as some
isotropy assessments.

\cite{M&V2010} also worked on wind tunnel turbulence
generated by space-filling low-blockage fractal square grids. They
introduced the wake-interaction length-scale $x_*$ which is defined in
terms of the largest length and thickness on the grid and they showed
from their data that $x_{peak} \approx 0.5 x_*$. They documented how
very inhomogeneous and non-Gaussian the turbulent velocity statistics
are in the production region near the grid and how homogeneous and
Gaussian they appear by comparison beyond $0.5 x_*$. They confirmed
the findings of \cite{H&V2007} and \cite{S&V2007} and 
added the observation that both $Re_{\lambda}$ and
$L/\lambda$ are increasing functions of the inlet velocity
$U_{\infty}$. Thus, the value of $L/\lambda$ seems to be set by the
inlet Reynolds number, in this case defined as $Re_{0}=U_{\infty}
x_{*}/\nu$ for example.

Finally, \cite{M&V2010} brought the two different single-scale
turbulence decay behaviours of \cite{George1992} and
\cite{George&Wang2009} into a single framework which they used to
analyse the turbulence decay in the downstream region beyond $x_{peak}
\approx 0.5x_*$. This allowed them to introduce and confirm against
their data the notions that, in the decay region, the fast turbulence
decay observed by \cite{H&V2007} and \cite{S&V2007} may not be
exponential but a fast decaying power-law and that $L$ and $\lambda$
are in fact increasing functions of $x$ which keep $L/\lambda$
approximately constant.

The results of \cite{H&V2007}, \cite{S&V2007} and \cite{M&V2010} suggest that, 
in the decay region downstream of space-filling low-blockage fractal square grids, high
Reynolds number turbulence is such that
\begin{align}
L/\lambda \sim Re_{0}^{\alpha} A({x-x_{0}\over x_{*}})
\label{Eq:LlambdaX}
\end{align}
and 
\begin{align}
Re_{\lambda}\sim Re_{0}^{\beta} B({x-x_{0}\over x_{*}})
\label{Eq:ReLambdaX}
\end{align}
where $A$ is a slow-varying dimensionless function of ${x-x_{0}\over
  x_{*}}$ (in fact effectively constant), $B$ is a fast-decreasing
dimensionless function of ${x-x_{0}\over x_{*}}$ (perhaps even as fast as
exponential), and $\alpha$ and $\beta$ are positive real numbers.
 
Assuming that the dissipation-scale turbulence structure is approximately isotropic, 
we now use the relation $\varepsilon = 15 \nu
u'^{2}/\lambda^{2}$ which \cite{Taylor1935} obtained for isotropic
turbulence. With (\ref{Eq:DissipationCoeff}) this relation implies
\begin{align}
\frac{L}{\lambda} = \frac{C_{\varepsilon}}{15} Re_{\lambda}
\label{Eq:LOverLambda}
\end{align}
and, clearly, $C_{\varepsilon}$ cannot be constant (independent of $Re_0$
and $x$) with $Re_0$ and $x$ dependencies of $L/\lambda$ and
$Re_{\lambda}$ such as those observed in wind tunnel turbulence
generated by space-filling low-blockage fractal square grids. Instead, 
\begin{align}
C_{\varepsilon} = 15 Re_{0}^{\alpha - \beta} A({x-x_{0}\over
  x_{*}})/B({x-x_{0}\over x_{*}})
\label{Eq:CepsOverall}
\end{align}
which means that $C_{\varepsilon}$ should be increasing fast in the
downstream direction but which also means that a plot of
$C_{\varepsilon}$ versus $Re_{\lambda}$ can be quite different
depending on whether $Re_{\lambda}$ is varied by varying $Re_0$ whilst
staying at the same position $x$ or by moving along $x$ whilst keeping
$Re_0$ constant. This is a point which we discuss and attempt to bring
out clearly in the present paper.

Relations \eqref{Eq:LlambdaX} and \eqref{Eq:ReLambdaX} and their 
consequent decoupling of $L/\lambda$ and $Re_{\lambda}$ were observed at moderate 
to high values of $Re_{\lambda}$ where \cite{S&V2007} and \cite{M&V2010} also
observed a well-defined broad power-law energy spectrum. Indeed
$Re_{\lambda}$ needs to be large enough for the study of fully
developed turbulence. Active grids were introduced by \cite{Makita91} to
improve on the Reynolds number values achieved by regular grids in
conventional wind tunnels. Fractal square grids achieve comparably
high values of $Re_{\lambda}$ but also a far wider range of
$Re_{\lambda}$ values along the streamwise direction. This makes if
much easier to study $Re_{\lambda}$-dependencies, a point which we
make and discuss in some detail in the present paper.

In this paper we report an experimental assessment of turbulent flows
generated by a low-blockage space-filling fractal square grid (see
figure \ref{Fig:Grid}) and a regular square-mesh grid. The main focus
of this paper is to complement former research on fractal-generated
turbulence by extending the assessed decay region and using the new
data to re-address the previously reported dramatic departure from
$C_{\varepsilon} = Const$ and $A({x-x_{0}\over x_{*}})= B({x-x_{0}\over
  x_{*}})$ and the abnormally high decay exponents \cite[]{H&V2007,
  S&V2007, M&V2010}. We provide estimates of these exponents, and also show 
  that $\alpha \approx \beta$ and that our fractal-generated turbulence
  behaves in a way which is very close to
self-preserving single-length scale turbulence \cite[]{M&V2010},
particularly if the turbulence anisotropy is taken into account when
calculating 3D energy spectra. We also show that, even though previous 
studies by \cite{S&V2007} and  \cite{M&V2010} found that the mean flow
and turbulence profiles are approximately homogeneous in much of the decay
region, there nevertheless remains significant transverse turbulent transport
of turbulent kinetic energy and turbulent transport of pressure.
The decaying turbulence is therefore not homogeneous
and isotropic in terms of third order one point statistics even though it more closely
is in terms of lower order one point statistics.
Whenever possible a comparison between fractal-generated and
non-fractal-generated turbulence is made emphasising similarities and
differences.

In the following section we describe the experimental apparatus as
well as the anemometry systems, probes and the details of the data
acquisition. The experimental results are presented in
Sec. \ref{sec:Results} and are organised in four subsections. In
Sec. \ref{subsec:xstar} it is suggested that the wake-interaction
length-scale introduced by \cite{M&V2010} to characterise the extent
of the production region in the lee of the fractal grid is also
meaningful for regular static grids. In
Sect. \ref{subsec:HomogeneityAndIsotropy} the homogeneity and isotropy of the
fractal-generated flow is investigated following the methodology used
by \cite{CorrsinHandbook} and \cite{CC66} for regular static grids.
In Sec. \ref{Sec:Ceps} \& \ref{Sec:Decay} the normalised energy dissipation rate 
and the decay law are re-assessed using the new data. 
In Sec. \ref{subsec:Collapse} we investigate the
possibility of a self-similar, single-length-scale behaviour by
collapsing the 1D energy spectra and the $2^{nd}$-order structure
functions using large-scale variables; also the 3D energy spectrum
function is calculated to provide isotropy corrections on the
collapse. In Sec. \ref{sec:concl} we end this paper by highlighting
the main conclusions drawn from the present measurements and discuss
some of the questions raised.

\section{The experimental setup}\label{sec:Setup}
\subsection{Experimental hardware} \label{Subsec:Hardware&Wires}
The experiments are performed in the $T=0.46m$ wind-tunnel described
in some detail in \cite{M&V2010} and sketched in figure
\ref{Fig:WTSketch} ($T$ is the lateral width of the tunnel's square test section).
 The inlet velocity $U_{\infty}$ is imposed and
stabilised with a PID feedback controller using the static pressure
difference across the 8:1 contraction and the temperature near the
inlet of the test section which are measured using a Furness Controls
micromanometer FCO510.

All data are taken with one- and two-component hot-wire anemometers
operating in constant-temperature mode (CTA).
 The hot-wires are driven by a DANTEC StreamLine CTA system with
an in-built signal conditioner. We use both square- and sine-wave
testing to measure the cut-off frequency at the verge of attenuation
($f^{0dB}_{cut-off}$) and at the standard \mbox{'-3dB'} attenuation
level ($f^{-3dB}_{cut-off}$). In table \ref{WireDetails} we present
the results from the electronic testing of our anemometry
system. Further information concerning electronic testing of thermal
anemometers and a discussion of the consistency between the square and
sine-wave tests can be found in \cite{F77}.

For the single component measurements three different single-wires
(SW) are used with a sensing length ($l_{w}$) of $1mm$, $0.45mm$ \&
$0.2mm$ respectively. For the two component measurements two
cross-wires (XW) with sensing lengths of $l_{w}=0.5$ \& $1mm$
respectively are used, but for both the separation between the wires
is around $1mm$. All the sensors except the $l_{w}=1mm$ XW are based
on Dantec probes modified to use in-house etched
Platinum-(10\%)Rhodium Wollaston wires soldered to the prongs (further
details can be found in table \ref{WireDetails}). The $l_{w}=1mm$ XW
is a Dantec 55P51 tungsten probe. It should be noted that the
$l_{w}=0.2mm$ single-wire, which has a diameter of $d_{w}=1.27\mu m$, is
operated in the limit of the bridge stability, on the verge of having
non-damped oscillations. Nonetheless, the sine-wave test indicated
that $f^{0dB}_{\mathrm{cut-off}}$ was about $40kHz$.  The hot-wires
are calibrated at the beginning and at the end of each measurement
campaign using a $4^{th}$-order polynomial in the SW case and a
velocity-pitch map in the XW case. Note that, unless otherwise stated,
the data shown are acquired with the $l_{w}\approx 1mm$ SW hot-wire
probe. All the two-component data presented are acquired with the
$l_{w}\approx 0.5mm$ XW except the spanwise traverse data presented in
Sec. \ref{subsec:Homogeneity}.

Note that two other anemometry systems have been used as well in order
to compare with previous experimental results, but these results are
not included here. The other anemometry systems are: the AALab AN-1005
CTA system used in \cite{S&V2007} and \cite{H&V2007} and the DISA
55M10 CTA bridge with a DISA 55D26 signal conditioner used in
\cite{M&V2010}. It is found that the results obtained with the DISA
55M10 CTA unit closely match those obtained with the StreamLine CTA
system, when the same hot-wire probe is used, except at very high
frequencies where the higher noise floor of the DISA CTA system buries
the velocity signal. On the other hand it is found that the
measurements taken with the AALab AN-1005 CTA system are significantly
different at frequencies above $6$kHz and therefore the turbulence
statistics involving velocity derivatives are significantly
different. This is likely the reason for the difference between the
normalised energy dissipation rate $C_{\varepsilon}$ results reported
in \cite{S&V2007} and the ones presented in Sec. \ref{Sec:Ceps} of
this paper. The comparison between the results of the different
anemometry systems will be presented elsewhere.

\begin{table}
\centering
\begin{tabular*}{0.9\textwidth}{@{\extracolsep{\fill}}ccccccccc}
SW/XW & $l_{w}$ & $d_{w}$ & $l_{w}/d_{w}$ & Hot-wire  & $U_{\infty}$ & $f^{-3dB}_{\mathrm{cut-off}}$ & $f^{0dB}_{\mathrm{cut-off}}$ &  $l_{w}/\eta$  \\
 & $(mm)$ & $(\mu m)$ &  & probe & $(ms^{-1})$ & $(kHz)$ & $(kHz)$ &   \\
\midrule
\multirow{2}{*}{SW} & \multirow{2}{*}{$\sim 1$} & \multirow{2}{*}{$5.1$} & \multirow{2}{*}{$196$} & \multirow{2}{*}{55P16} & 10 & $\sim 25$ & $\sim 12$ & 7-3 \\
& & & & & 15 & $\sim 32$ & $\sim 16$ & 9-5\\
\multirow{2}{*}{SW} & \multirow{2}{*}{$\sim 0.45$} & \multirow{2}{*}{$2.5$} & \multirow{2}{*}{$180$} & \multirow{2}{*}{55P16} & 10 & $\sim 45$ & $\sim 21$  & 3-2 \\
& & & & & 15 & $\sim 45$ & $\sim 23$ & 4-2\\
\multirow{2}{*}{SW} & \multirow{2}{*}{$\sim 0.2$} & \multirow{2}{*}{$1.27$} & \multirow{2}{*}{$157$} & \multirow{2}{*}{55P11} & 10 & $>50$ & $\sim 40$ & $\sim 1$ \\
& & & & & 15 & $>50$ & $\sim 40$ & 2-1\\
\multirow{2}{*}{XW} & \multirow{2}{*}{$\sim 0.5$} & \multirow{2}{*}{$2.5$} & \multirow{2}{*}{$200$} & \multirow{2}{*}{55P51} & 10 & $\sim 45$ & $\sim 21$  & 3-2 \\
& & & & & 15 & $\sim 45$ & $\sim 23$ & 4-2\\
XW & $ 1.0$ & $5$ & $200$ & 55P51 & 15 & $\sim 30$ & $\sim 14$ & 9-5\\
\end{tabular*}
\caption{Details on the hot-wires, cut-off frequencies \& resolution. $l_{w}$ and $d_{w}$ are the sensing length and diameter of the wires, $l_{w}/\eta$ is the ratio between the sensing length and the Kolmogorov inner length-scale and $U_{\infty}$ is the inlet velocity. $f^{-3dB}_{\mathrm{cut-off}}$ is the cut-off frequency corresponding to $-3dB$ signal attenuation and $f^{0dB}_{\mathrm{cut-off}}$ is the highest frequency with negligible attenuation.}
\label{WireDetails}
\end{table}

\subsection{Data acquisition and signal processing}
The pressure and temperature measurements are digitally transferred to
the computer using a parallel port. The analogue signal from the
anemometers is sampled using a 16-Bit National Instruments
NI-6229(USB) card, at a sampling frequency set to be higher than twice
the analogue low-pass filtering frequency ($30$kHz). The data acquisition and
signal processing are performed with the commercial software
MATLAB\tm.

The turbulent velocity signal was acquired for 9min corresponding to
more than 100,000 integral-time scales. This was confirmed to be
sufficient for converged measured statistics of interest such as the
integral scale, the first four moments of the velocity signal and the
2$^{nd}$ moment of the velocity derivative signal. The time-varying
turbulent signal was converted into spatially-varying by means of a
local Taylor's hypothesis following the algorithm proposed in
\cite{KMGC98}. Before Taylor's hypothesis is used the signal is
digitally filtered at a frequency corresponding to $k_{1}\eta\sim 1.1$
(where $\eta\equiv (\nu^3/\varepsilon)^{1/4}$ is the Kolmogorov inner length-scale  and $k_1$ the
wavenumber) using a $4^{th}$-order Butterworth filter to eliminate
higher frequency noise.
 
The integral scale $L_u$ is estimated as 
\begin{equation*}
L_u = \int_0^{r_L} f(r) \, dr ,
\end{equation*}
where $f(r) \equiv \overline{u(x)u(x+r)}/\overline{u(x)^2}$ is the
auto-correlation function of the streamwise velocity fluctuations for
streamwise separations $r$ and $r_L$ is maximum integration range
taken to be about $10$ times the integral length scale. It was checked
that (i) changing the integration limit $r_L$ by a factor between
$2/3$ and $2$ has little effect on the numerical value of the integral
scale and (ii) the choice of $r_L$, if large enough, does not
influence the way that $L_u$ varies with downstream distance. The
transverse integral scale is estimated in a similar way. The
longitudinal and transverse spectra are calculated using an FFT based
periodogram algorithm using a Hanning window with 50\% overlap and window
length equivalent to at least $180$ integral length scales. The
dissipation $\varepsilon$ is estimated from the longitudinal
wavenumber spectra $F_{11}$ as
\begin{equation*}
\varepsilon = 15\nu \int_{k_{min}}^{k_{max}}k_{1}^2\,F_{11}(k_{1})\, dk_{1},
\end{equation*}
where $k_{min}$ and $k_{max}$ are determined by the window length and
the sampling frequency respectively. To reduce the unavoidable
contamination of noise at high frequencies (which can bias the dissipation
estimate) we follow \cite{antoniadissipation} and fit an exponential
curve to the high frequency end of the spectra which we then
integrate. We checked that calculating the dissipation with and
without Antonia's (2003) method changes the
dissipation by less than $4\%$ in the worst case.

It might be worth mentioning that the measurements of the fractal
grid-generated turbulence posed a lesser challenge to hot-wire
anemometry than the regular grid-generated turbulence quite simply
because the turbulent signal to anemometry noise ratio is
higher in the former case, but nonetheless the Kolmogorov microscales
(which influence the maximum frequency to be measured) for the highest
$Re_{\lambda}$ measurement location ($Re_{\lambda}\approx 350$ and
$Re_{\lambda}\approx 150$ respectively) are roughly the same
($\eta\approx 0.11mm$ and $\eta\approx 0.13mm$ respectively).

\subsection{Turbulence generating grids}
\begin{figure} 
\centering
\includegraphics[height=60mm]{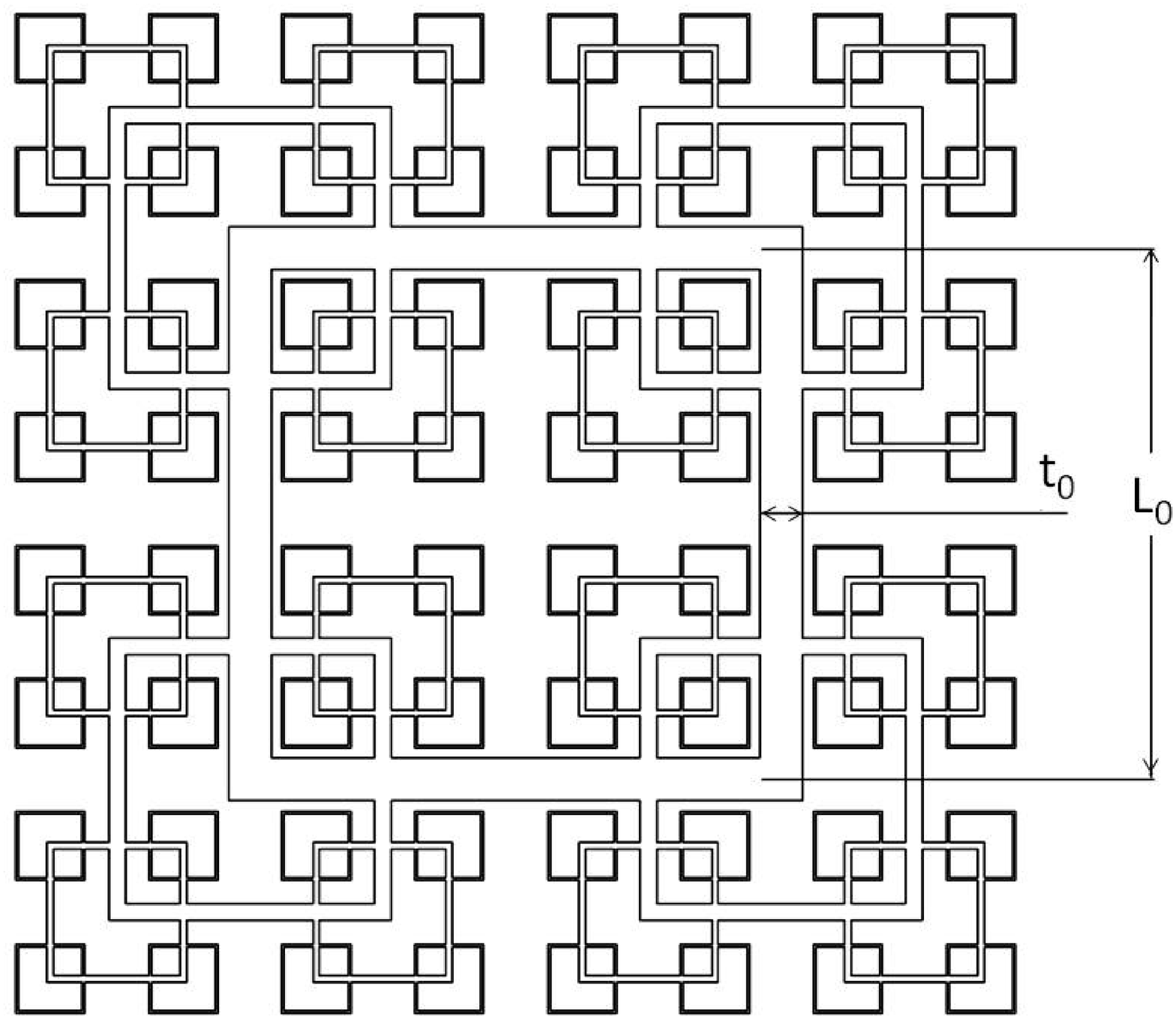}
\caption{Low-blockage space-filling fractal square grid (SFG). The
  grid is space-filling because the fractal dimension of its
  delimiting line takes the maximum value of $2$ over the range of
  scales on the grid. In the limit of infinite number of fractal
  iterations the blockage ratio will tend to unity, without taking bar
  thickness into account. However with only four iterations and the
  bar thickness in the figure the grid's blockage ratio is only 25\%.}
\label{Fig:Grid}
\end{figure}

The bulk part of the measurements are performed on turbulence
generated by a low-blockage space-filling fractal square grid (SFG)
with 4 'fractal iterations' and a thickness ratio of $t_{r}=17$, see
figure \ref{Fig:Grid}. It is one of the grids used in the experimental
setup of \cite{M&V2010} where further details of the fractal grids and
their design can be found. Measurements of turbulence generated by a
regular bi-plane grid (RG) with a square mesh and composed of square
rods are also performed. The summary of the relevant grid design
parameters is given in table \ref{Table:GridDetails}.

\begin{table}
\centering
\begin{tabular*}{0.9\textwidth}{@{\extracolsep{\fill}}cccccccccc}
Grid & N & $L_{0}$ & $t_{0}$ & $L_{r}$ & $t_{r}$ & $R_{L}$ & $R_{t}$ & $\sigma$ & $M_{eff}$  \\
 & & $(mm)$ & $(mm)$ &  & & & & & $(mm)$   \\
\midrule
SFG & 4 & 237.8 & 19.2 & 8 & 17 & 0.5 & 2.57 & 0.25 & 26.2\\
RG  & 1 & 60   & 10   & 1 & 1  & 1 & 1 & 0.32 & 60  \\
\end{tabular*}
\caption{Details of the regular grid (RG) and the low-blockage
  space-filling fractal square grid (SFG). $N$ is the number of
  fractal iterations of the grids (for a regular grid $N=1$). $L_{0}$
  is the centreline distance separating the largest bars and $t_{0}$
  their lateral thickness, see figure \ref{Fig:Grid}.  $L_{r}$ and
  $t_{r}$ are respectively the length and thickness ratio between the
  largest and smallest bars. $R_{L}$ and $R_{t}$ are the length and
  thickness ratio between two consecutive fractal iterations. $R_{t}$
  is related to $t_{r}$ and $N$ via $t_{r}=R_{t}^{-N+1}$. The blockage
  ratio $\sigma$ is defined as the fraction of area occupied by the
  grid and $M_{eff}$ is the effective mesh size as defined in
  \cite{H&V2007} which reverts to the definition of mesh size for a
  regular grid.}
\label{Table:GridDetails}
\end{table}

\section{Results}\label{sec:Results}
The turbulent field in the lee of the space-filling fractal square
grids can be considered to have two distinct regions
\cite[]{H&V2007,M&V2010}: a production region where the turbulent
kinetic energy (on the centreline) is increasing and the flow is being
homogenised, and a decay region where the energy of the turbulent
fluctuations are rapidly decreasing and the flow is roughly
homogeneous with an isotropy factor around $u'/v'\sim1.1 - 1.25$, where $u'$
and $v'$ are the longitudinal and transverse r.m.s. velocities
respectively.

\subsection{The wake-interaction length-scale}\label{subsec:xstar}

\cite{M&V2010} introduced the wake-interaction length-scale
$x_{*}=L_{0}^{2}/t_{0}$ (see definitions of $L_{0}$ \& $t_{0}$ in
figure \ref{Fig:Grid} and in the caption of table
\ref{Table:GridDetails}) to characterise the extent of the turbulence
production region in the lee of the space-filling fractal square
grids. This length-scale is based on the largest square of the grid
since the wakes it generates are the last to interact, although there
is a characteristic wake-interaction length-scale for each grid
iteration \cite[for a schematic of the wake interactions occurring at
  different streamwise locations refer to figure 4a
  in][]{M&V2010}. They then related the wake interaction length-scale
with the location of the maximum of the turbulence intensity along the
centreline $x_{peak}$, which marks the end of the production region
and the start of the decay region and found that
$x_{peak}/x_{*}\approx0.45$. Note that this is not the only peak in
turbulence intensity in the domain nor is it the overall maximum, but
it is the last peak occurring furthest downstream before decay. This
can be seen for example in figure 9 in \cite{M&V2010}, where the
streamwise variations of the turbulence intensity both along the
centreline and along an off-centre parallel line are shown. The
turbulence intensity along this particular off-centre line peaks much
closer to the grid and at a higher intensity value than the turbulence
recorded on the centreline.

The wake-interaction length-scale can also be defined for a regular
grid, where the mesh size and the bar thickness are now the relevant
parameters, $x_{*}=M^{2}/t_{0}$. \cite{J&W1992} measured the
turbulence intensity very near the grid, $1<x/M<30$ and observed two
different regions, a highly inhomogeneous region up to $x/M\approx3$
which is a production region where the turbulence intensity increases
along a streamwise line crossing half distance between grid bars and a
decay region beyond that. Note that $x/M\approx3$ corresponds to
$x_{peak}/x_{*}\approx0.55$ close to $x_{peak}/x_{*}\approx0.45$
encountered by \cite{M&V2010} for the fractal square grids. A
qualitatively similar conclusion can be drawn from the direct
numerical simulation of turbulence generated by a regular grid
presented in \citet*{Ertunc2010}. In their figure 16 one can find the
development of the turbulent kinetic energy very close to the grid
$0.5<x/M<10$ along three straight streamwise lines located,
respectively, behind a grid bar, half-distance between bars and
in-between the other two traverses. It can be seen that the turbulence
intensity peaks first directly behind the grid bar at
$x_{peak}/M\approx 1$ and lastly behind the half-distance between grid
bars (somewhat equivalent to the centreline in the square fractal
grid) at $x_{peak}/M\approx 2.5$. This latter streamwise location
corresponds to $x_{peak}/x_{*}\approx 0.68$, once more not far from
$x_{peak}/x_{*}\approx0.45$. Note nonetheless that this simulation was
performed at very low Reynolds numbers, $Re_{\lambda}<17$, so care
must be taken in quantitative comparisons.

Note that $x_{peak}/x_{*}$ appears to be slightly higher for the
regular static grids than for the fractal square grids. This is likely
due not only to the typically low Reynolds numbers generated by the
regular grids but also to the characteristic production mechanism of
the fractal square grids, i.e. before the larger wakes interact all
the smaller wakes have already interacted and generated turbulence
that increases the growth rate of the larger wakes, thus making them
meet closer to the grid and therefore causing a smaller value of
$x_{peak}/x_{*}$.

The fact that the fractal grid has multiple wake-interaction
length-scales, for the present fractal square grid ranging from a few
centimetres to more than a meter, is precisely part of what makes the
fractal grid generate turbulence that is qualitatively different from
regular grid-generated turbulence. Consequently one could expect that
a fractal grid designed so that it produces a narrow range or a single
dominant wake-interaction length-scale, will lead to turbulence that
is similar to regular grid-generated turbulence. \cite{H&V2007}
included in their study the assessment of fractal cross grids, which
resemble regular grids but with bars of varying thicknesses. The
ratios between the thickest and the thinnest bars of their fractal
cross grids ranged from 2.0 to 3.3, thus yielding a narrow span of
wake-interaction length-scales. Furthermore, the wake interaction
pattern of the fractal cross grids, as designed and studied in
\cite{H&V2007}, is considerably different from the wake interaction
pattern of their fractal square grids. In the fractal square grids
case, the main interaction events occur when similar sized wakes meet,
whereas in the fractal cross grids the main interaction events occur
between adjacent wakes, which may or may not be of similar
size. Therefore one could expect the results obtained with fractal
cross grids, for example the power-law turbulence decay exponent, not
to be very different from the typical results found for regular
grid-generated turbulence. In fact, examining figure 10 in
\cite{H&V2007} one can see that the turbulence decays as
$(x-x_{0})^{-n}$ with $1 < n < 1.5$ for $x_{0} \approx 0$, although
they encounter a general difficulty of finding the appropriate virtual
origin. We will return to the problem of finding the appropriate
virtual origin and the power-law decay exponent in
Sec. \ref{Sec:Decay} where we present different power-law decay
fitting methods applied to our data.

\subsection{Homogeneity, isotropy and wall interference} \label{subsec:HomogeneityAndIsotropy}

\subsubsection{Homogeneity} \label{subsec:Homogeneity} 
Previous experimental investigations on the turbulence generated by
space-filling fractal square grids, \eg \cite{M&V2010}, reported that
the flow field close to the grid is highly inhomogeneous. It was also
observed that during the process of turbulent kinetic energy build up
the turbulent flow is simultaneously homogenised by turbulent
diffusion, and by the time it reaches a peak in turbulence intensity
(what they considered to be the threshold between the production and
decay regions) the flow has smoothed out most
inhomogeneities. \cite{S&V2007} measured the turbulent kinetic energy
production in various planes perpendicular to the mean flow along the
centreline and observed that the turbulent production decreases
rapidly just after the peak, i.e. where $ 0.45(\approx x_{peak}/x_{*})
<x/x_{*}<0.75$ and that the turbulent energy production typically
represents less than $30\%$ of the dissipation and never exceeds
$20\%$ beyond this region.

\cite{M&V2010} compared the characteristic time scales of the mean
velocity gradients $({\partial U\over \partial x})^{-1}$ and
$({\partial U\over \partial y})^{-1}$ (where $U$ is the streamwise
mean velocity and $y$ is a coordinate along the horizontal normal to
the streamwise direction) with the time scale associated with the
energy-containing eddies and reached the conclusion that beyond the
peak the mean gradient time scale is typically one to two orders of
magnitude larger. Consequently the small-scale turbulence dynamics are
not affected by large-scale mean flow inhomogeneities.

Here we complement the previous analyses by following the approach of
\cite{CorrsinHandbook} and \cite{CC66} and using some of their
homogeneity criteria, as they did for regular grids. The commonly
accepted 'rule-of-thumb' for the regular grids is that the turbulent
flow can be considered statistically homogeneous in transverse planes
for $x/M>30$ and the unavoidable inhomogeneity along the mean flow
direction becomes relatively unimportant for $x/M>40$
\cite[]{CorrsinHandbook}.

\begin{figure} 
\centering
\includegraphics[trim=20 5 35 20, clip=true, width=62mm]{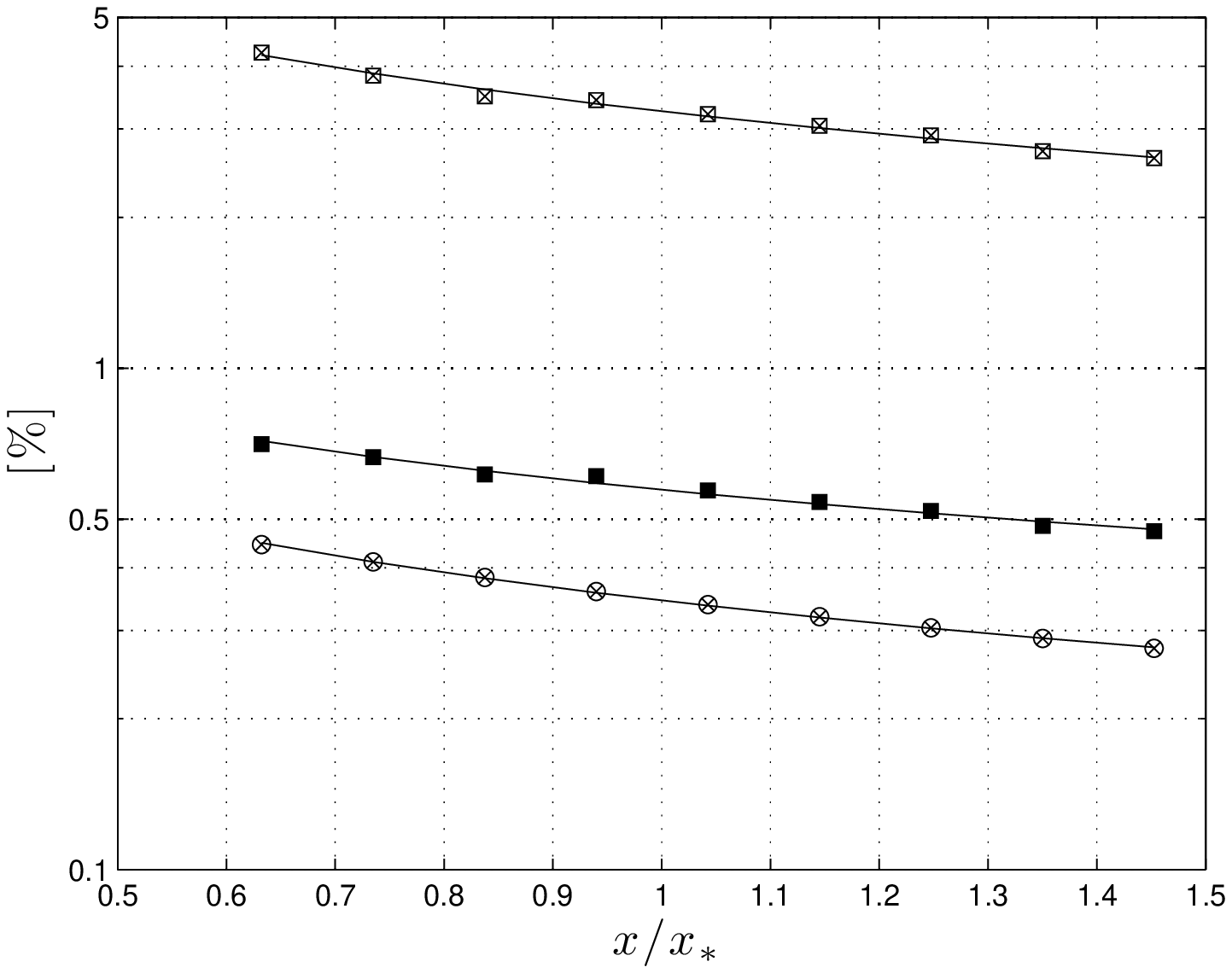} 
\includegraphics[trim=20 5 35 20, clip=true, width=62mm]{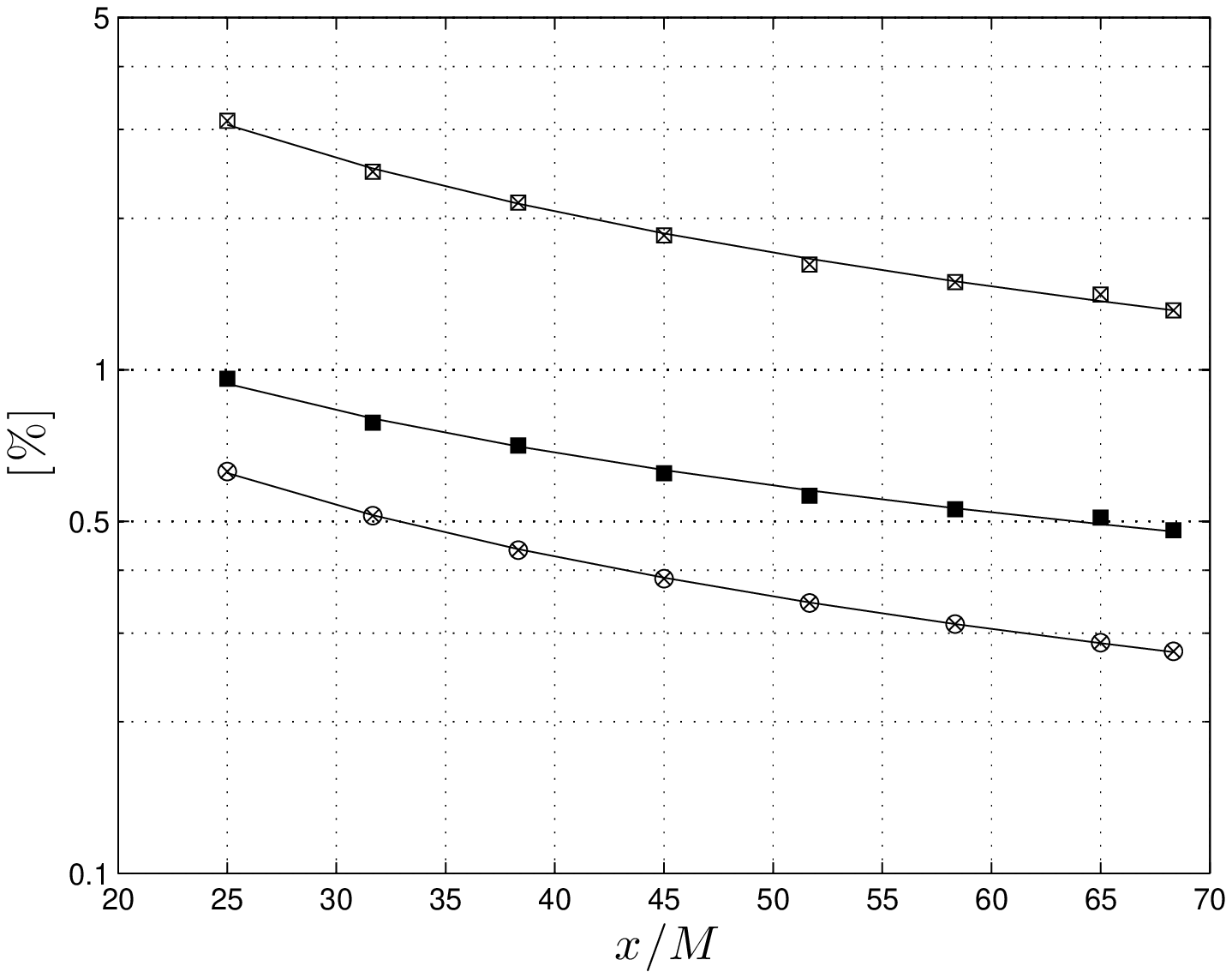} 
\includegraphics[trim=80 20 80 30, clip=true, width=130mm]{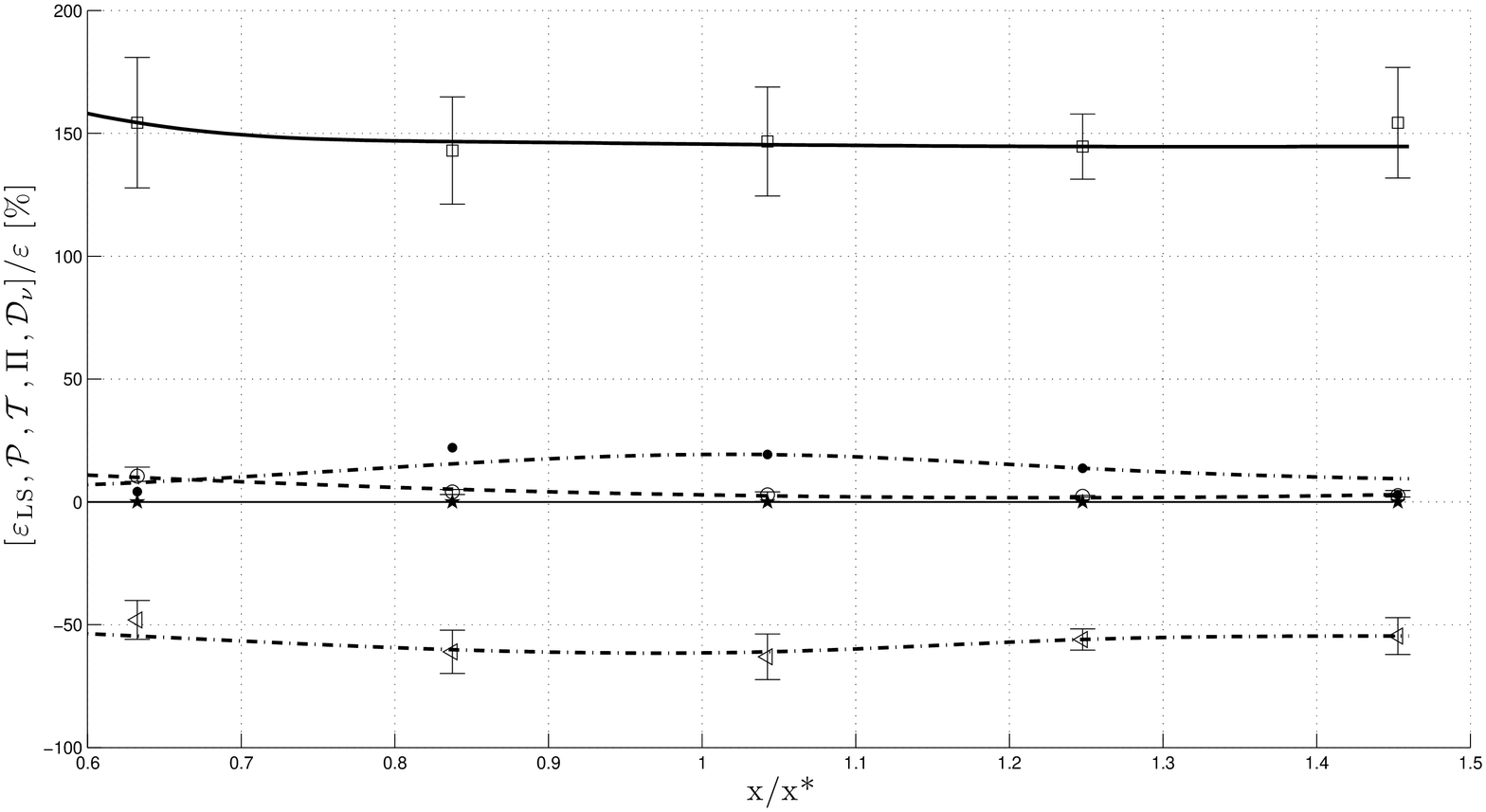}
\caption{Homogeneity assessment for the fractal square grid in the decay region around the centerline for $U_{\infty}=15ms^{-1}$. Top figures: dimensionless streamwise inhomogeneity measures (\rlap{\SmallSquare}\SmallCross) $\frac{L_{u}}{\overline{u^{2}}}\frac{\partial \overline{u^{2}}}{\partial x}$, (\FilledSmallSquare) $\frac{L_{u}}{\lambda}\frac{\partial \lambda}{\partial x}$,  (\rlap{\SmallCircle}\SmallCross)  $\frac{\partial L_{u}}{\partial x}$ for the (a) fractal square grid (SFG), (b) regular grid (RG). Bottom figure: (c) T.K.E budget \eqref{QuasiHomogeneousEq} normalised by the dissipation for the SFG at the centreline, (\SmallSquare)  $\varepsilon_{LS}$ - advection, (\SmallCircle) $\mathcal{P}$ - production, (\SmallTriangleLeft) $\mathcal{T}$ -  triple-correlation transport, (\FilledSmallCircle) $\Pi$ - pressure transport, ($\star$) $\mathcal{D}_{\nu}$ - viscous diffusion.}
\label{Fig:CorrsinHomogeneityTKEBudget}
\end{figure}

For the downstream decaying turbulence to be considered a good
approximation to spatially homogeneous decaying turbulence two
criteria must be met, (i) the eddy turn-over time $L_u/u'$ must be
small compared to the time-scale associated with the velocity
fluctuation decay rate $(\partial u'/\partial x)^{-1}$ \cite[see also Sec. 3.3 of][]{Townsend:book} and (ii)
the rate of change of the turbulent length-scales must be small
compared to the length-scales themselves. Following
\cite{CorrsinHandbook} we measure
\[ 
\frac{L_{u}}{\overline{u^{2}}}\frac{\partial
  \overline{u^{2}}}{\partial x},\hspace{2mm}
\frac{L_{u}}{\lambda}\frac{\partial \lambda}{\partial x},\hspace{2mm}
\frac{\partial L_{u}}{\partial x},
\]
and confirm that these quantities are small for the entire decay
region assessed here, i.e $x/x_{*}>0.6$ (figure
\ref{Fig:CorrsinHomogeneityTKEBudget}a) and comparable with those
obtained for a regular grid (figure \ref{Fig:CorrsinHomogeneityTKEBudget}b).
Note that the 'rule-of-thumb' $x/M>40$ suggested by \cite{CorrsinHandbook} 
was based on the streamwise location where his regular grid data yielded these dimensionless
quantities to be below 4\%, so for our regular grid data this 'rule-of-thumb' translates to $x/M>25$ and for 
our fractal square grid data to $x/x_*>0.7$.

A thorough assessment of the inhomogeneity of the flow can be made by
using the statistical equations and measuring the terms that should be
zero in a statistically homogeneous flow field. Starting with
single-point statistics, \eg the turbulent kinetic energy equation
(here $U_{1}=U$, $U_{2}=V$ \& $U_{3}=W$ denote mean flow speeds,
$u_{1}=u$, $u_{2}=v$, $u_{3}=w$ \& $p$ are zero mean fluctuating
velocities and pressure, and $x_{1}=x$, $x_{2}=y$ \& $x_{3}=z$ are the
components of a coordinate system aligned with the respective velocity
components),
\begin{align}
\frac{U_{k}}{2}\frac{\partial\, \overline{q^{2}}}{\partial x_{k}} = -
\overline{u_{i} u_{j}}\, \frac{\partial U_{i}}{\partial x_{j}} -
\frac{\partial}{\partial x_{k}}\left( \frac{\overline{u_{k} q^{2}}}{2}
+\frac{\overline{u_{k} p}}{\rho} \right) + \frac{\nu}{2}
\frac{\partial^{2} \overline{q^{2}}}{\partial x_{m} \partial x_{m}} -
\nu \overline{\frac{\partial u_{i}}{\partial x_{k}} \frac{\partial
    u_{i}}{\partial x_{k}}},
\end{align}
where use is made of Einstein's notation and $\overline{q^2}\equiv
\overline{u^2} + \overline{v^2} + \overline{w^2}$ ($K\equiv {1\over 2}
\overline{q^2}$), over-bars signifying averages over an infinite
number of realisations (here, over time).

The flow statistics inherit the grid symmetries,
i.e. reflection symmetry around the y \& z axes (as well as diagonal
reflection symmetry) and symmetry with respect to discrete
$90^{\circ}$ rotations and therefore the transverse mean velocities are negligibly
small, $V=W\approx 0$, and the turbulent kinetic energy equation at
the centreline reduces to:
\begin{equation}
\begin{aligned}
\frac{U}{2}\frac{\partial\, \overline{q^{2}}}{\partial x} =
\overbrace{-\left(\overline{u^{2}}\, \frac{\partial U}{\partial x} +
  2\overline{uv}\, \frac{\partial U}{\partial y}
  \right)}^{\mathcal{P}} & \overbrace{-\left( \frac{\partial}{\partial
    x} \frac{ \overline{u q^{2}}}{2} + 2\frac{\partial}{\partial y}
  \frac{ \overline{v q^{2}}}{2} \right)}^{\mathcal{T}}
\\ \underbrace{-\left( \frac{\partial}{\partial x}\frac{ \overline{u p}}{\rho} + 
2  \frac{\partial}{\partial y}\frac{ \overline{v p}}{\rho}\right)}_{\Pi} & \underbrace{+\frac{\nu}{2}
  \left( \frac{\partial^{2} \overline{q^{2}}}{\partial x^{2}} +
  2\frac{\partial^{2} \overline{q^{2}}}{\partial y^{2}}
  \right)}_{\mathcal{D}_{\nu}} \underbrace{-\nu
  \overline{\frac{\partial u_{i}}{\partial x_{k}} \frac{\partial
      u_{i}}{\partial x_{k}}}}_{\varepsilon},
\label{QuasiHomogeneousEq}
\end{aligned}
\end{equation}
where $\mathcal{P},\,\mathcal{T},\,\Pi,\,\mathcal{D}_{\nu}$ and
$\varepsilon$ are the production, triple-correlation transport,
pressure transport, viscous diffusion and dissipation terms
respectively.

Data from both single- and cross-wire measurements are used to
estimate all the terms in \eqref{QuasiHomogeneousEq} (except the
pressure-velocity correlations) along the centreline in the decay
region for $U_{\infty}=15ms^{-1}$ (see table
\ref{Table:HomogeneityParameters}). The pressure transport is
indirectly estimated from the balance of
\eqref{QuasiHomogeneousEq}. The last term in
\eqref{QuasiHomogeneousEq} is evaluated assuming isotropy: for the
single-wire measurements $\varepsilon_{iso}^{SW}\equiv
15\nu\overline{\left(\partial u /\partial x \right)}$; for the
cross-wire measurements one can impose one less isotropy constraint
and estimate $\varepsilon$ from $\varepsilon_{iso}^{XW}\equiv
3\nu\overline{\left(\partial u /\partial x
  \right)}+6\nu\overline{\left(\partial v /\partial x \right)}$
\citep*{SSG73}. It should be noted that the separation between the
cross-wires is about 1mm and is almost 10 times the Kolmogorov
length-scale so caution should be taken interpreting the direct
measurements of dissipation using the cross-wires as they may be
underestimated. On the other hand the isotropic estimate of the
dissipation using single-wire measurements is likely to be
overestimated since we show that $\overline{(dv/dx)^2} /
\overline{(dv/dx)^2}<2$. In figure
\ref{Fig:CorrsinHomogeneityTKEBudget}c the mean between the single-
and cross-wire dissipation estimates is used as the normalising
quantity and the error (taken as the difference between the two
estimates) contributes to the error bar of the normalised quantities.
The advection $1/2\, U \partial \overline{q^{2}}/\partial x$ is estimated from the non-linear
least-squares power law fit of $\overline{q^{2}}$ (see
Sec. \ref{Sec:Decay} for further details) and $\overline{q^{2}}$ is
estimated as $\overline{q^{2}}=\overline{u^{2}}( 1 +
2\overline{v^{2}}/\overline{u^{2}} )$ with $\overline{u^{2}}$ from the
single-wire data and $\overline{v^{2}}/\overline{u^{2}}$ from the
cross-wire data; for the advection as well, the error is taken to be
the difference between the single-wire (no anisotropy correction) and
cross-wire estimate. The ratio between advection and dissipation can
be seen (figure \ref{Fig:CorrsinHomogeneityTKEBudget}c) not to be
unity but tending to be approximately 1.5 beyond $x/x_*\approx0.8$; we
will return to this issue at the end of this subsection and in
Sec. \ref{Sec:Decay} where we estimate the decay rate of our
turbulence.

The longitudinal production terms are calculated from the single wire
data (finer streamwise resolution), whereas the transverse production
terms are estimated using the cross-wire spanwise traverse data. The
latter contribution $\overline{uv} \,\partial U/\partial y$ is
approximately zero at the centreline (due to the reflexion symmetry),
so it is preferred to estimate it just off the centreline around
$y\pm10mm \approx L_u/5 \approx L_v/2$, to infer on its contribution
in this region of the flow. The total contribution from the production
terms around the centreline can be seen (figure
\ref{Fig:CorrsinHomogeneityTKEBudget}c) to be less than $10\%$ of the
estimated dissipation (in agreement with \cite{S&V2007}) and beyond
$x>x_{*}$ they become negligible (there is a residual production of
2-4\% of the dissipation due to non-vanishing streamwise mean velocity
gradients). The viscous diffusion, as expected, is always negligibly
small (table \ref{Table:HomogeneityParameters}). The longitudinal
triple-correlation transport (table \ref{Table:HomogeneityParameters})
shows a trend not dissimilar to that of the production terms, it is
less than $10\%$ closer to the kinetic energy peak ($x/x_* < 0.8$) and
becomes vanishingly small beyond $x>x_{*}$.

\begin{figure} \centering
\includegraphics[trim=1mm 0mm 10mm 0mm, clip=true,width=67mm]{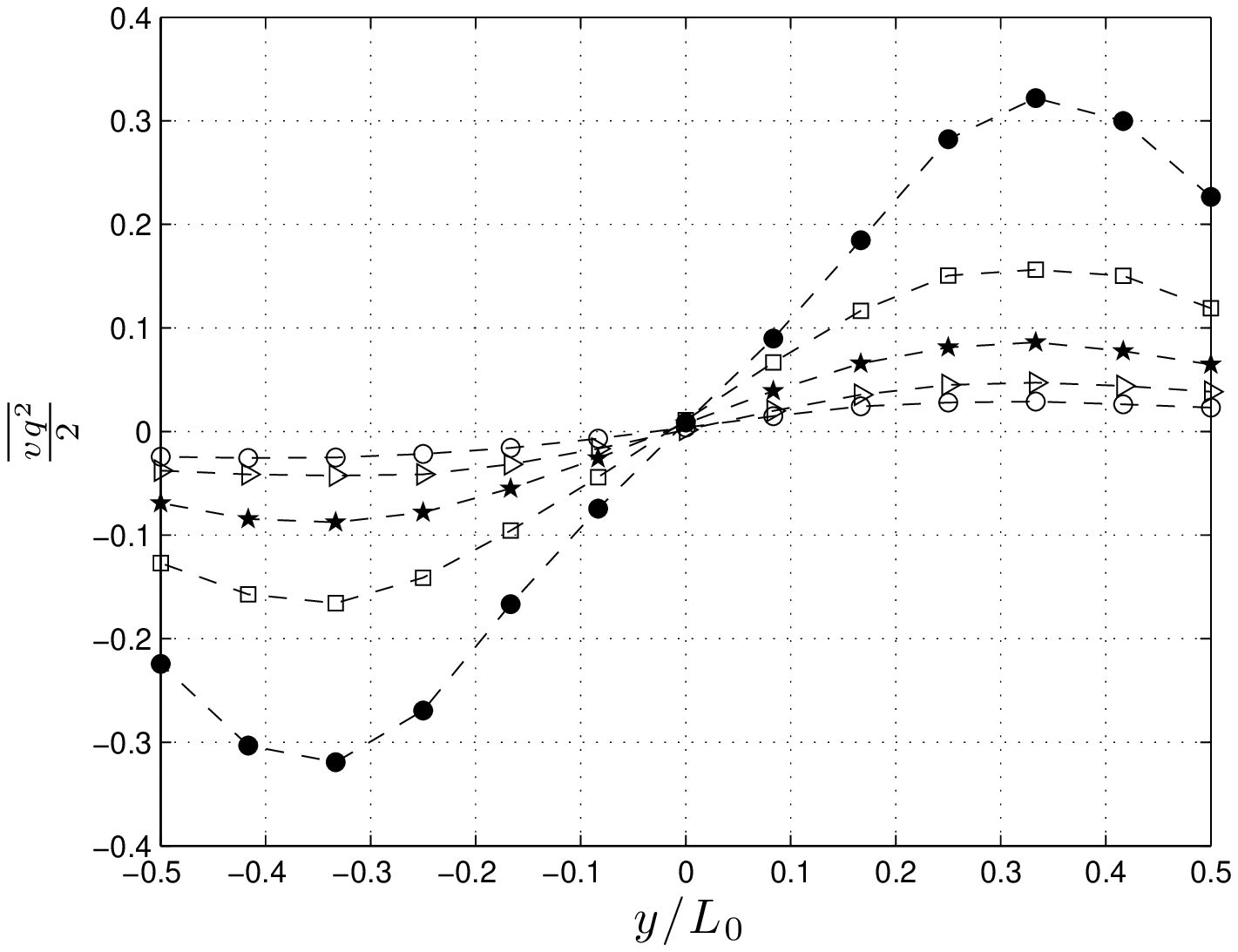}
\includegraphics[trim=1mm 0mm 10mm 0mm, clip=true,width=67mm]{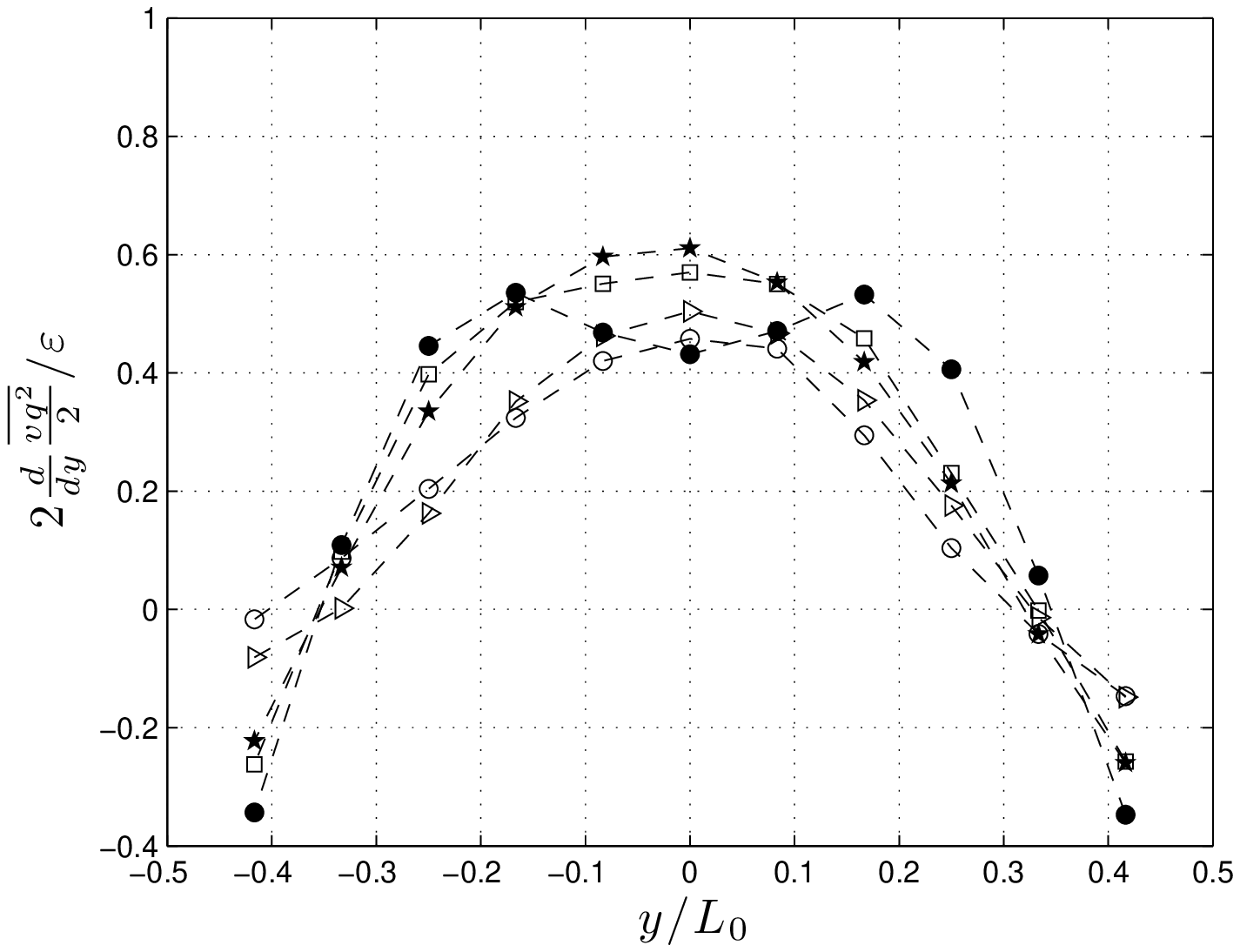}
\caption{Transverse profiles of (a) $\frac{\overline{vq^2}}{2}$, (b) $2\frac{d}{dy}\frac{\overline{vq^2}}{2}/\varepsilon$ at five streamwise downstream positions and $U_{\infty}=15ms^{-1}$: (\FilledSmallCircle) $x/x_{\star}=0.63$, (\SmallSquare) $x/x_{\star}=0.83$,  ($\star$) $x/x_{\star}=1.04$, (\SmallTriangleRight)  $x/x_{\star}=1.24$, (\SmallCircle) $x/x_{\star}=1.44$ }
\label{vq2} \end{figure}

The transverse triple-correlation transport was assessed by measuring
the triple correlation $\overline{vq^2}/2$ (figure \ref{vq2}a) along
the vertical symmetry plane of the grid ($z=0$) for the five
streamwise downstream locations specified in table
\ref{Table:HomogeneityParameters}. The transverse measurements ranged
from the lower to the upper largest bars of the fractal grid ($ -120mm <
y < 120\,mm$) and were recorded with a spacing of $20mm$. The total
transverse triple-correlation transport $d\overline{vq^2}/dy$
(i.e. twice the transport at each transverse direction $y$ and $z$) decreases 
together with the dissipation and not faster as the other measured inhomogeneity terms
(figure \ref{Fig:CorrsinHomogeneityTKEBudget}c). It typically amounts to
40-60\% of the dissipation (at the centreline) and perhaps
surprisingly, it stays nearly the same fraction for all the
assessed decay region. This seems to be the case not only along the
centreline but for all the transverse measurement locations as well
(figure \ref{vq2}b), although the ratio between the transport and
dissipation are different for different $y$ locations and can if fact
be zero and negative (at $y/L_0 \approx 0.35$ and beyond that
respectively) .

In Sec. \ref{Sec:Decay} we argue that this persistent
spanwise energy transport has no significant effect on the power law
exponent of the turbulence energy decay because the dissipation and
the lateral transport remain roughly proportional throughout the part
of the decay region explored here.

\begin{table}
\caption{Turbulence statistics for five streamwise locations recorded at $U_{\infty}=15ms^{-1}$.} \vspace{2mm}
\centering
\begin{tabular*}{0.9\textwidth}{@{\extracolsep{\fill}}rccccc}
Position(mm) & 1850  &  2450  &  3050  &  3650 & 4250 \\
\midrule
$x/x_{*}$ &   0.63 & 0.83 & 1.04 & 1.24 & 1.44 \vspace{1mm} \\
$Re_{\lambda}$ & 352 & 292 & 253 & 226 & 210  \vspace{1mm} \\
$\sqrt{\overline{u^{2}}}$ [ms$^{-1}$] & 1.28 & 0.99 & 0.79 & 0.65 & 0.56 \vspace{1mm} \\ 
$L_{u}$     [mm] & 45.7 & 47.6 & 50.0 & 50.7 & 53.6 \vspace{1mm}\\
$\lambda$ [mm] & 4.1   & 4.4  & 4.8    & 5.2 & 5.6 \vspace{1mm}\\
$\eta$        [mm] & 0.11 & 0.13& 0.15 & 0.18& 0.2 \vspace{1mm}\\
$\varepsilon_{LS} \equiv -\frac{U}{2}\frac{\partial}{\partial x} \overline{u^{2}} \left(1 + 2\frac{\overline{v^{2}}}{\overline{u^{2}}} \right)$ [m$^{2}$s$^{-3}$] & 28.6 & 13.9 & 7.7 & 4.7 & 3.0\vspace{1mm} \\
$\varepsilon_{iso}^{SW}\equiv 15\nu \overline{\left( \frac{\partial u}{\partial x} \right)^{2}}$ [m$^{2}$s$^{-3}$] & 21.6 & 11.1 & 6.0 & 3.5 & 2.2 \vspace{1mm}\\
$\varepsilon_{iso}^{XW}\equiv 3\nu \overline{\left( \frac{\partial u}{\partial x} \right)^{2}}+6\nu \overline{\left( \frac{\partial v}{\partial x} \right)^{2}}$ [m$^{2}$s$^{-3}$] &16.6 & 8.7 & 4.7 & 3.1 & 1.7 \vspace{1mm}\\
-$\overline{u^{2}} \frac{\partial U}{\partial x}$ [m$^{2}$s$^{-3}$] & 0.62 & 0.27 & 0.11 & 0.07 & 0.05 \vspace{1mm} \\
$-2\overline{uv} \frac{\partial U}{\partial y}$ [m$^{2}$s$^{-3}$] &1.32 & 0.12 & 0.04 & 0.004 & 0.0007 \vspace{1mm} \\
$\frac{\partial}{\partial x} \frac{\overline{uq^{2}}}{2}$ [m$^{2}$s$^{-3}$] &0.69 & 0.39 & 0.06 & -0.01 & -0.004 \vspace{1mm}\\
$2\frac{\partial}{\partial y} \frac{\overline{vq^{2}}}{2}$ [m$^{2}$s$^{-3}$] &8.22 & 5.53 & 3.25 & 1.83 & 1.08    \vspace{1mm}\\
$\nu\frac{\partial^{2} \overline{q^{2}}}{\partial x^{2}}${\scriptsize $(\times 10^{5})$} [m$^{2}$s$^{-3}$] & 4.23 & 1.72 & 0.81 & 0.42 & 0.024 \vspace{2mm} \\
$-\nu\frac{\partial^{2} \overline{q^{2}}}{\partial y^{2}}${\scriptsize $(\times 10^{3})$} [m$^{2}$s$^{-3}$] & 3.24 & 1.55 & 1.29 & 1.08 & 0.45 \vspace{2mm} \\
$\sqrt{\overline{u^{2}}/\overline{v^{2}}}$ & 1.15 & 1.13 & 1.11 & 1.13 & 1.10 \vspace{1mm} \\ 
$\frac{\overline{(dv/dx)^{2}}}{\overline{(du/dx)^{2}}}$& 1.39 & 1.40 & 1.42 & 1.44 & 1.46 \\
$L_{u}/L_{v}$  & 3.7 & 3.2 & 3.0 & 2.7 & 3.0 \\
\end{tabular*}
\label{Table:HomogeneityParameters}
\end{table}

\subsubsection{Isotropy} \label{Sec:Isotropy}
The simplest assessment of large-scale anisotropy is achieved by
comparing the ratio of streamwise and transverse r.m.s. velocity
components, sometimes referred to as isotropy factor. The results of
such measurements at the centreline are presented in table
\ref{Table:HomogeneityParameters} and show a fair agreement with
\cite{H&V2007} for the same set-up, confirming that the flow is
reasonably isotropic for all the assessed decay region,
$u'/v'\approx1.1-1.25$. The range of isotropy factors encountered in our
flow are comparable to those obtained by \cite{M&W1996} for their
active grids, although further research shows it is possible to tune
the active grid to decrease the anisotropy of the flow
\citep*{KCM02}. Similarly it should be possible to further optimise
the design of the fractal grids to increase isotropy, \eg by
increasing the thickness ratio as is suggested by the data presented
by \cite{H&V2007}.
\cite{H&V2007} also reported the ratio between the longitudinal and
transversal integral length-scales ($L_{u}$ and $L_{v}$) for the same
low-blockage space-filling fractal square grid to be
$L_{u}/L_{v}\approx 2$, but this is not confirmed by the present data
where the integral scales ratio is larger than 2 as shown in table
\ref{Table:HomogeneityParameters}, even though this ratio decreases
further downstream. This discrepancy is likely due to the calculation
method of the transversal integral scales; integrating the transverse
correlation function to the first zero crossing as \cite{H&V2007} (incorrectly) did
we recover an integral scale ratio closer to 2.

\begin{figure} 
\centering
\includegraphics[trim=10 0 30 18, clip=true,
  height=50mm, width=65mm]{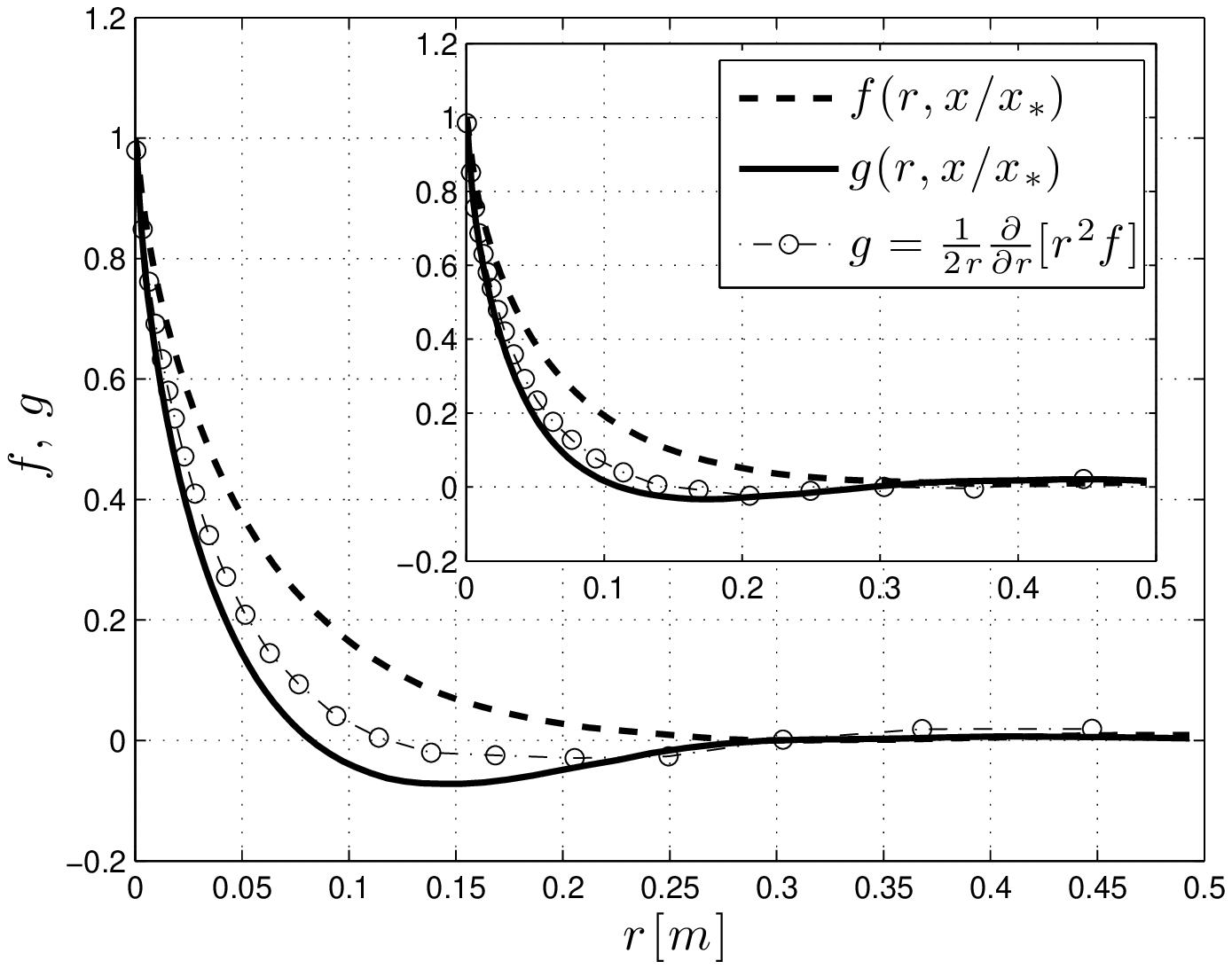}
\includegraphics[trim=10 5 30 18, clip=true,
  height=50mm,width=65mm]{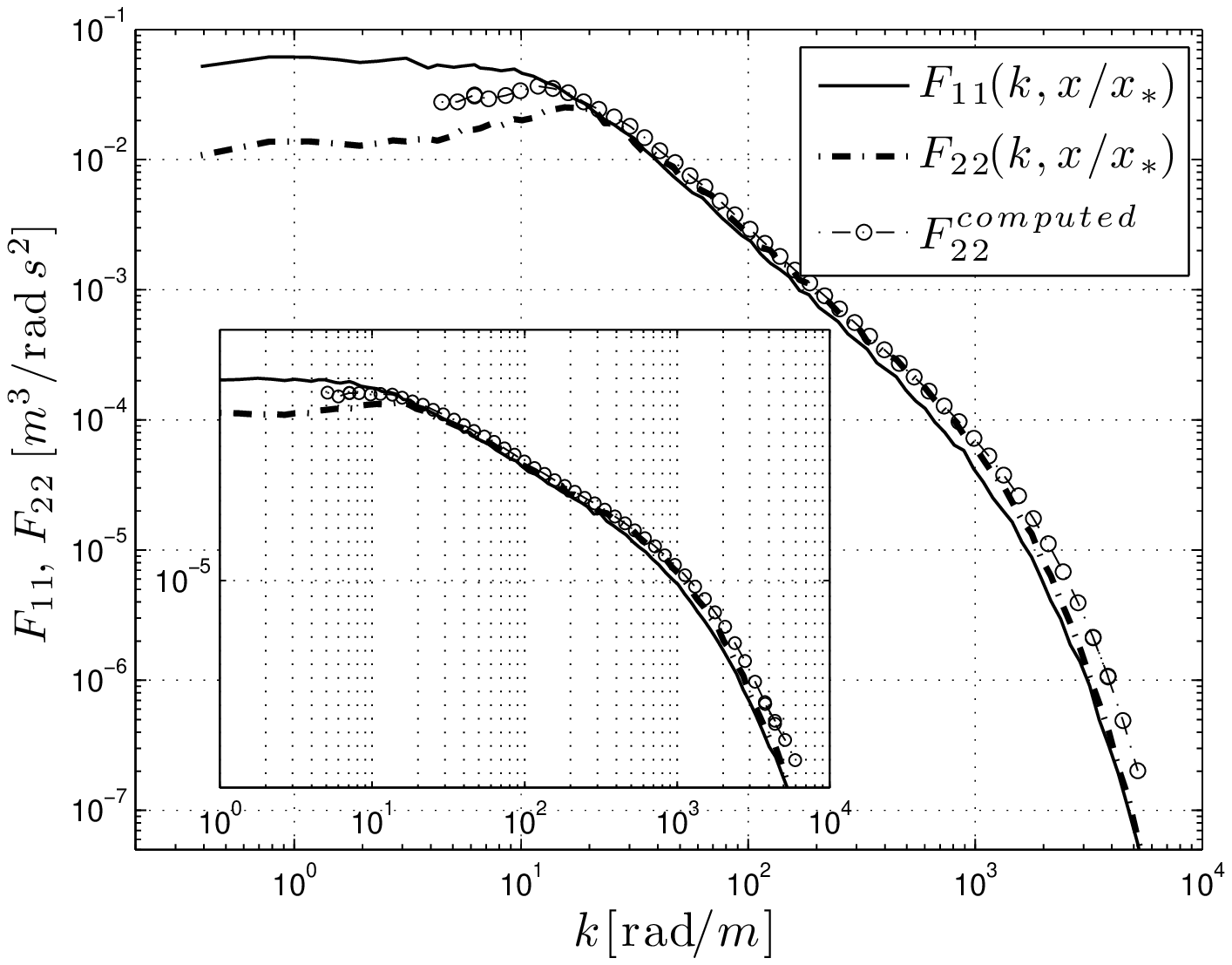}
\includegraphics[trim=15 0 30 20, clip=true, width=65mm,
  height=50mm]{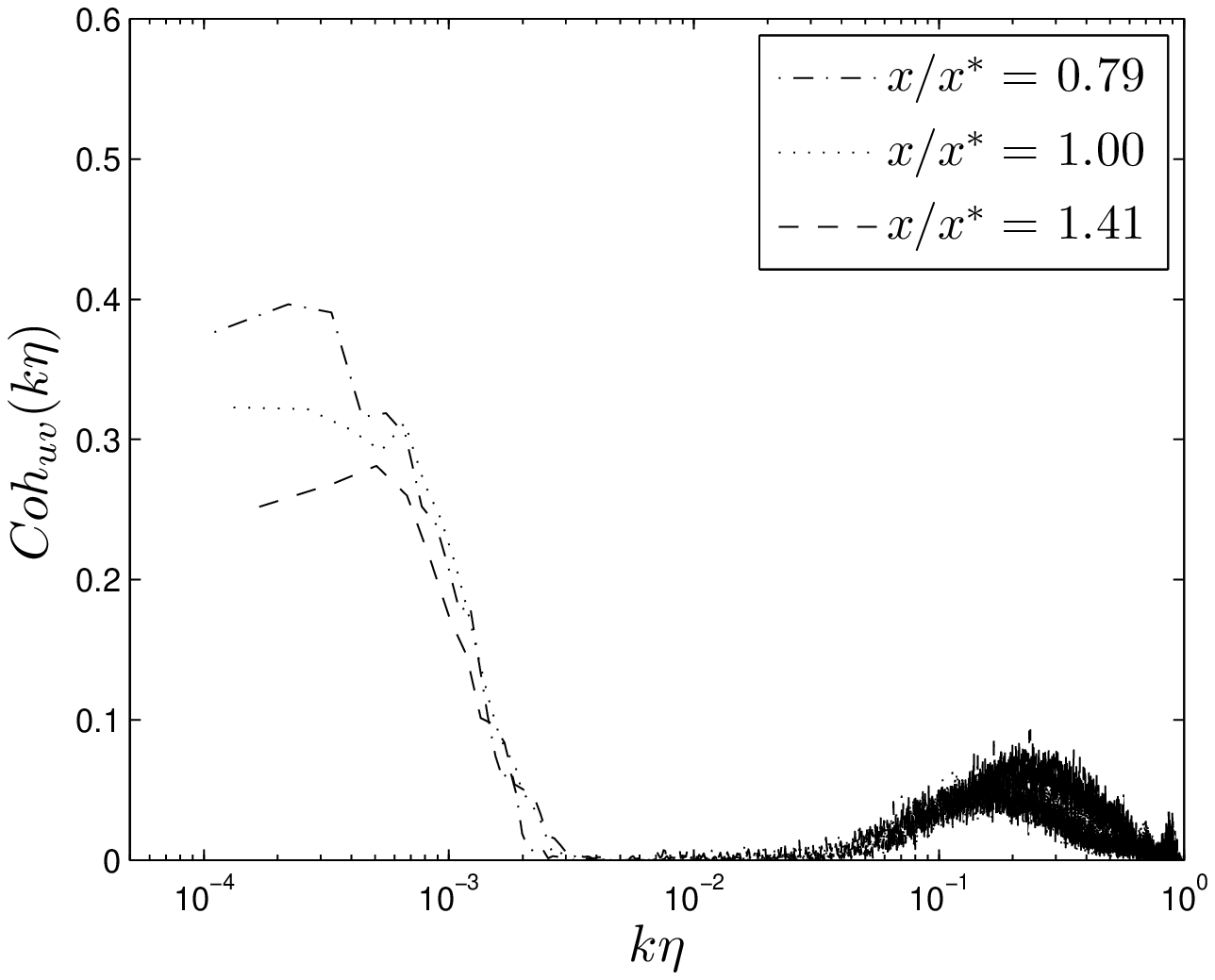}
\caption{Longitudinal and transversal one-dimensional (a) correlation
  function (b) energy spectra for $x/x_{*}=0.63$ (and for $x/x_{*}=1.4$ in
  the insert); (c) coherence spectra in the coordinate system rotated
  $45^{\circ}$ with respect to the flow direction. $U_{\infty}=15ms^{-1}$.}
\label{Fig:IsoCorrSpectra}
\end{figure}

A complementary assessment of isotropy is obtained by computing the
longitudinal and transversal correlation functions, $f(r,x)\equiv
\overline{u(x)u(x+r)}/\overline{u(x)^{2}}$ and $g(r,x)\equiv
\overline{v(x)v(x+r)}/\overline{v(x)^{2}}$ , and comparing $g(r,x)$
with $g^{iso}(r)=\frac{1}{2r}d[r^{2}f(r)]/dr$ which is the relation
between the two correlation functions in the presence of isotropy. The
comparison is shown in figure \ref{Fig:IsoCorrSpectra}a for two
downstream locations and it can be seen that there is a modest
agreement between the measured and computed transverse correlation
functions, although the agreement improves downstream. A similar
comparison in spectral space is shown in figure
\ref{Fig:IsoCorrSpectra}b, where the isotropic relation between the
longitudinal and transversal one-dimensional spectra is:
$F_{22}^{iso}=\left[F_{11} + k_{1} d F_{11}/dk_{1} \right]/2$. There
is a fair agreement between the measured and computed transverse
one-dimensional spectra in the 'inertial region', but not at the low
wave-numbers (which is consistent with $L_{u}/L_{v}>2$) nor at high
wave-numbers (reflecting that
$\overline{(dv/dx)^{2}}/\overline{(du/dx)^{2}}<2$). This lack
of small-scale isotropy was not reported by \cite{S&V2007} nor by
\cite{M&W1996} in their active-grid experiments because they filtered
out the highest frequencies where their cross wire measurements could
not be trusted. Note that in agreement with the latter experiments the
coherence spectra (figure \ref{Fig:IsoCorrSpectra}c) show that the
anisotropy (inferred by the cross-correlation of the velocity
components in a coordinate system rotated by $45^{\circ}$) is mostly
contained in the large scales. The cause for this, perhaps apparent, small-scale
anisotropy in figure \ref{Fig:IsoCorrSpectra} and in our values of
$\overline{(dv/dx)^2}/\overline{(du/dx)^2}$ in table
\ref{Table:HomogeneityParameters} is most probably the separation
between the cross-wires ($\approx 1mm$) being up to ten times the
Kolmogorov length-scale. It should be noted that, precisely because of
this problem, the velocity derivative ratios in \cite{S&V2007} were
obtained for a low-pass filtered velocity signal at $k\eta\approx
0.1$. This way, these authors obtained
$\overline{(dv/dx)^2}/\overline{(du/dx)^2}\simeq 2$ even though
strictly speaking
$\overline{(du_{i}/dx)^{2}}=\int_0^{\infty}\!k_{1}^{2}F_{ii}(k_{1})\,dk_{1}$,
where contributions coming from $k\eta>10^{-1}$ cannot necessarily be
written off as negligible.

\subsubsection{Wind-tunnel confinement} \label{Sec:WTConf}
\begin{figure} 
\centering
\vspace*{10mm}
\includegraphics[trim=0 0 1 0, clip=true, angle=90,width=140mm]
{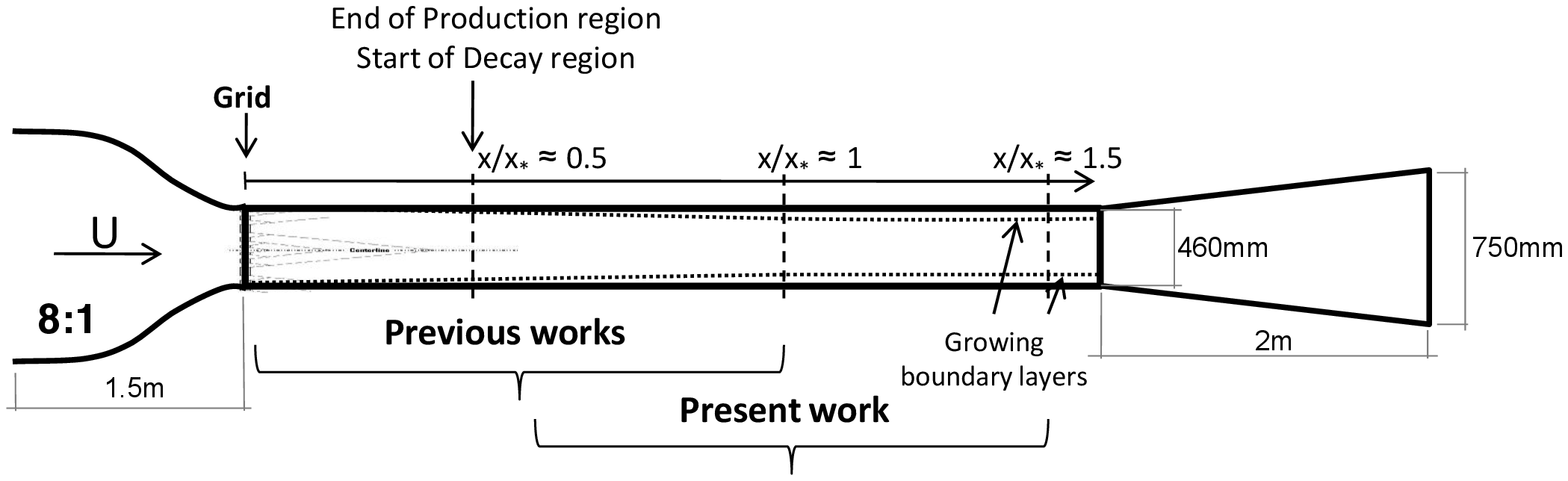} 
\caption{Sketch of the wind-tunnel where the decay of turbulence
  generated by regular and fractal square grids was measured. This
  wind-tunnel is a modified version, with an extended test section, of
  the wind-tunnel used by \cite{H&V2007}, \cite{S&V2007} and
  \cite{M&V2010} in their experimental investigations of fractal
  generated turbulence. The boundary layers developing at the wall
  were estimated to have a displacement thickness of $\delta_1\approx
  4mm$ at $x=2m$ ($x/x_{*}=0.7$), $\delta_1\approx 8mm$ at $x=3.5m$
  ($x/x_{*}=1.2$) and $\delta_1\approx 10mm$ at $x=4.5m$
  ($x/x_{*}=1.5$).}
\label{Fig:WTSketch}
\end{figure}

A qualitative assessment of the effect of flow confinement in
wind-tunnel experiments can be made by comparing the tunnel's
height/width with the flow's integral scale and comparing the ratio
with similar experiments and with DNS. For
simplicity we take the longitudinal integral scale\footnote{For an
  isotropic flow the longitudinal integral-scale and the one obtained
  using the 3D energy spectrum (
  $L=\pi/\overline{u^{2}}\int_0^{\infty}\!E(k,t)/k\, dk $) coincide}
at the centreline to be the representative scale for each transverse
section and it is typically 8.5 to 10 times smaller than the
wind-tunnel width. This is just about in-line with what is typically
used in DNS of decaying homogeneous turbulence 
\cite*[]{Kaneda2006,Wang&George2002}, considering the
boundary-layers on the wind-tunnels walls which reduce the effective
transverse size of the tunnel down to 8 times the integral scale
(based on the displacement thickness of the boundary-layers) very far
downstream. The active-grid experiments by \cite{M&W1996} were
performed at equivalent $Re_{\lambda}$ in a similar sized wind-tunnel
and produced larger integral-scales\footnote{Note \cite{M&W1996} used
  a different definition of integral-scale, but \cite{G&G2000} used
  the same data to extract the integral-scale as defined here.} but
were in line with typical decay properties and did not observe any of
the outstanding features of our flow reported in the Subsections
\ref{Sec:Ceps}, \ref{Sec:Decay} and \ref{subsec:Collapse} below. It is
therefore unlikely that our results, namely the abnormally high decay
exponent and the proportionality between the integral and the Taylor
micro-scale, may be due to confinement. However it is conceivable that
the effective choking of the tunnel by the growing boundary layers
very far downstream does have some effect on the larger turbulence
scales at these very far distances (see figure \ref{Fig:WTSketch}).

\subsection{Normalised energy dissipation rate} \label{Sec:Ceps}

\begin{figure} 
\centering
\includegraphics[trim=20 5 30 20, clip=true,
  width=90mm]{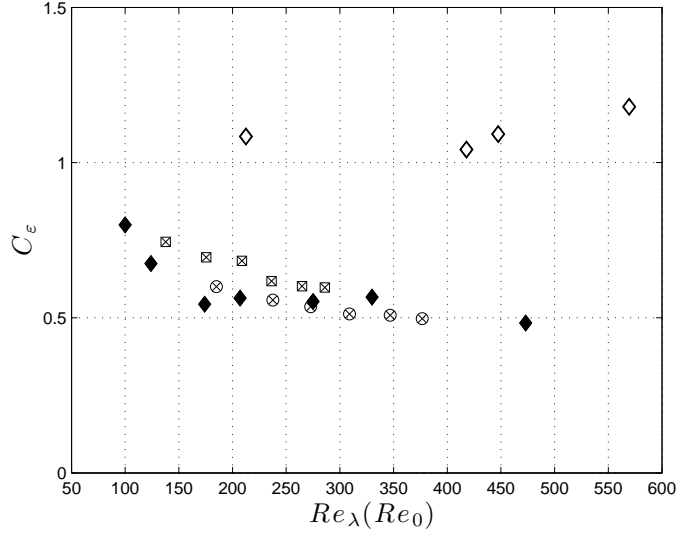}
\caption{Normalised energy dissipation rate $C_{\varepsilon}$ versus
  $Re_{\lambda}$ with $Re_{\lambda}$ changing as a function of the
  inlet Reynolds number $Re_{0}$ for a fixed streamwise downstream
  location for fractal square grid-, active grid- and regular
  grid-generated turbulence. For the fractal square grid data the
  inlet Reynolds number is changed by varying the free-stream speed
  between $5ms^{-1} < U_{\infty} < 17.5 ms^{-1}$ and is measured with a 
  $l_{w}=0.45mm$ sensing length single-wire at
  two streamwise downstream positions:
  (\rlap{\SmallCircle}\SmallCross) $x/x_*=0.63$ and
  (\rlap{\SmallSquare}\SmallCross)
  $x/x_*=1.04$. (\FilledSmallDiamondshape) Active grid data is taken
  from table 1 of \cite{G&G2000} which is based on the experimental
  data by \cite{M&W1996} (\cite{G&G2000} computed the longitudinal and
  the transverse integral scales from the spectra, but their latter estimate
  yielded less scatter, hence we assume isotropy and use twice the transverse integral
  scale). (\SmallDiamondshape) Regular grid data from the data
  compilation by \cite{Sreeni84}, figure 1 (only data by
  \cite{Kistler} is used since no other experiment with more than one
  data point had $Re_{\lambda}>100$). }
\label{Fig:CepsRe0}
\end{figure}

It follows from this paper's introduction that for fully-developed
turbulence generated by at least some space-filling low blockage
fractal square grids, the normalised energy dissipation rate
$C_{\varepsilon}$ depends both on an initial conditions/global
Reynolds number $Re_{0}$ (e.g. $Re_{0} \equiv U_{\infty} x_{*}/\nu$)
and on a local Reynolds number ($Re_{\lambda}(x)$). This distinction
between two different Reynolds number dependencies follows from
equations (1.5) and (1.7) and does not need to be made in the context
of the Richardson-Kolmogorov phenomenology where the functions $A$ and
$B$ are identical and the exponents $\alpha$ and $\beta$ are both
equal to $1/2$.

The present measurements of the normalised energy dissipation rate
$C_{\varepsilon}$ for different $Re_{0}$ (by varying $U_{\infty}$) at
two fixed streamwise downstream positions from the fractal grid
(figure \ref{Fig:CepsRe0}) suggest that $C_{\varepsilon}(Re_0)$ is
roughly constant beyond $Re_{\lambda}(Re_0) \approx 200$ (figure
\ref{Fig:CepsRe0}). From (1.7), this observation implies that, at high
enough values of $Re_{0}$, $\alpha = \beta$ and
\begin{align}
C_{\varepsilon} = 15 A({x-x_{0}\over x_{*}})/B({x-x_{0}\over x_{*}})
\label{Eq:what?what?}
\end{align}
irrespetive of $Re_{0}$. The facts that $A$ is a slow-varying whereas
$B$ is fast varying function of ${x-x_{0}\over x_{*}}$ is reflected in
the steep increase of $C_{\varepsilon}$ with $x$ (see figure 7a). This
is fundamentally different from the cornerstone assumption that
$C_{\varepsilon}$ is constant, an assumption which is approximately
verified by the turbulence generated by our regular grid provided
$Re_0$ is large enough (see figure 7a).

The high $Re_0$ behaviour of $C_{\varepsilon}(Re_{0})$ is very
comparable to that found with regular and active-grids at similar
Reynolds numbers (figure \ref{Fig:CepsRe0}) and more generally with
other boundary-free turbulent flows such as various wakes \cite[see
  e.g.][]{Burattini2005, Pearson} and DNS of forced stationary
homogeneous turbulence \cite[see data compilations by][]{Sreeni98,
  Burattini2005}. However, the fundamental difference with the present
fractal square grid-generated turbulence is that the $C_{\varepsilon}$
asymptote for high $Re_0$ is different for different streamwise
downstream locations. This is high Reynolds number
non-Richardson-Kolmogorov behaviour

\begin{figure} 
\centering
\includegraphics[trim=20 5 35 20, clip=true,
  width=65mm]{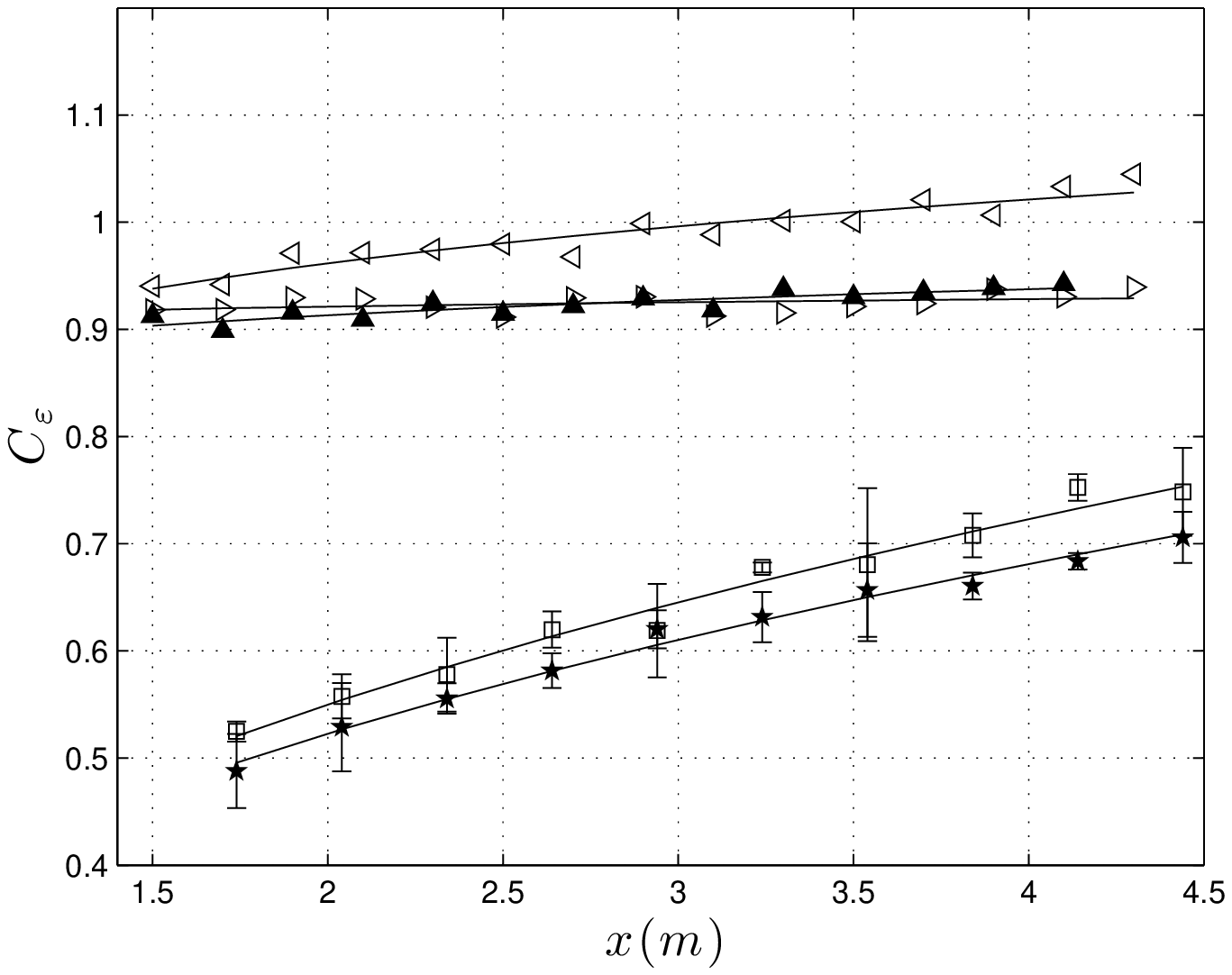}
\includegraphics[trim=20 5 35 20, clip=true,
  width=65mm]{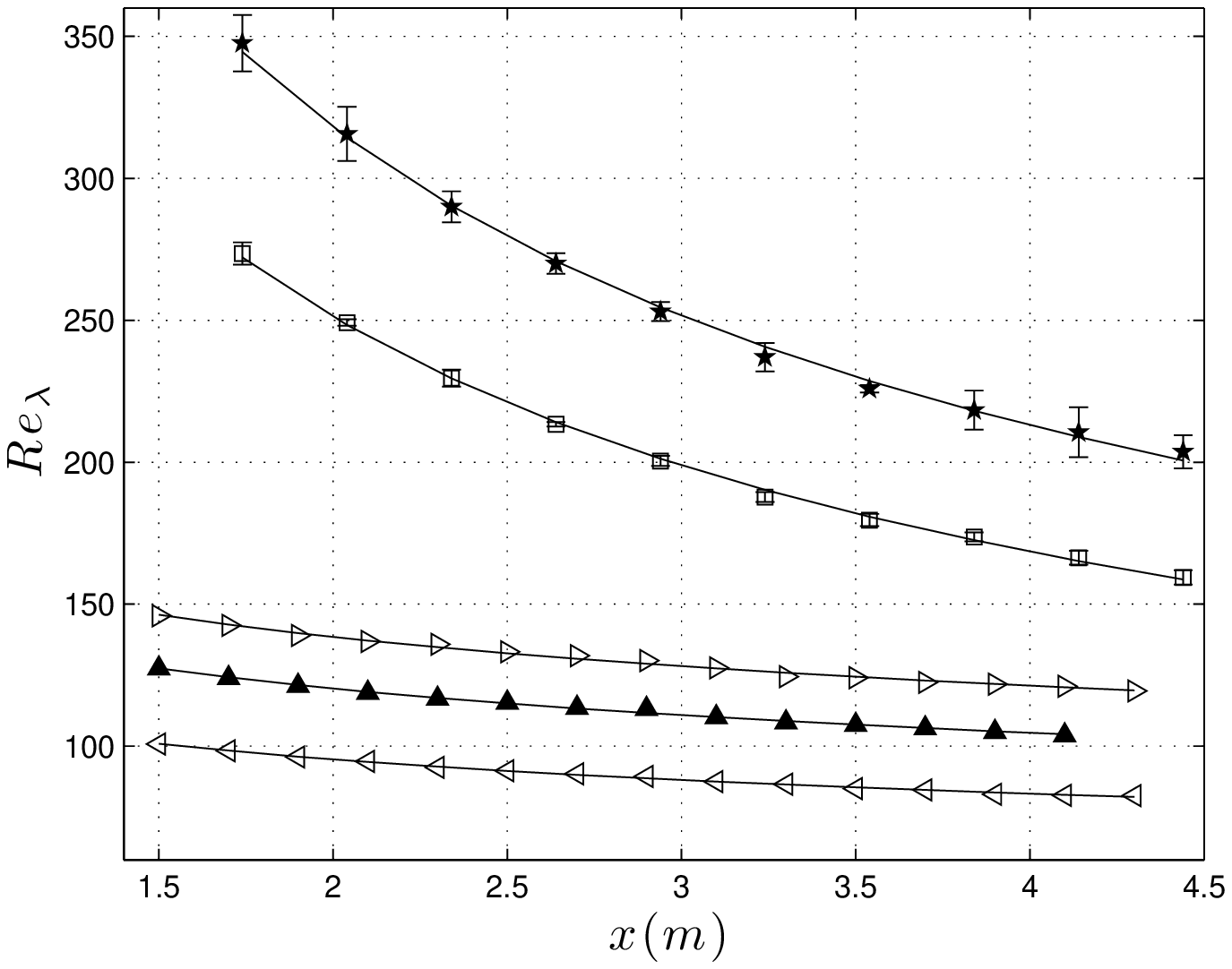}
\caption{Normalised energy dissipation rate $C_{\varepsilon}$ and
  Reynolds number $Re_{\lambda}$ versus streamwise downstream location
  $x$ for both the square fractal grid (SFG) and regular grid (RG)
  data recorded at different inlet velocities: (\SmallSquare) SFG at
  $U_{\infty}=10ms^{-1}$, ($\star$) SFG at $U_{\infty}=15ms^{-1}$,
  (\SmallTriangleLeft) RG at $U_{\infty}=10ms^{-1}$,
  (\FilledSmallTriangleUp) RG at $U_{\infty}=15ms^{-1}$,
  (\SmallTriangleRight) RG at $U_{\infty}=20ms^{-1}$. Since the square
  fractal grid data is acquired with three different wire resolutions
  (see Sec. \ref{Subsec:Hardware&Wires}) we plot the arithmetic mean plus
  error bars.}
\label{Fig:CepsAndReLambdavsX}
\end{figure}

\begin{figure} 
\centering
\includegraphics[trim=20 5 25 20, clip=true, width=120mm]{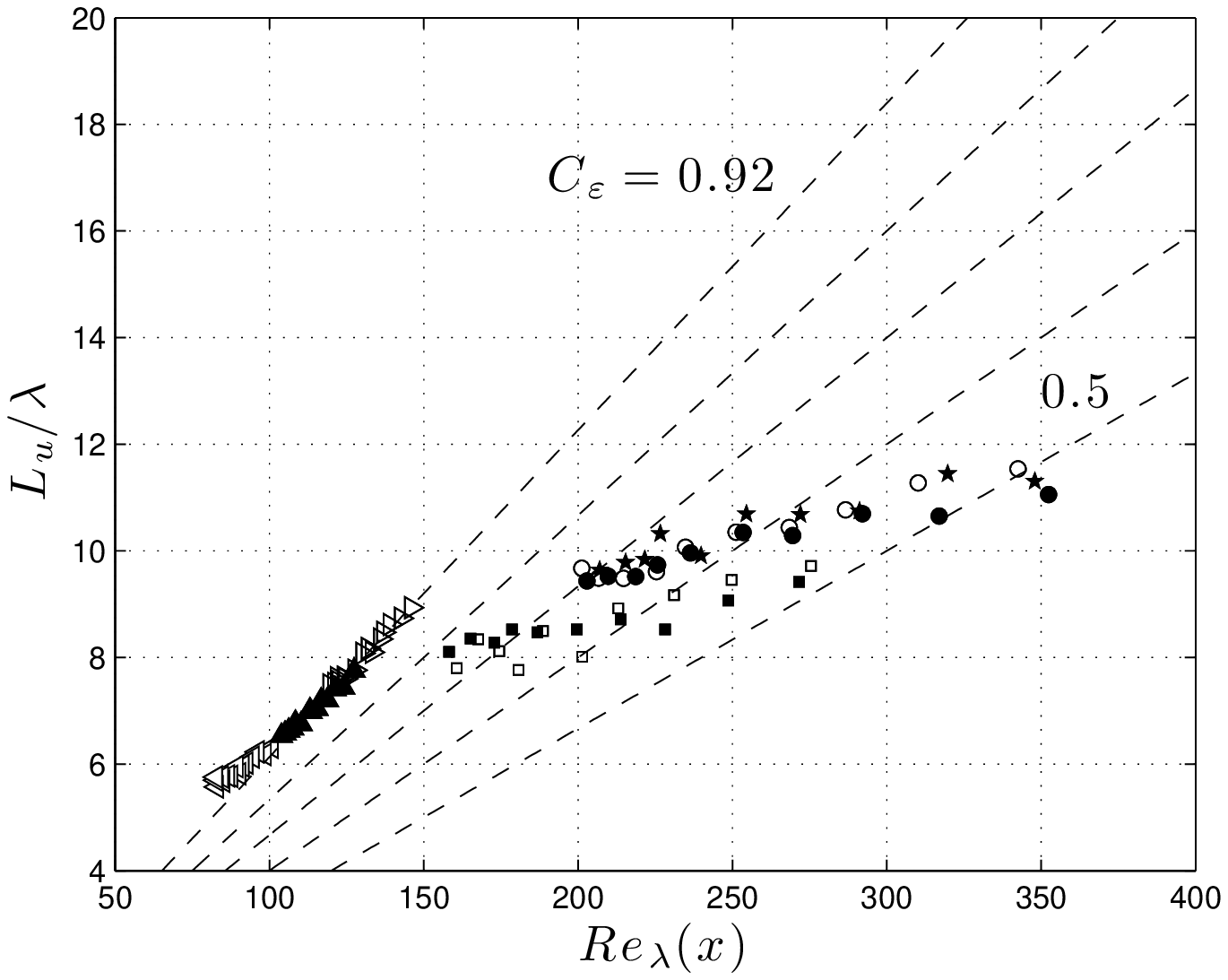} 
\includegraphics[trim=20 5 25 20, clip=true, width=120mm]{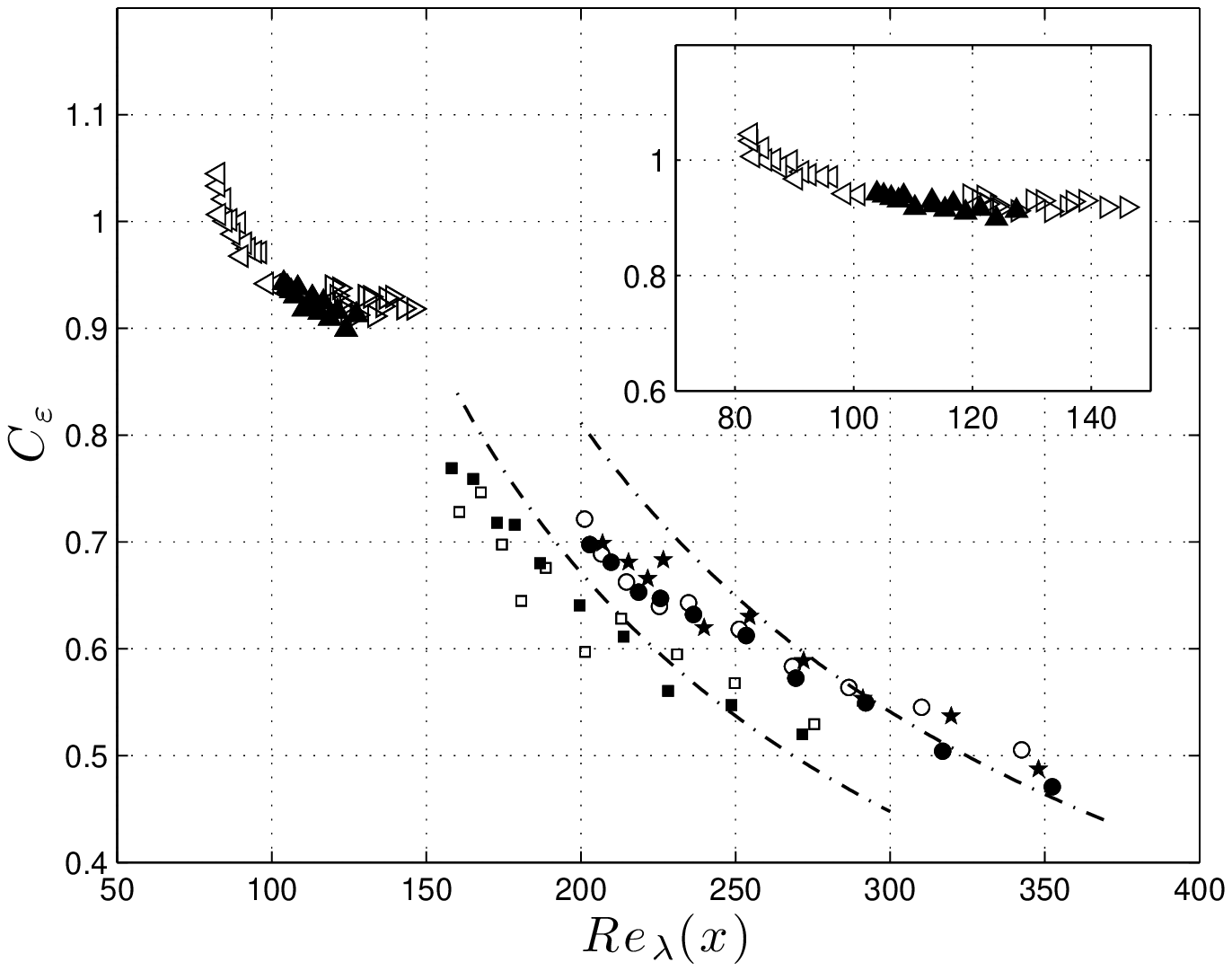}
\caption{Local Reynolds number dependence $Re_{\lambda}(x)$ of (a)
  Integral length scale to Taylor micro-scale ratio $L_u/\lambda$ and
  (b) normalised energy dissipation rate $C_{\varepsilon}$, for both
  the square fractal grid (SFG) and regular grid (RG) data recorded at
  different inlet velocities with single-wire (SW): (\FilledSmallSquare) SFG recorded at $U_{\infty}=10ms^{-1}$ with a
  $l_w=1mm$ sensing length SW,  (\SmallSquare) SFG at $U_{\infty}=10ms^{-1}$, $l_w=0.45mm$, (\FilledSmallCircle) SFG at $U_{\infty}=15ms^{-1}$, $l_w=1mm$, (\SmallCircle) SFG  at $U_{\infty}=15ms^{-1}$, $l_w=0.45mm$, ($\star$) SFG at $U_{\infty}=15ms^{-1}$, $l_w=0.2mm$, (\SmallTriangleLeft) RG at $U_{\infty}=10ms^{-1}$, $l_w=0.45mm$, (\FilledSmallTriangleUp) RG at
  $U_{\infty}=15ms^{-1}$, $l_w=0.45mm$, (\SmallTriangleRight) RG at  $U_{\infty}=20ms^{-1}$, $l_w=0.45mm$. The dashed-dot lines follow the form $\propto Re_{\lambda}^{-1}$. The insert of the second figure is a zoomed plot of the RG data.}
\label{Fig:LOverLambdaCeps}
\end{figure}

The key departure behind the present fractal square grid-generated
turbulence behaviour lies in the difference between the streamwise
dependencies of $L_u/\lambda$ and $Re_{\lambda}$ ($A(x)\neq B(x)$, see
equations 1.4 and 1.5). For steady initial conditions (fixed $Re_0$)
there is a significant $Re_{\lambda}$ decrease during decay (figure
\ref{Fig:CepsAndReLambdavsX}b), whereas $L_u/\lambda$ stays
approximately constant (figure \ref{Fig:LOverLambdaCeps}a), leading to
a steep monotonic downstream increase in the normalised dissipation
rate $C_{\varepsilon}$ (figure \ref{Fig:CepsAndReLambdavsX}a) which
follows approximately the form $C_{\varepsilon}\propto
(L_u/\lambda)/Re_{\lambda} \sim Re_{\lambda}^{-1}$ (figure
\ref{Fig:LOverLambdaCeps}b). Note, in particular, how the
$C_{\varepsilon}$ versus $Re_{\lambda}$ curve shifts to the right as
$Re_{0}$ increases, which is clear evidence of the two independent
dependencies that $C_{\varepsilon}$ has on $Re_{\lambda}$ and $Re_0$
in this fractal-generated turbulence.

Data were taken with probes of different spatial resolutions to
confirm that these results are not meaningfully biased by the
resolution of the measurements yielding figure
\ref{Fig:LOverLambdaCeps} (see Sec. \ref{Subsec:Hardware&Wires} for
details). Nonetheless it can be seen that the lesser resolution probe
($d_{w} \approx 5\mu m$, $l_{w} \approx 1mm$) has a slightly lower
$L_{u}/\lambda$ ratio due to the underestimation of
$\overline{(\partial u /\partial x)^{2}}$, but it does not change the
main observation that $L_{u}/\lambda$ is effectively roughly constant,
at least compared to the wide variation of $Re_{\lambda}$, during
decay.

We now contrast the behaviour of our fractal square grid-generated
turbulence behaviour with that of turbulence generated by
regular-grid. Such turbulence is thought to follow
Richardson-Kolmogorov phenomenology, although it's usually difficult
to exceed Reynolds numbers beyond $Re_{\lambda}\approx 150$ in
typically sized laboratory wind-tunnels (at least if Corrsin's
restriction $x/M>30$ is applied, \cite{CorrsinHandbook}) and therefore
the regular grid experiments are commonly at the lower end of the
range of validity of the Richardson-Kolmogorov phenomenology.
Nevertheless, our regular grid data for $U_{\infty}=20ms^{-1}$ appear
to have sufficiently high Reynolds numbers to support
$C_{\varepsilon}\approx const$ (figure \ref{Fig:LOverLambdaCeps}b) and
related $L_u/\lambda \propto Re_{\lambda}$ (figure
\ref{Fig:LOverLambdaCeps}a). Furthermore it can be seen that the
$Re_{\lambda}$ dependence of $C_{\varepsilon}$ falls on the same curve
regardless of how $Re_{\lambda}$ is varied, whether by varying $Re_0$
or by varying the streamwise position of the measurement. The same
observation can be made for the curve $L_{u}/\lambda$ versus
$Re_{\lambda}$. This is well-defined Richardson-Kolmogorov behaviour
where $A(x)=B(x)$, $\alpha = \beta =1/2$ and consequently no
distinction between local and global Reynolds number exists. Below
$Re_{\lambda} \approx 120$ direct dissipation becomes
noticeable and causes a departure from $C_{\varepsilon} \approx
const$, presumably due to an insufficiently large separation between
outer and inner scales \cite[]{Dimotakis2000}.

Summarising, the present fractal square grid-generated decaying
turbulence is fundamentally and qualitatively different from regular
grid-generated decaying turbulence. 
The $C_{\varepsilon}\approx const$ behaviour is not observed in figure
\ref{Fig:LOverLambdaCeps} for the fractal square grid despite the
moderately large turbulent Reynolds numbers $Re_{\lambda}$ (around
three times the $Re_{\lambda}$ necessary for the regular grid to
exhibit $C_{\varepsilon}=const$ on this plot) and the evidence that
the global/inlet Reynolds number $Re_0$ is sufficiently large for
$C_{\varepsilon}$ to be independent of $Re_0$ (figure
\ref{Fig:CepsRe0}). In fact the normalised dissipation rate is closer
to $C_{\varepsilon}\sim Re_{\lambda}^{-1}$ and $L_{u}/\lambda \approx
const$, which is in line with the previous experiments by
\cite{M&V2010}, although the larger length of the present wind-tunnel
brings to evidence that $L_{u}/\lambda$ and $C_{\varepsilon}
Re_{\lambda}$ are not exactly constant in this tunnel, but are only
roughly so for all the assessed decay region. This might be an effect
brought about, perhaps paradoxically, by the eventual low (though not too low) values of
$Re_{\lambda}$ far downstream. Or it might be due to a decrease in the
growth of $L_{u}$ because of the boundary layers at the tunnel walls
which begin to have a significant thickness very far downstream in
this longer wind-tunnel. As this wall effect might not affect the
growth of $\lambda$, $L_{u}/\lambda$ would monotonically decrease
downstream. Nevertheless, as we show further down in this paper, the
downstream evolutions of $L_{u}/\lambda$ and $C_{\varepsilon}$ are
consistent with a self-preserving evolution of energy spectra which
can be made to collapse with a single-scale reasonably well, as
opposed to the two different inner and outer scales required by
Richardson-Kolmogorov phenomenology.

\begin{figure} 
\centering
\includegraphics[trim=10 5 25 20, clip=true, width=90mm]{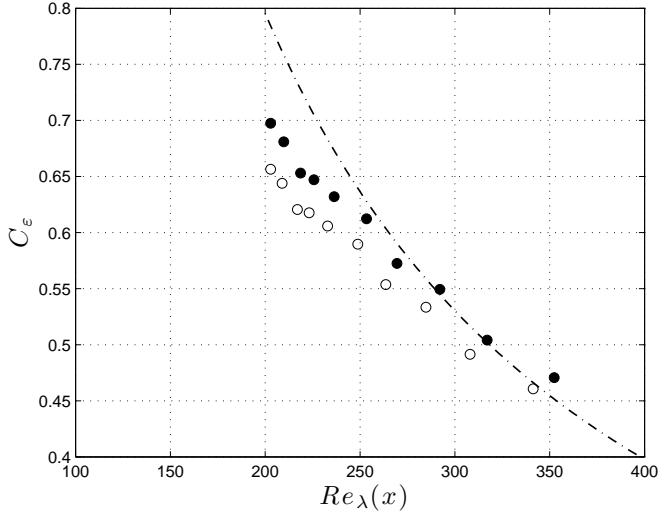} 
\caption{ Uncertainty \& bias due to large and small scale (an)isotropy in the observed normalised energy dissipation rate $C_{\varepsilon}$ behaviour. In this specific plot we re-define $u'^2$ and $\varepsilon$ to be $u'^2 \equiv \overline{u^2}(1+2 \overline{v^2}/\overline{u^2})$, $\varepsilon \equiv \nu\, \overline{(du/dx)^2}[3+6 \overline{(dv/dx)^2}/\overline{(du/dx)^2}]$ and take(\FilledSmallCircle) the isotropic estimates $\overline{v^2}/\overline{u^2}=1$ and $\overline{(dv/dx)^2}/\overline{(du/dx)^2}=2$ and (\SmallCircle) the anisotropy estimates of $\overline{v^2}/\overline{u^2}$ and $\overline{(dv/dx)^2}/\overline{(du/dx)^2}$ from table \ref{Table:HomogeneityParameters}. The dashed-dot line follows $\propto Re_{\lambda}^{-1}$}
\label{Fig:CepsAni}
\end{figure}

Finally note that the large and small scale anisotropy ( characterised by the ratios $u'/v'\approx 1.1$ and $\overline{(dv/dx)^2} / \overline{(du/dx)^2} \approx 1.5$, see Sec. \ref{Sec:Isotropy} and table \ref{Table:HomogeneityParameters}) 
change the exact numerical values of $C_{\varepsilon}$ and $Re_{\lambda}$ for each measurement location (see figure \ref{Fig:CepsAni}) and can be considered a source of uncertainty. Nevertheless, the main difference is an offset of the $C_{\varepsilon}$ versus $Re_{\lambda}$ curve and there is no meaningful change of its functional form.  

\subsection{Energy decay} \label{Sec:Decay}
The functional form of the turbulent kinetic energy decay is usually
assumed to follow a power-law, which is mostly in agreement with the 
large database of laboratory and numerical experiments for both
grid-generated turbulence and boundary-free turbulent
flows
\begin{align}
\overline{u^{2}}\sim (x-x_{0})^{-n}
\label{Eq:powerlaw}
\end{align}
where $\overline{u^{2}} \equiv u'^{2}$. 

\cite{M&V2010} proposed a convenient alternative functional form for
the kinetic energy decay (and for the evolution of $\lambda$ when
$U_{\infty} {d\over dx} \overline{u^{2}} \propto \nu
\overline{u^{2}}/\lambda^{2}$ is a good approximation) that is both
consistent with the power-law decay and the exponential decay law
proposed by \cite{George&Wang2009}:
\begin{equation}
\left\lbrace
\begin{aligned}
\lambda^2 = \lambda_{0}^2 & \left[1+\frac{4\nu a |c|}{l^2(x_{0}) U_{\infty}}(x-x'_{0})\right]\\
\overline{u^{2}}=\frac{2\, u'^2_{0}}{3} & \left[1+\frac{4\nu a |c|}{l^2(x_{0}) U_{\infty}}(x-x'_{0})\right]^{(1-c)/2c}
\end{aligned}
\right.
\label{Eq:MVEquations}
\end{equation}
where $c<0$. In the limit of $c \rightarrow 0$ it asymptotes to an
exponential decay with constant length-scales throughout the decay,
but otherwise it is a power-law decay where $x_{0}$ is not the
conventional virtual origin where the kinetic energy is singular. The
two equations \eqref{Eq:powerlaw} \& \eqref{Eq:MVEquations} are
equivalent with $n=(c-1)/2c$ and $x_{0} = x'_{0} - l^{2}_{0}\,
U_{\infty}/(4\nu a c)$.

\begin{figure}
\centering
\includegraphics[trim=10 0 30 10, clip=true,width=65mm, height=60mm]{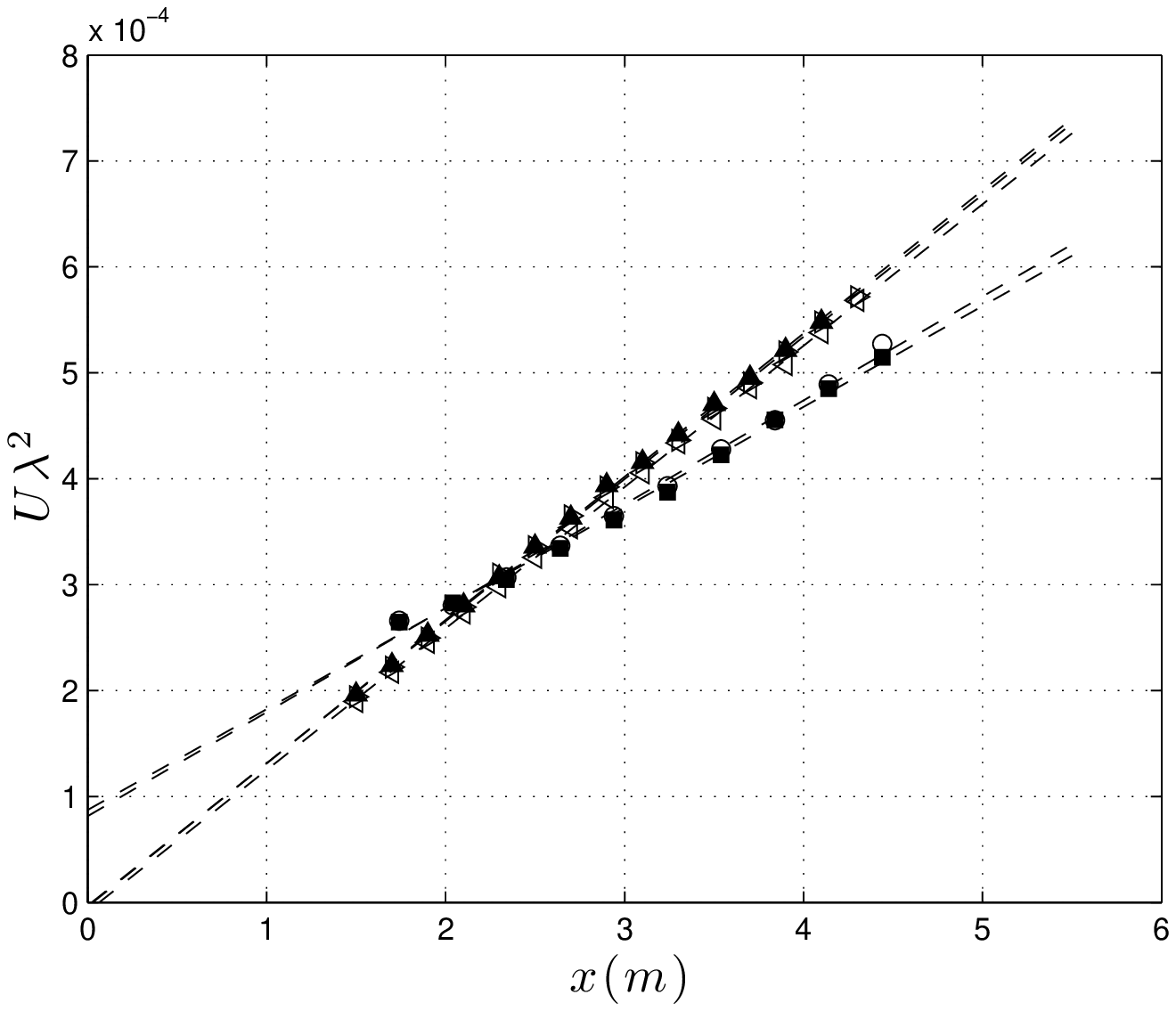} 
\includegraphics[trim=10 0 30 10, clip=true,width=65mm, height=60mm]{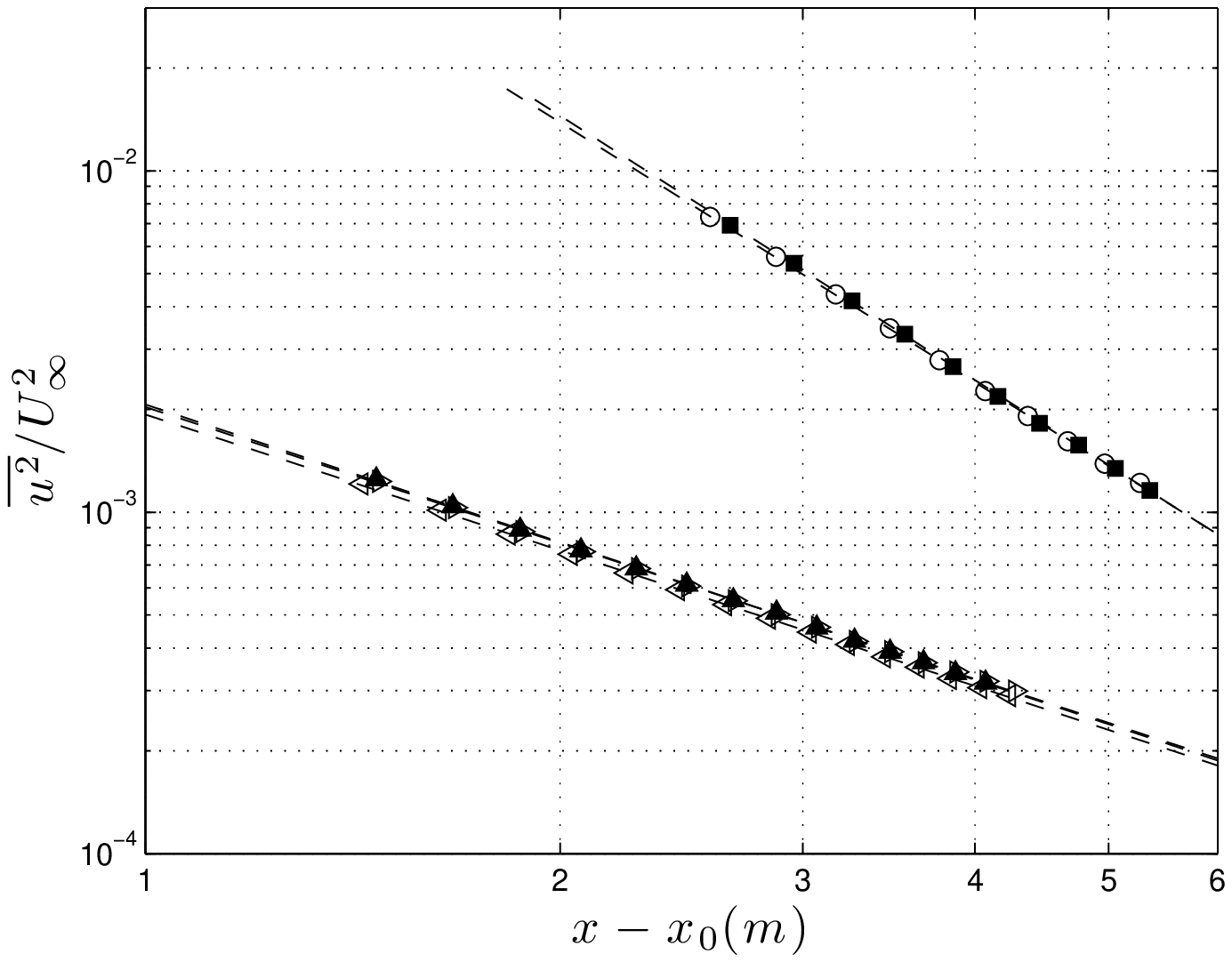}
\includegraphics[trim=10 0 30 10, clip=true, width=65mm,height=60mm]{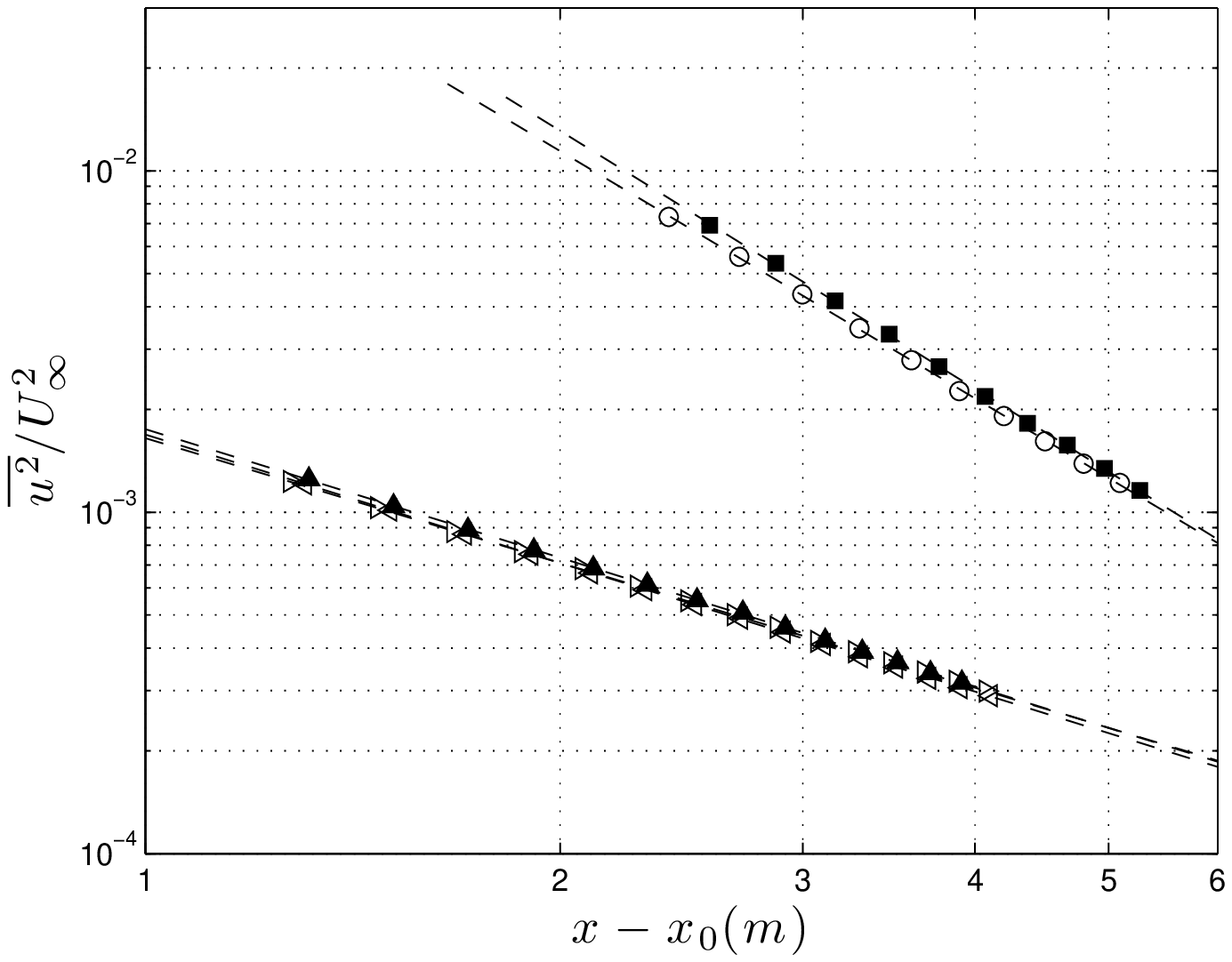} 
\includegraphics[trim=10 0 30 10, clip=true, width=65mm,height=60mm]{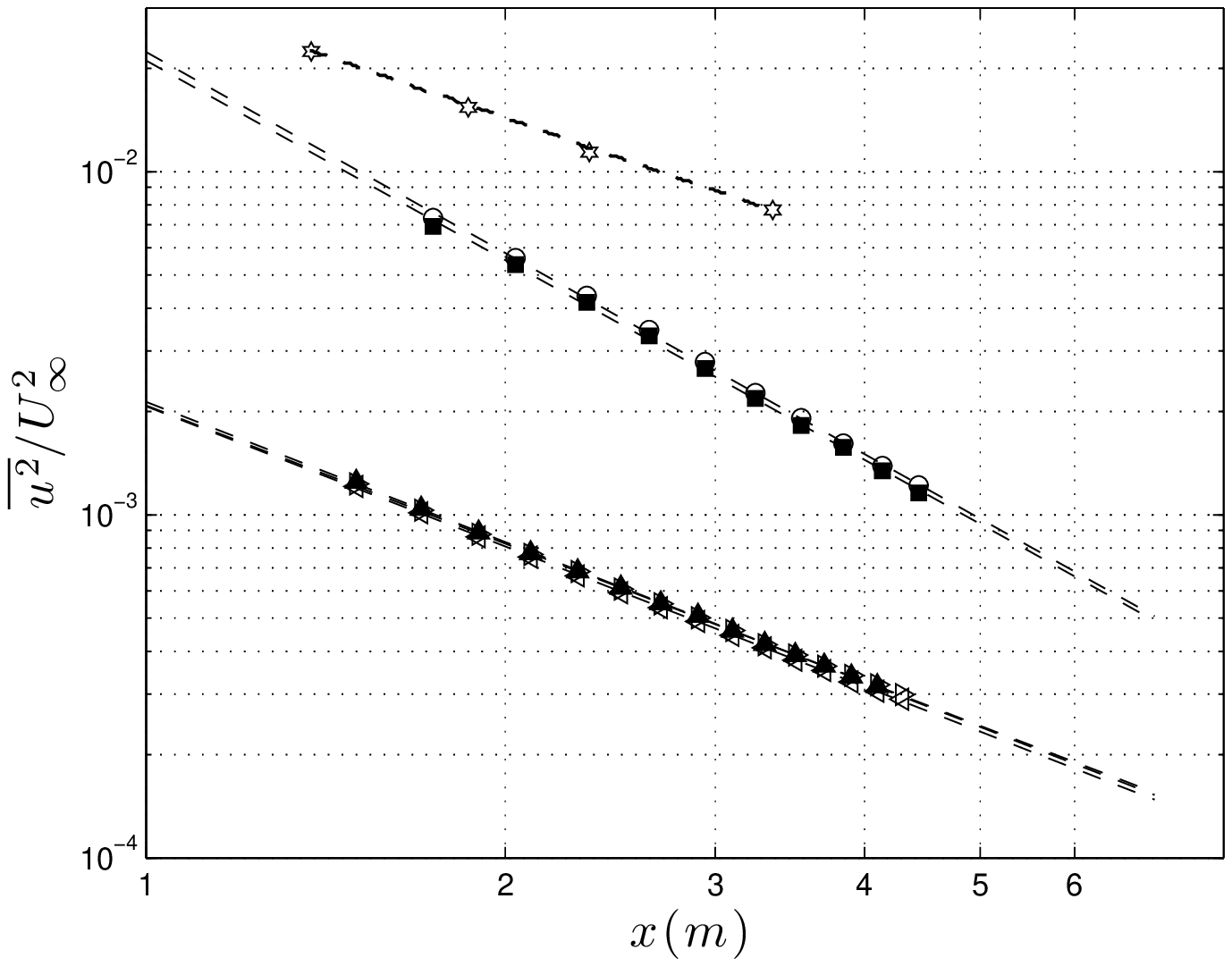}
\caption{Decay of turbulence generated by the regular (RG) and the fractal square (SFG) grid: (a) linear growth of $U\lambda^2$ (b) power-law fit using method I, (c) power-law fit using method III (d) power-law fit using method IV. (\FilledSmallSquare) SFG at $U_{\infty}=10ms^{-1}$, (\SmallCircle) SFG at $U_{\infty}=15ms^{-1}$, (\SmallTriangleLeft) RG at $U_{\infty}=10ms^{-1}$, (\FilledSmallTriangleUp) RG at $U_{\infty}=15ms^{-1}$, (\SmallTriangleRight) RG at $U_{\infty}=20ms^{-1}$, ($\smallstar$) data from the Active-grid experiment by \cite{M&W1996}.}
\label{Fig:Decay}
\end{figure}

Determining the decay exponent directly from \eqref{Eq:powerlaw} is
difficult, although feasible, since a non-linear fit is generally
needed to determine $n$ and $x'_{0}$ simultaneously. For homogeneous
(isotropic) turbulent decaying flow where advection balances dissipation 
it is possible to obtain a linear
equation for the Taylor micro-scale that can be used to determine the
virtual origin, thus simplifying the task of determining the decay
exponent. Using $\lambda^2=15\nu \overline{u^2}/\varepsilon$ in
conjunction with the advection dissipation balance characteristic of
homogeneous isotropic turbulence ($3/2 \, U \partial
\overline{u^{2}}/\partial x=-\varepsilon$) and assuming power-law
energy decay \eqref{Eq:powerlaw} we get
\begin{align}
\lambda^{2}=\frac{10 \, \nu}{n \, U}(x-x_{0}).
\label{Eq:lambda}
\end{align}
Note that for $\lambda^{2}$ to be linear the mean velocity has to be
constant otherwise the linear relation holds for $U\lambda^2$.
 Even though advection does not balance dissipation in our
fractal grid-generated decaying turbulence because of the significant presence
of transverse energy transport as shown in
Sec. \ref{subsec:Homogeneity}, transverse energy transport and
dissipation remain approximately proportional to each other throughout the assessed
decay region and for the range of values of $U_{\infty}$ tried
here. This suggests that 
\begin{equation}
U {d\over dx} \overline{u^{2}} \propto \nu \overline{u^{2}}/\lambda^{2}
\label{Eq:AdvectionTaylor}
\end{equation}
might be a good approximation for the decay region of our
fractal-generated turbulence as is indeed supported by our data which
show that U$\lambda^2$ grows linearly with downstream location and even
that $U\lambda^{2}$ versus $x$ collapses the data well for different
inlet velocities $U_{\infty}$ (see figure \ref{Fig:Decay}a).

\begin{figure}
\centering
\includegraphics[trim=22 5 30 20, clip=true, width=75mm, height=60mm]{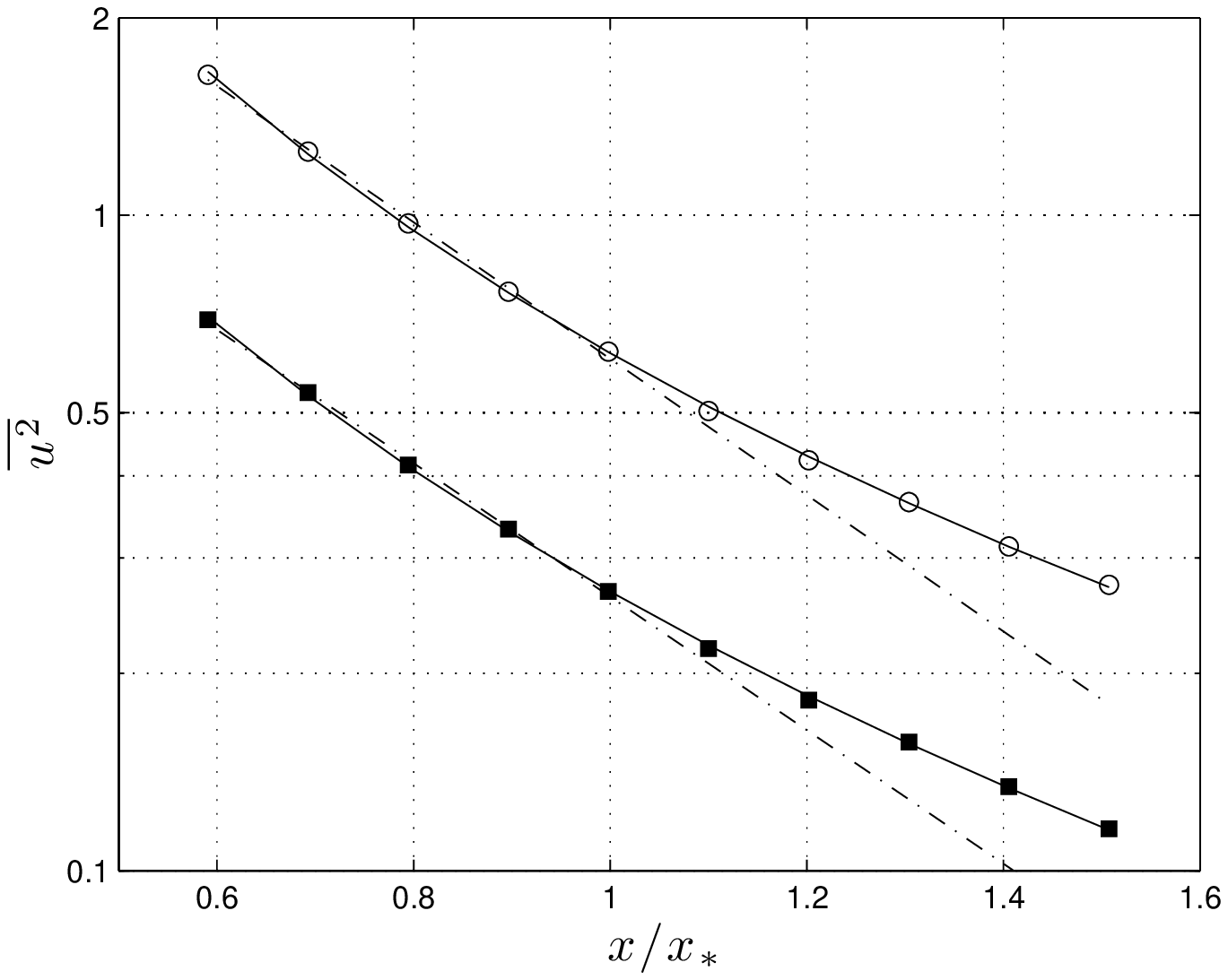}
\caption{Turbulent kinetic energy decay of turbulence generated by the fractal square grid fitted to \eqref{Eq:MVEquations} using method II (dashed-dot line) and method III (solid line). The data range used in method II is $0.6 < x/x_{*} < 1.1$, which corresponds to the streamwise region assessed by \cite{M&V2010}. Streamwise data was taken at two fixed inlet velocities: (\FilledSmallSquare) $U_{\infty}=10ms^{-1}$, (\SmallCircle) $U_{\infty}=15ms^{-1}$. Notice that for $0.6 < x/x_{*} < 1.1$ the two methods appear to fit the data reasonably well, but further downstream the differences become evident.}
\label{Fig:Decay3}
\end{figure}
 
The decay exponents of \eqref{Eq:powerlaw} and \eqref{Eq:MVEquations}
are estimated using four alternative methods: 
\begin{itemize}
\item Method I: linear fit to $U\lambda^2$ \eqref{Eq:lambda} to determine the
  virtual origin followed by a linear fit to the logarithm of
  \eqref{Eq:powerlaw} to determine the exponent $n$, as done by
  \cite{H&V2007}. \cite{antonia2003similarity} determined the virtual
  origin in a similar fashion by plotting $\lambda^2/(x-x_{0})$ for
  different $x_{0}$ and choosing the virtual origin yielding the
  broadest plateau (which for their regular grid experiment was
  $x_{0}\approx 0$). 
\item Method II: the linearised logarithm method proposed in
  \cite{M&V2010} to determine the unknowns in \eqref{Eq:MVEquations}.

\item Method III: direct application of a non-linear least-squares
  regression algorithm ('NLINFIT' routine in MATLAB\tm) to determine
  the decay exponent and virtual origin simultaneously. This is
  related to the method used by \cite*{lavoie2007effects}, but
  allowing the virtual origin to be determined by the algorithm as
  well. This method can be applied to \eqref{Eq:powerlaw} as well as
  to \eqref{Eq:MVEquations}. Note that if applied to
  \eqref{Eq:powerlaw} as we do here, this fitting method does not
  necessarily yield a virtual origin compatible with
  \eqref{Eq:lambda}.

\item Method IV: assume the virtual origin coincides with the grid
location and linearly fit the logarithm of \eqref{Eq:powerlaw}. This
crude method typically yields biased estimates of the decay exponent,
since there is no \emph{a priori} reason for the virtual origin to be
zero. Nevertheless this is a robust method typically used to get first
order estimates of power law decay exponents in many flows 
\cite[e.g. the active-grid data by][]{M&W1996}.  \\
\end{itemize} 
 
\begin{table}
\centering
\begin{tabular*}{0.9\textwidth}{@{\extracolsep{\fill}}cccccccc}
Grid & U &  \multicolumn{2}{c}{Method I} & Method II & \multicolumn{2}{c}{Method III} & Method IV  \\
 & $(ms^{-1})$ & n & $x_{0}/x_*$ &  $(1+c)/2c$ &  n & $x_{0}/x_*$    \\
\midrule
RG  & 10 & 1.32 & 0.18 & 4.34 & 1.25 & 0.53 & 1.36  \\
RG  & 15 & 1.34 & 0.08 & 5.04  & 1.25 & 0.52 & 1.36  \\
RG  & 20 & 1.32 & 0.06 & 5.47  & 1.21 & 0.63 & 1.33 \\
SFG  & 10 & 2.57 & -0.31 & 7.10  & 2.51 & -0.28 & 1.93 \\
SFG  & 15 & 2.53 & -0.28 & 8.01  & 2.41 & -0.22 & 1.95  \\
\end{tabular*}
\caption{Decay exponents and virtual origin estimation using different methods}
\label{Table:DecayExponents}
\end{table}

A main difference between these methods is the way of determining the
virtual origin, which has an important influence on the decay exponent
extracted. This inherent difficulty in accurately determining the
decay exponent is widely recognised in the literature \cite[see
  \eg][]{M&L1990}.

The decay data for the regular grid- and fractal square grid-generated
turbulence are well approximated by the curve fits obtained from
methods I \& III (see figures \ref{Fig:Decay}b \& \ref{Fig:Decay}c)
and the numerical values of the exponents change only marginally (see
table \ref{Table:DecayExponents}). On the other hand method IV also
seems to fit the data reasonably well (see figures \ref{Fig:Decay}d)
but the exponents retrieved for the fractal grid data are $n\approx
2$, slightly lower than the exponents predicted by the other methods
$n\approx 2.5$. The virtual origin which is forced to $x_0=0$ in
method IV leads to a slight curvature in the $\log(\overline{u^2})$
versus $\log( x)$ data (almost imperceptible to the eye, compare the
fractal grid data in figures \ref{Fig:Decay}c \& \ref{Fig:Decay}d)
and a non-negligible bias in the estimated exponents. Nevertheless the
difference in the power laws describing the measured regular grid- and
square fractal grid-generated turbulence is quite clear.  For
completeness, the results from the experimental investigation by
\cite{M&W1996} on decaying active grid-generated turbulence are added
in figure \ref{Fig:Decay}d. They applied a fitting method equivalent
to method IV and reported a power-law fit yielding a decay exponent
$n=1.21$. \cite{KCM02} employed the same method to their active
grid-generated turbulence data and retrieved a similar result,
$n=1.25$.

Note that there are residual longitudinal mean velocity gradients
(which cause a residual turbulence production of about $3\%$ of the
dissipation, see Sec \ref{subsec:Homogeneity}) and therefore it is
preferred to fit $\overline{u^2}$ data rather than
$\overline{u^2}/U^2$ data. Nevertheless we checked that fitting
$\overline{u^2}/U^2$ data does not meaningfully change the results nor
the conclusions.

Concerning method II it can be seen (table \ref{Table:DecayExponents})
to be the most discrepant of the four methods yielding a much larger
decay exponent. This method was proposed by \cite{M&V2010} to fit the
general decay law \eqref{Eq:MVEquations} and is based on the
linearisation of the logarithm appearing in the logarithmic form of
\eqref{Eq:MVEquations}, i.e.
\begin{equation}
\begin{aligned}
\log(u'^2) = \log\left(\frac{2\, u'^{2}_{0}}{3}\right) + \left[ -\frac{1+c}{2c} \right] \log\left(1+\frac{4\nu a c}{\lambda_{0}^2 U_{\infty}}(x-x'_{0})\right).
\end{aligned}
\label{Eq:logMVEq}
\end{equation}

Linearisation of the second logarithm on the right hand side of
\eqref{Eq:logMVEq} assumes 
$\frac{4\nu a c}{\lambda_{0}^2 U_{\infty}}(x-x'_{0})<<1$. This quantity, as we have confirmed in
our data, is indeed smaller than unity and for the farthest position
$4\nu a c/(l^2(x_{0}) U_{\infty})(x-x_{0})\approx 0.3$, but the fact
that this linearised method does not yield results comparable to
methods I and III suggests that the linearisation of the logarithm may
be an oversimplification. In figure \ref{Fig:Decay3} the kinetic
energy decay data of turbulence generated by the fractal square grid
is shown along with the fitted curves obtained from methods II and III
in a plot with a logarithmic ordinate and a linear abscissa. In figure
\ref{Fig:Decay3}a the data taken at positions beyond $x/x_{*}\approx
1.05$ are excluded in order to compare with the results presented in
\cite{M&V2010} where the data range was limited to
$0.5<x/x_{*}<1.05$. Visually, in figure \ref{Fig:Decay3}a, the two
different fitting methods appear to fit the data reasonably well and
thus the linearisation of the logarithm in \eqref{Eq:logMVEq} is
justified in this limited range. Note, however, that the two fitting
methods yield very different decay exponents because they also
effectively yield different virtual origins: for example at
$U_{\infty}=15ms^{-1}$ method III yields $(1+c)/(2c)\approx -2.4$
whereas method II yields $(1+c)/(2c)\approx -8.0$. In figure
\ref{Fig:Decay3}b, where no data is excluded, it can clearly be seen
that the two methods produce very different curves and very different
decay exponents (note however that the use of a longer test section,
which allows the assessment of the decay behaviour further downstream,
comes at the cost of having thicker boundary layers developing at the
walls which can have an increasing influence on the largest turbulent
eddies, as discussed in Sec. \ref{Sec:WTConf}).

\subsubsection{Influence of transverse transport on power-law decay exponent} \label{Sec:DecayAndDiffusion}

It is shown in Sec. \ref{subsec:Homogeneity} that dissipation does
not balance the advection but that the two are roughly proportional
throughout the measured decay region of the fractal square
grid-generated turbulence. It is also shown in that section that this
imbalance is mostly due to transverse triple-correlation transport
which remains roughly $50-60\%$ of the dissipation throughout the
measured region (with no clear increasing or decreasing trend),
whereas turbulence production and longitudinal triple-correlation
transport terms become negligible well before $x<x_*$. Pressure
transport, calculated from the kinetic energy balance, may also play a
noticeable role of countering a fraction (typically between $1/4$ and
$1/3$) of the triple-correlation transport. Based on these results,
equation \eqref{QuasiHomogeneousEq} which holds at the centreline reduces
to 
\begin{equation}
\frac{U}{2}\frac{\partial\, \overline{q^{2}}}{\partial x} =  -\varepsilon + \left[-2\frac{\partial}{\partial y} \frac{ \overline{v q^{2}}}{2} + \Pi \right] .
\label{AdvectionDissipationTransport}
\end{equation}
The decay rate of the kinetic energy as the turbulence is advected
downstream (effectively the advection term) is now determined both by
viscous dissipation and by a net effect of removing energy from the
centreline and transporting it to the sides. As in the portion of the decay region of the fractal-generated
turbulence where we take measurements this loss rate to the sides
remains approximately proportional to the dissipation rate, i.e. 
\begin{equation*}
\frac{U}{2}\frac{\partial\, \overline{q^{2}}}{\partial x} =  -\chi \, \varepsilon
\end{equation*}
where $\chi = 1 + [\, \partial \overline{v q^{2}}\partial y - \Pi\,
]/\varepsilon \approx 1.5$ (figure
\ref{Fig:CorrsinHomogeneityTKEBudget}c), we can expect the decay
exponent $n$ to be set by the dissipation rate $\varepsilon$
(irrespective of what sets the dissipation rate). Indeed, the higher
power law decay exponents exhibited by the fractal-generated
turbulence can be accounted for by the fact that $C_{\varepsilon}\sim
Re_{\lambda}^{-1}$ (see Sec. \ref{Sec:Ceps}) and consequently the
steep increase of $C_{\varepsilon} = \varepsilon L_u /u'^3$ with
streamwise location. In other words, an increasing proportion of
$u'^3/L_u$ is being dissipated at increasing streamwise locations
which leads to an increase in the power law decay exponent relative to
the $C_{\varepsilon} = const$ case.

\begin{figure} \centering
\includegraphics[trim=10mm 0mm 0mm 0mm, clip=true,width=3in]{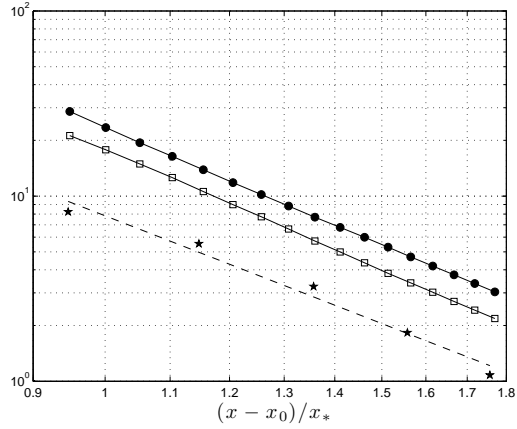}
\caption{Downstream decay of: (\FilledSmallCircle) kinetic energy $\frac{U}{2}\frac{\partial \overline{q^2}}{\partial x}$, (\SmallSquare) dissipation $\varepsilon$, ($\star$) transverse triple-correlation transport $2\frac{\partial}{\partial y} \left( \frac{\overline{vq^2}}{2} \right)$, for $U=15ms^{-1}$}
\label{DecayVsDissipationVsDiffusion} \end{figure}

In figure \ref{DecayVsDissipationVsDiffusion} we plot in logarithmic
axes the streamwise decay of the advection, the dissipation and the
transverse triple-correlation transport (which are all measured
independently) and they indeed seem to follow straight lines
(i.e. power laws) with the same slope (i.e. power law exponent), thus
supporting our argument.

To further substantiate our argumentation one more set of experiments
were conducted. Anemometry measurements at an inlet velocity of
$U_{\infty}=15ms^{-1}$ using a $l_w=0.5mm$ sensing length single-wire
were recorded between $0.63<x/x_*<1.44$ along four parallel lines
aligned with the mean flow and crossing the grid at
$z=0,\,0,\,20,\,20\, mm$ and $y=0,\,40,\, 80,\, 120\, mm$ ($z=0$ is
the vertical plane of symmetry of the grid). From the transverse
triple-correlation transport measurements for $z=0$ (figure
\ref{vq2}b) we expect the contribution from this term to be very
different at the centreline (where it is maximal) and off the
centreline where is can be roughly zero ($y\approx 80$) or negative
($y > 100$). However, if a value of $\chi$ can be defined that is
constant throughout the streamwise decay range assessed here for each transverse
$(y,z)$ position, then the argument outlined in the previous two
paragraphs will hold even if $\chi$ varies with transverse positions,
as indeed it does. 
The consequence is that, in the decay region assessed, the decay
exponent n should remain about the same at all these transverse
positions and also remain unusually large due to the $C_{\varepsilon}$
behaviour.
The data for the different transverse locations are fitted using method III and the 
 results (see table \ref{Table:DecayExponentsII}) are encouraging. In spite of some
variation in the best fit power-law decay exponents, the numerical
values of these exponents are all relatively close to each other
ranging between $2.3$ and $2.6$. We note that these exponents are
larger than all boundary-free turbulent flows listed in table \ref{Table:OtherFlows}.

\begin{figure}
\centering
\includegraphics[trim=10mm 0mm 0mm 0mm, clip=true,width=65mm]{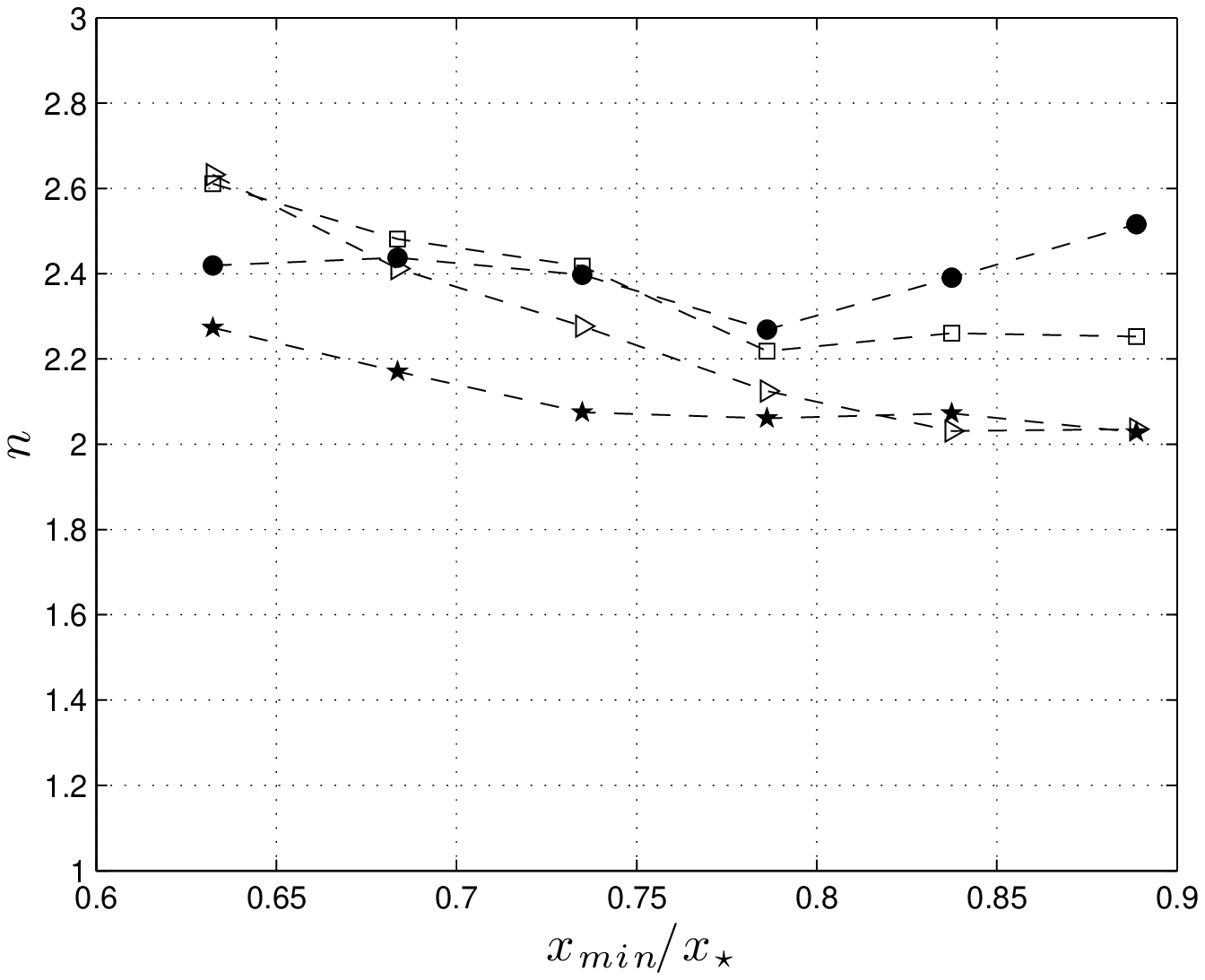}
\includegraphics[trim=0mm 0mm 10mm 0mm, clip=true,width=65mm]{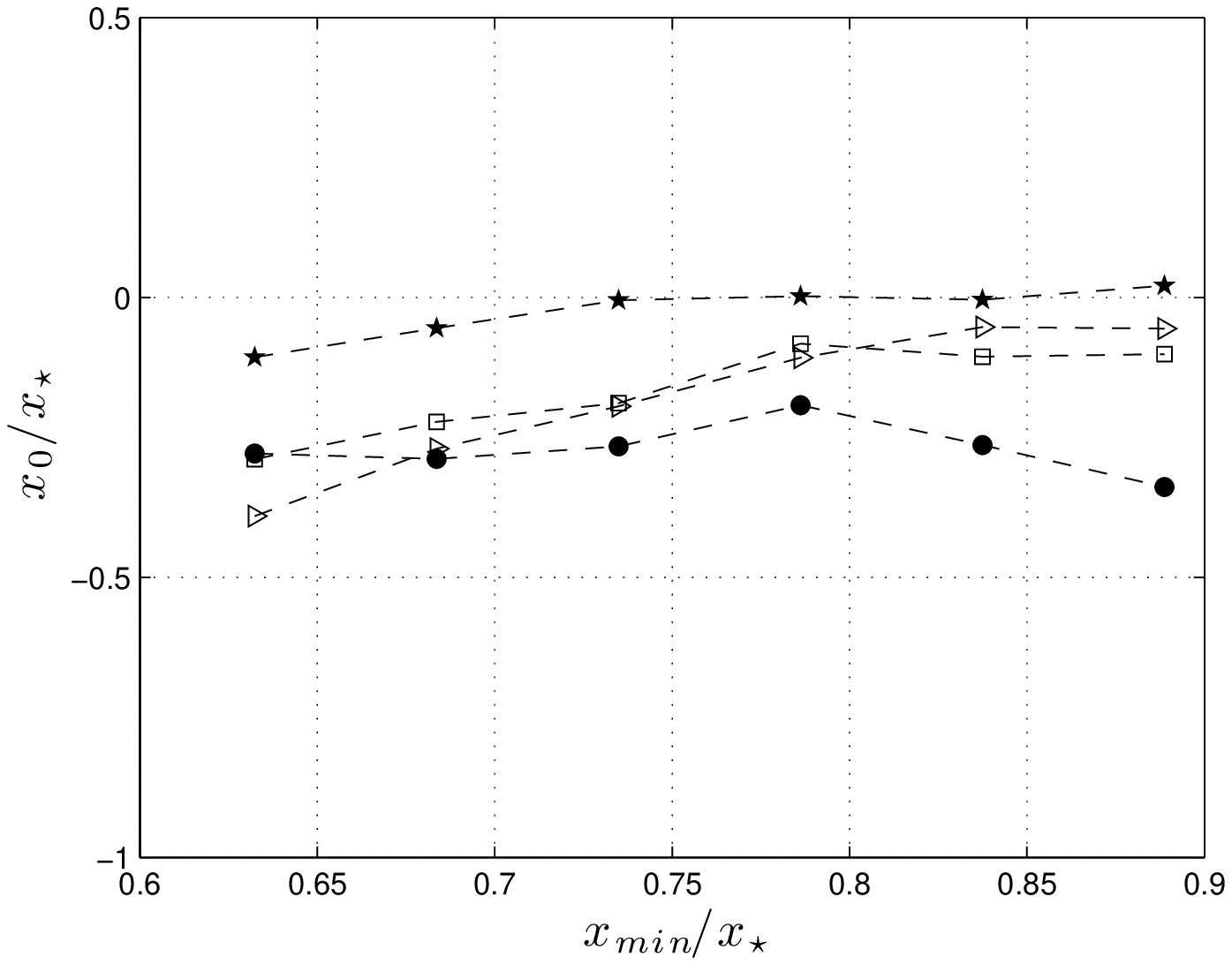}
\caption{ Decay law estimates for different data ranges $ x_{min} < x/x_{\star} < 1.40$, $U=15ms^{-1}$: (\FilledSmallCircle) $y=0mm$, (\SmallSquare) $y=40mm$, ($\star$) $y=80mm$, (\SmallTriangleRight)  $y=120mm$. (a) decay exponent $n$ and (b) virtual origin $x_0/x_{\star}$ obtained from method III}
\label{n_xmin}
\end{figure}

Finally, as some presence of turbulence production and longitudinal
transport remains for some distance downstream of $x_{peak} \approx
0.45 x_{*}$ (though not in any significant way beyond $x_*$) we explore
how the power-law fits of the turbulence energy decay change when the
smallest streamwise location considered in the fit is increased. We do
this both for centreline and off-centreline data and report our
results in figure \ref{n_xmin}. On the centreline the decay exponent and virtual origin remain
approximately the same within the scatter 
($n\approx2.4,\,x_0/x_*\approx -0.3$), but they show a respectively 
decreasing/increasing tendency off-centreline up to
$x/x_*\approx0.8$. At any rate, the decay exponents $n>2.0$ for all
our data.\\

\begin{table}
\caption{Decay law estimates along four parallel streamwise oriented lines at the centreline and off the centreline between $ 0.63 < x/x_{\star} < 1.40$  obtained from method III. }
\centering
\begin{tabular*}{0.4\textwidth}{@{\extracolsep{\fill}}ccc}
y (mm) &  n & $x_{0}/x_{\star}$   \\
\midrule
0    & 2.42 &  -0.27 \\
40  & 2.61 &  -0.29 \\
80  & 2.27 &  -0.11 \\
120& 2.63 &  -0.39 \\
\end{tabular*}
\label{Table:DecayExponentsII}
\end{table}

In conclusion the decay exponents for the present fractal-generated
turbulence measured both at the centreline and off the centreline in
the region $0.6 < x/x_{*} <1.5$ are consistently higher than those in
all boundary-free turbulent flows listed in table 1 and much higher
(by a factor between 4/3 and 2) than those of decaying turbulence
generated by regular and active grids \cite[]{M&W1996,KCM02}. It might be interesting to note that in many boundary-free turbulent flows a conserved quantity such as $u'^2L^{M+1}=const$ exists. Look at table \ref{Table:OtherFlows} and note that $M=1,\,3,\,5,\,7$ for the four wakes, $M=-1$ for the mixing layer, $M=0,\,1$ for the jets and $M \ge 2$ for regular-grid turbulence. If the flow is also such that $Udu'^2/dx\propto -\varepsilon$ then $C_{\varepsilon}=const$ implies
 \begin{equation*}
 n=\frac{2(M+1)}{M+3} 
 \end{equation*}
 and $C_{\varepsilon}\sim Re_{\lambda}^{-1}$ implies 
 \begin{equation*}
 n=\frac{M+1}{2}
 \end{equation*}
 (which is larger than $n=2(M+1)/(M+3)$ provided that $M>1$). Considering, for example, the range $ M \geq 2$, the exponent $n$ corresponding to $C_{\varepsilon}\sim Re_{\lambda}^{-1}$ is at least $5/4$ times larger than the exponent $n$ corresponding to $C_{\varepsilon} = const$, and is generally much larger. If $M=3$ or $M=4$ then $C_{\varepsilon} \sim Re_{\lambda}^{-1}$ implies $n=2$ or $n=2.5$, close to what is observed here, whereas $C_{\varepsilon} \sim const$ implies $n=4/3$ or $n=10/7$.
 
At this stage we do not have any proof that a conserved quantity such
as $u'^{2}L^{M+1}=const$ exists for our fractal-generated
turbulence. The previous paragraph is therefore only indicative and
serves to illustrate how a $C_{\varepsilon}$ which is a decreasing
function of $Re_{\lambda}$ can cause the decay exponent to be
significantly larger than a $C_{\varepsilon}$ which is constant during
decay and can even return decay exponents comparable to the ones
observed here. Of course the decaying turbulence we study in this work
is not perfectly homogeneous and isotropic because of the presence of
transverse turbulent transport of turbulent kinetic energy and
therefore significant gradients of third-order one-point velocity
correlations. As a consequence, a conserved quantity such as
$u'^{2}L^{M+1}=const$, if it exists, cannot result from a two-point
equation such as the von-K\'arm\'an-Howarth equation for homogeneous
turbulence \cite[see][]{V2011}. We leave the investigation of
conserved quantities in third-order inhomogeneous decaying turbulence
such as the present one for the future (we include gradients of
pressure-velocity correlations in the term "third-order
inhomogeneous").

Nevertheless, it is clear that the dissipation rate of kinetic energy
is increasingly larger than $u'^{3}/L$ as the turbulence moves further
downstream in cases such as the present one where $C_{\varepsilon}$
increases in approximate proportion to $1/Re_{\lambda}$ as the
turbulence and $Re_{\lambda}$ decay. In the absence of any other type
of loss or gain of kinetic energy, and assuming no counter-effect of
$C_{\varepsilon}$ on the integral scale, a much steeper decay (e.g. much
larger exponent $n$) will result than if $C_{\varepsilon}$ was constant
during decay. In the present case where loss of energy also occurs by
turbulent transport, see equation \eqref{AdvectionDissipationTransport}, this conclusion can remain
the same in the region assessed only if, in that region, the loss of
energy by turbulent transport remains proportional to the loss of
energy by dissipation, as indeed observed.

The question then naturally arises whether this balance between
turbulent transport and dissipation persists for the entire decay range
all the way to very large values of $x/x_*$, much larger than those
accessible here. If it does, then the implication is that perfectly
homogeneous isotropic turbulence is impossible at any stage of the
decay. If it does not and if turbulent transport starts to decay much
faster than dissipation beyond a certain $x/x_*$, then a turbulence
that is third-order homogeneous and isotropic may well appear if it
has the time to do so before the final stages of decay. If
$C_{\varepsilon}$ continues to increase nearly as $1/Re_{\lambda}$ in
such a third-order homogeneous isotropic turbulence then the decay
will remain exceptionally fast with values of $n$ such as the present
ones. However, it may be that the unusual behaviour observed here for
the dissipation rate $\varepsilon$ (a two-point statistic) is in fact the
result of gradients in particular one-point statistics such as
third-order velocity correlations and pressure-velocity correlations,
i.e. inhomogeneities. Either way, the consequences can be far reaching
and call for much future research, in particular re-examinations of
Reynolds number dependencies of $C_{\varepsilon}$ in all manner of
turbulent flows, in particular boundary-free turbulent flows such as
those listed in table \ref{Table:OtherFlows}.

As a final remark, note that the data points for transverse turbulent
transport in figure \ref{DecayVsDissipationVsDiffusion} seem to curve downwards at high
$x/x_*$. However we cannot extrapolate much from this observation as
we do not measure pressure directly and we do not know how gradients
of pressure-velocity correlations curve at high $x/x_*$.

\subsection{Collapse of the energy spectra and structure functions}\label{subsec:Collapse}
As explained in \cite{S&V2007} and \cite{M&V2010} single-length-scale self-preserving energy spectra can allow for $L_u/\lambda=const$ during decay. This can be assessed by plotting the normalised energy spectra for different positions along the mean flow direction and evaluating the collapse of the data or the lack thereof. 
It should be mentioned that the three-dimensional energy spectrum and one-dimensional energy spectra can be shown to be equivalent for an isotropic flow. It should also be  noted that the flow is not exactly isotropic as discussed in Sect. \ref{Sec:Isotropy}, so we might expect some effect on the spectral collapse.

\subsubsection{One-dimensional energy spectra}
\begin{figure} 
\centering
\includegraphics[trim=15mm 0mm 5mm 0mm, clip=true,width=65mm]{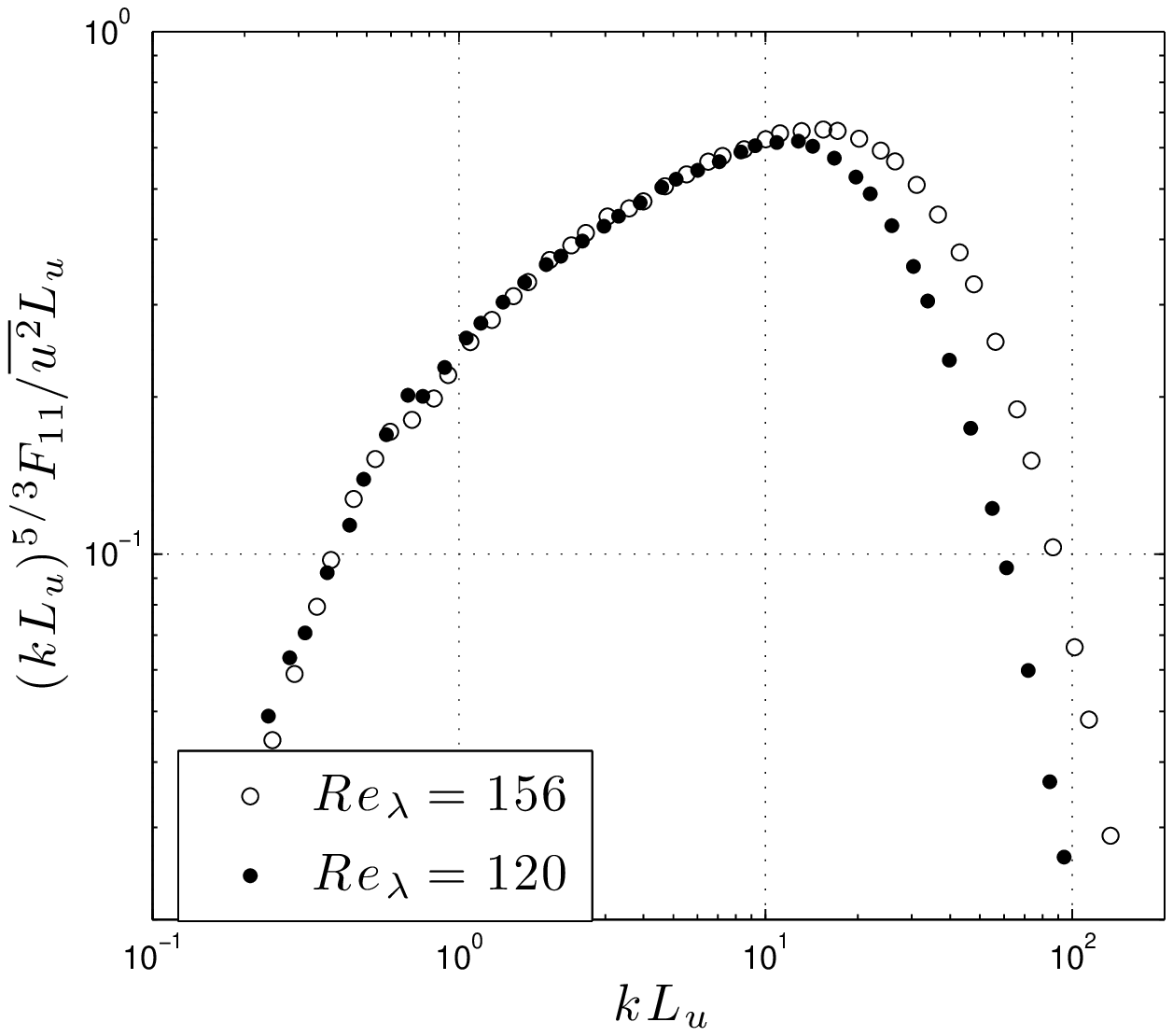}
\includegraphics[trim=15mm 0mm 5mm 0mm, clip=true,width=65mm]{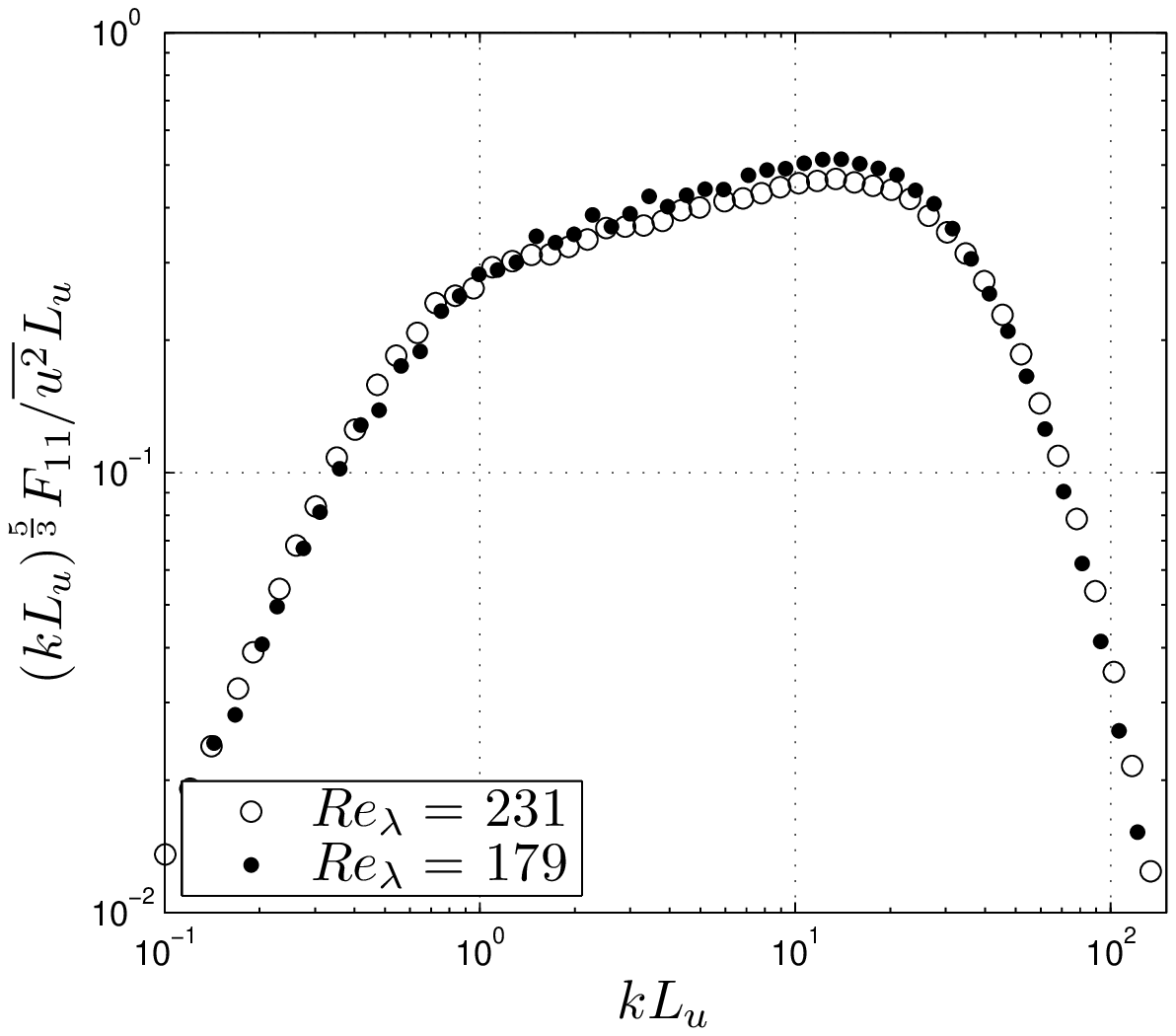}
\caption{Compensated 1D energy spectra, normalised with $\overline{u^{2}},\,L_{u}$ at two streamwise locations for the (a) regular grid- and (b) fractal square grid-generated turbulence data. Both plots have roughly the same Reynolds number ratio, $Re_{\lambda_1}/Re_{\lambda_2}\approx 1.3$ (see Appendix \ref{ap:appendixB}). The data are recorded at $U_{\infty}=20ms^{-1}$ and $U_{\infty}=10ms^{-1}$ respectively.} 
\label{Fig:ComparisonRG_SFG}
\end{figure}

We begin by illustrating the qualitative difference between the collapse of the normalised energy spectra (using large scale variables: $u'^{2}$, $L_{u}$) of turbulence generated by the regular grid and by the fractal square grid,  see figure \ref{Fig:ComparisonRG_SFG}. The data for the regular grid are taken in a region where $L_{u}/\lambda \propto Re_{\lambda}$ and $C_{\varepsilon} \approx const$, see figure \ref{Fig:LOverLambdaCeps}. The normalised spectra measured in the lee of the regular grid show a good collapse at the low frequencies but not at the high frequencies, which is in-line with Kolmogorov's theory. On the other hand it can be seen that the normalised turbulence spectra generated by the fractal square grid appears to collapse at all frequencies, in-line with the single-length-scale assumption as previously observed by \cite{M&V2010}.

\begin{figure} 
\centering
\includegraphics[trim=8mm 0mm 5mm 0mm, clip=true,width=65mm]{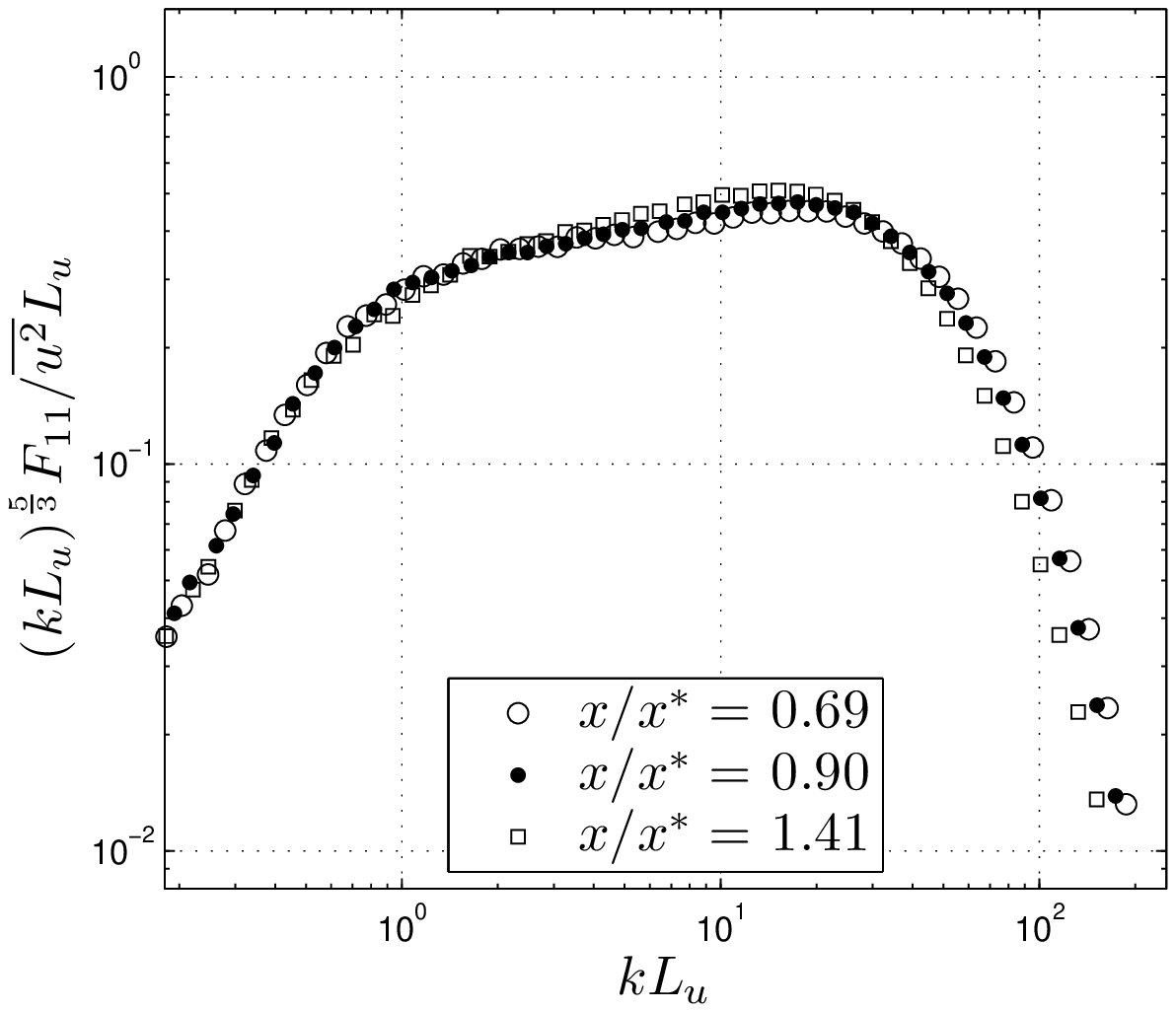}
\includegraphics[trim=10mm 0mm 5mm 0mm, clip=true,width=65mm]{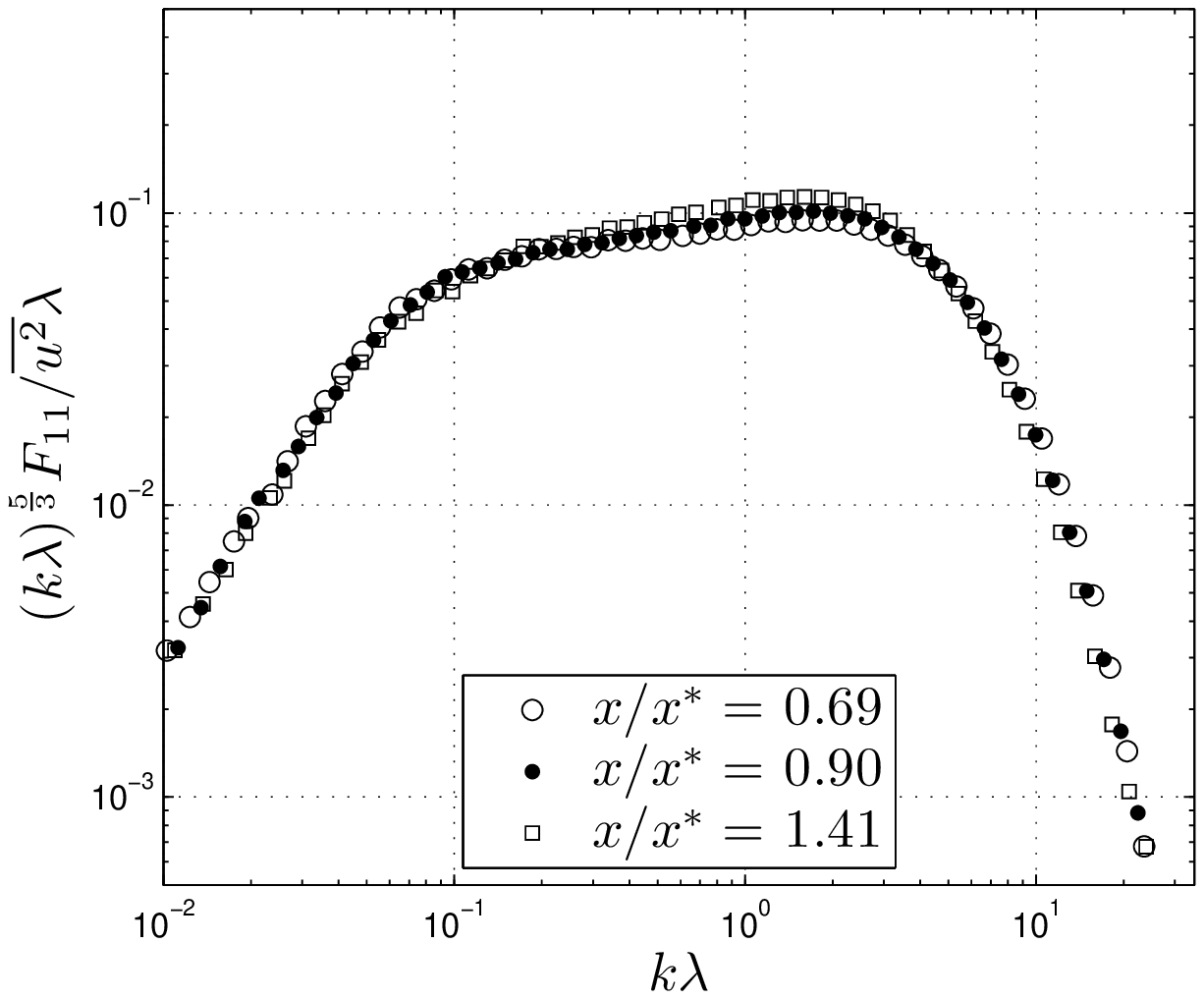}
\includegraphics[trim=15mm 0mm 5mm 2mm, clip=true,width=65mm]{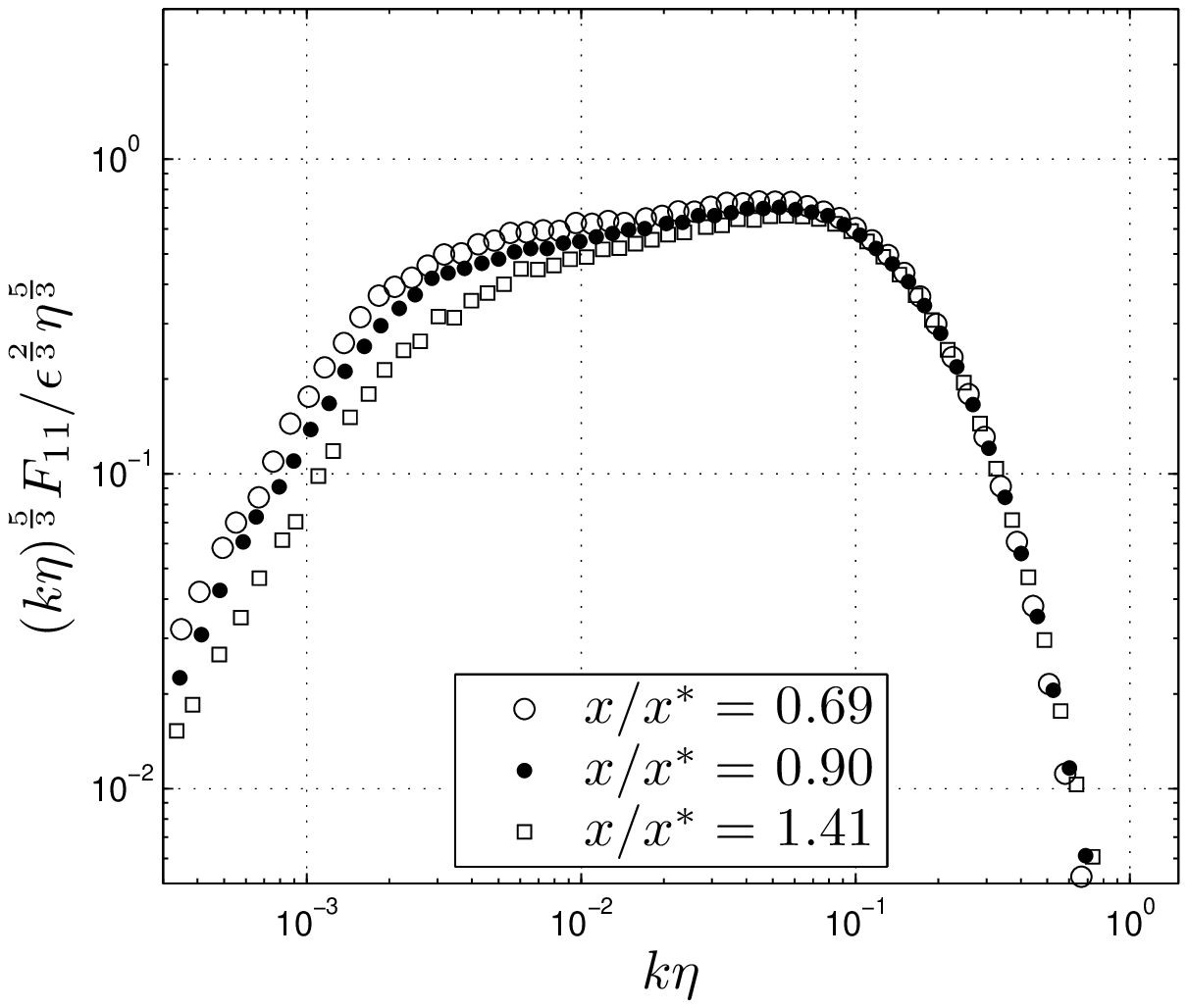}
\caption{Compensated 1D energy spectra  of turbulence generated by the fractal square grid at three streamwise downstream locations ($Re_{\lambda}=324,\,273,\,210$) at $U_{\infty}=15ms^{-1}$, normalised by (a) $\overline{u^{2}}$ and $L_{u}$ (b) $\overline{u^{2}}$ and $\lambda$ (c) $\nu$ and $\varepsilon$. }
\label{Fig:CollapseI} 
\end{figure}
\begin{figure} 
\centering
\includegraphics[trim=8mm 0mm 5mm 2mm, clip=true,width=65mm]{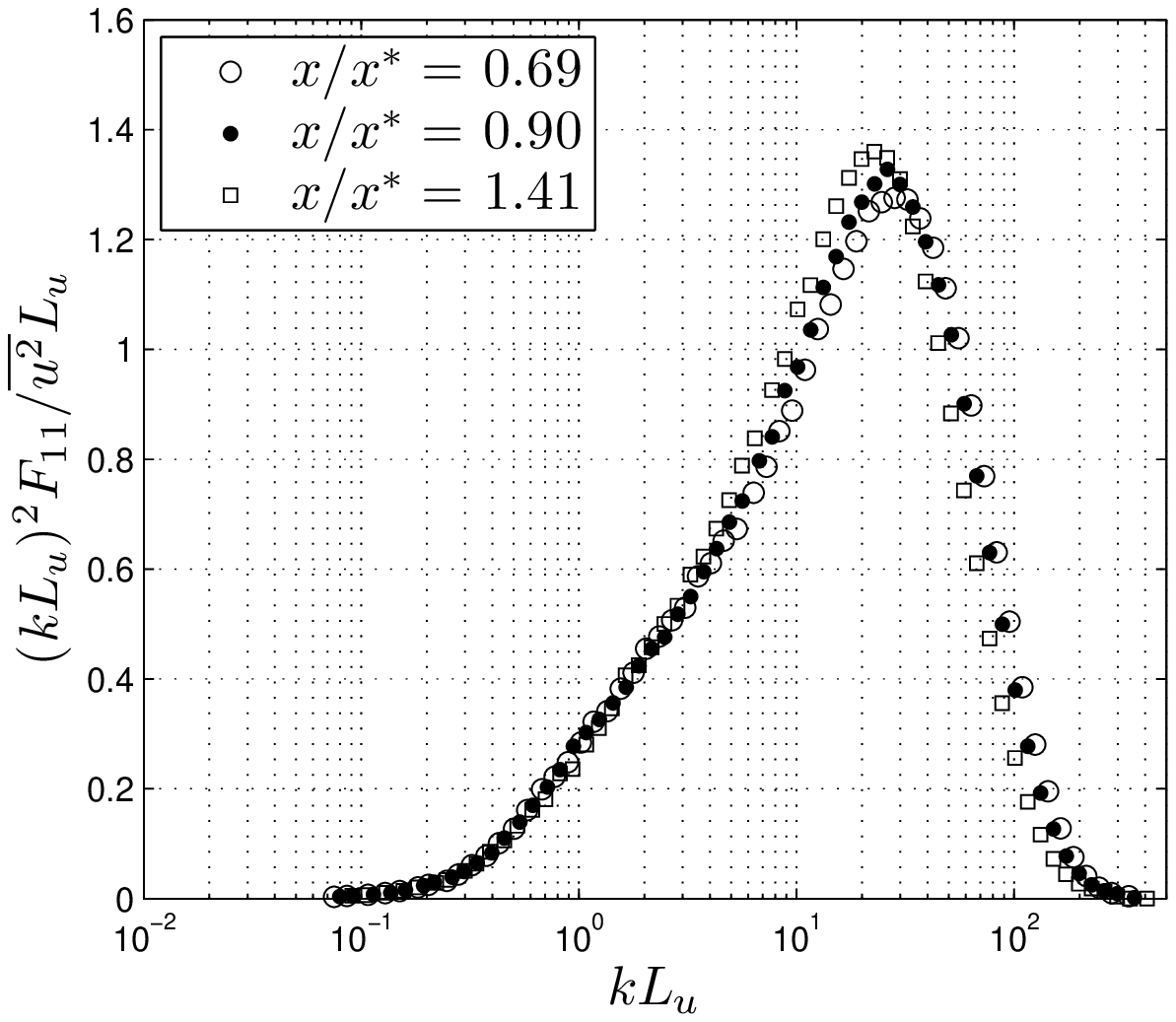}
\includegraphics[trim=10mm 0mm 5mm 2mm, clip=true,width=65mm]{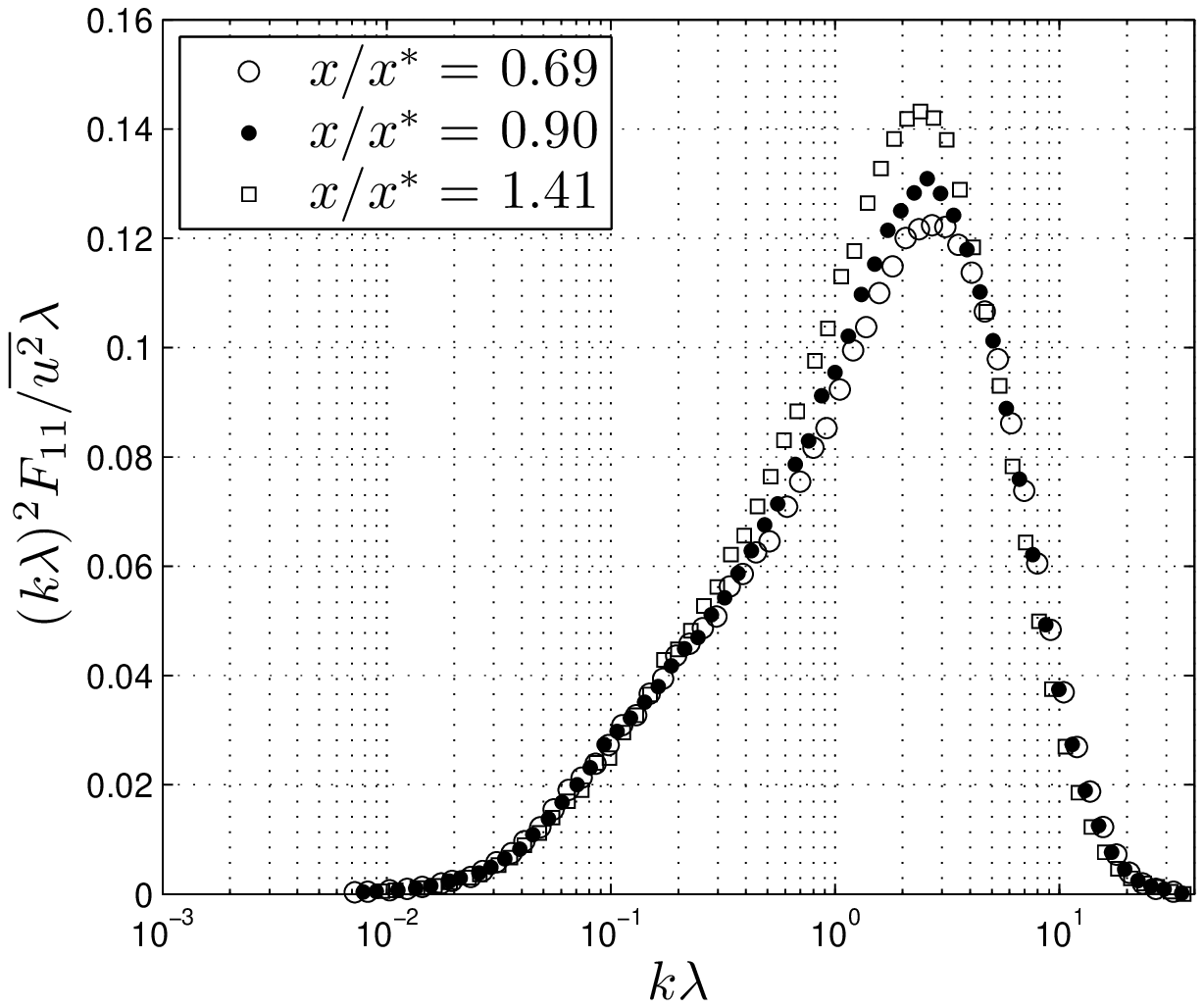}
\caption{Compensated 1D energy spectra of turbulence generated by the fractal square grid at three streamwise downstream locations corresponding to $Re_{\lambda}=324,\,273,\,210$ at $U_{\infty}=15ms^{-1}$, normalised by (a) $\overline{u^{2}}$ and $L_{u}$ (b) $\overline{u^{2}}$ and $\lambda$.}
\label{Fig:CollapseII} 
\end{figure}

In order to complement the previous results, the assessed decay region is extended allowing to further test the single-length-scale assumption. The normalised spectra of decaying turbulence downstream of our fractal square grid are shown in figure \ref{Fig:CollapseI} using both the integral-scale and the Taylor micro-scale. For the extended region it can be seen that the normalised spectra using the Taylor micro-scale do collapse for the entire frequency range, although the collapse using the integral-scale at high frequencies is modest for $k L_{u}>40$ where the furthermost point ($x/x_{*}=1.41$) is taken into account. However the discrepancy between data at $Re_{\lambda}=324$ and $Re_{\lambda}=210$ is much too small compared to the lack of collapse which would occur if the data obeyed Richardson-Kolmogorov scaling as in figure \ref{Fig:ComparisonRG_SFG}a. It should be noted that in theory the collapses with $L_{u}$ or with $\lambda$ should be identical if $L_{u}\propto \lambda$, but as was seen in figure \ref{Fig:LOverLambdaCeps} this is not verified exactly in our wind-tunnel's extended test section. 

Nevertheless, in Appendix \ref{ap:appendixB} we propose a methodology for making a rough estimate of the 
quality of collapse of normalised spectra at high frequencies and we find that it depends on the logarithm of the Reynolds number ratio $Re_{\lambda_1}/Re_{\lambda_2}$ at two streamwise distances $x=\xi_{1}$ and $x=\xi_{2}$ with a pre-factor which depends on the behaviour of $L_{u}/\lambda$ during decay. In the Appendix, we apply this methodology to the active grid data of \cite{KCM02}, for which there is evidence of a Richardson-Kolmogorov cascade, and show how spectral collapse with outer variables can be misleading because the Reynolds number ratio is small. The same methodology applied to our data shows that we are not fully able to conclude on the very high frequency end of fractal grid-generated energy spectra. 

Pre-multiplying the 1D energy spectra by the square of the frequency yields the Fourier spectrum of $\overline{(du/dx)^{2}}$, so a second test to the single-length-scale assumption is to assess the collapse of this isotropic equivalent of the dissipation spectra. The data, plotted in figure \ref{Fig:CollapseII} show a reasonable collapse onto a single curve using both length-scales, though it can be seen that the peak of the pre-multiplied spectra does not collapse perfectly. This may, to some extent, be an effect of the slight anisotropy of the flow, since it affects the large scale variables the normalisation is based on. It is shown in the following section how it is possible to partly account for this effect by computing the three-dimensional energy spectrum. 

\subsubsection{Three-dimensional energy spectra}

The 3D energy spectrum is computed using the two-component velocity signal from the cross-wire measurements, with a similar algorithm to the one presented in \cite{H&V77}. The central assumption of the algorithm is isotropy in order to relate the one-dimensional total energy spectrum $F_{ii}(k_{1})=F_{11}(k_{1})+F_{22}(k_{1})+F_{33}(k_{1})$ with the three-dimensional spectrum $E(k)$,
\begin{align}
E(k)=-k_{1} \frac{d F_{ii}}{d k_{1}}
\end{align}
where the transverse one-dimensional spectra are considered to be approximately the same, i.e. $F_{22}(k_{1})\approx F_{33}(k_{1})$. The first derivative of the spectrum is computed using the logarithmic derivative proposed by \cite{U63}:
\[
E(k)=-F_{ii} \frac{d \ln F_{ii}}{d \ln k_{1}}
\]
The 3D energy spectrum was evaluated at 50 logarithmically spaced frequencies, and a $2^{nd}$-order polynomial was fitted between two neighbouring frequencies using a least-squares-fit in order to obtain a smooth derivative of the spectrum.

From the 3D energy spectrum the integral scale $L$, the turbulent kinetic energy and the Taylor micro-scale can be recovered. The difficulty in accurately determining the low frequency range of the energy spectra and consequently estimating the integral length scale should be noted. For this reason, the assessment of the spectrum's slope near $k\rightarrow 0$ was not possible.

\begin{figure} 
\centering
\includegraphics[trim=10mm 0mm 5mm 2mm, clip=true,width=65mm]
{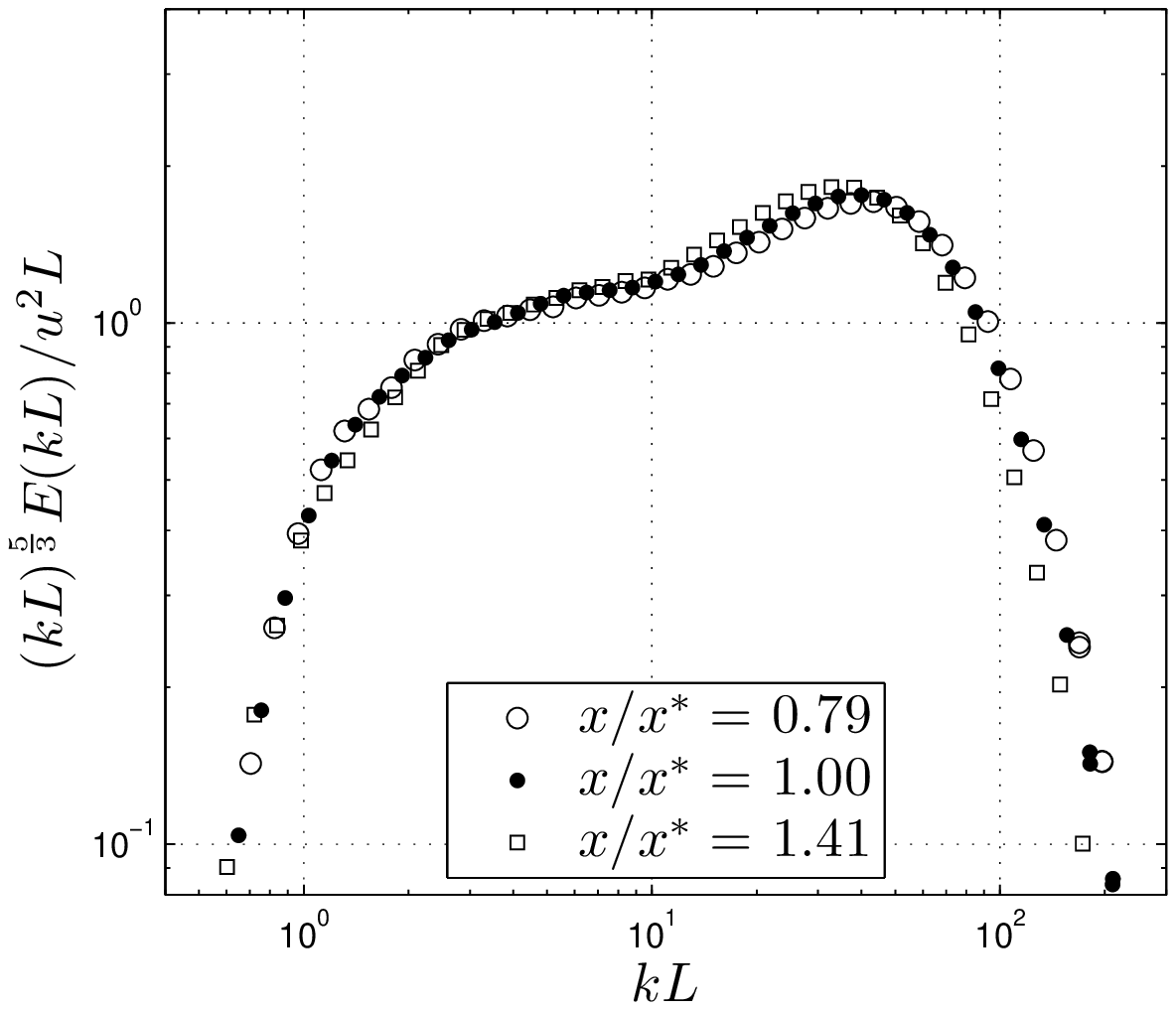}
\includegraphics[trim=10mm 0mm 5mm 2mm, clip=true,width=65mm]
{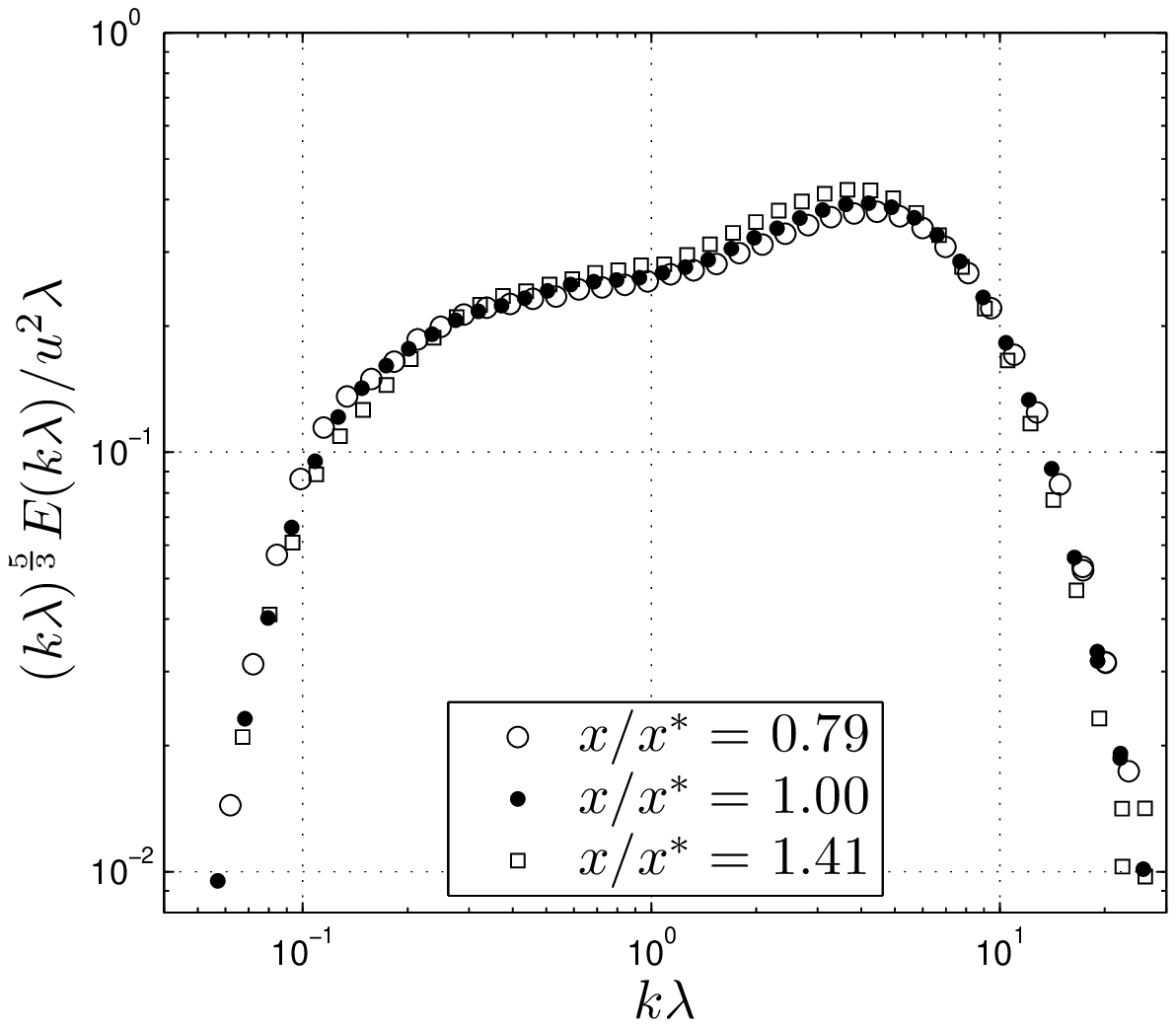}
\caption{3D energy spectra of turbulence generated by the fractal square grid at three streamwise downstream locations corresponding to $Re_{\lambda}=300,\,238,\,210$ at $U_{\infty}=15ms^{-1}$  and normalised by (a) $\overline{u^{2}}=2/3\overline{q^{2}}$ and $L$ (b) $\overline{u^{2}}$ and $\lambda$.} 
\label{Fig:CollapseIII} 
\end{figure}

\begin{figure} 
\centering
\includegraphics[trim=10mm 0mm 5mm 2mm, clip=true,width=65mm]
{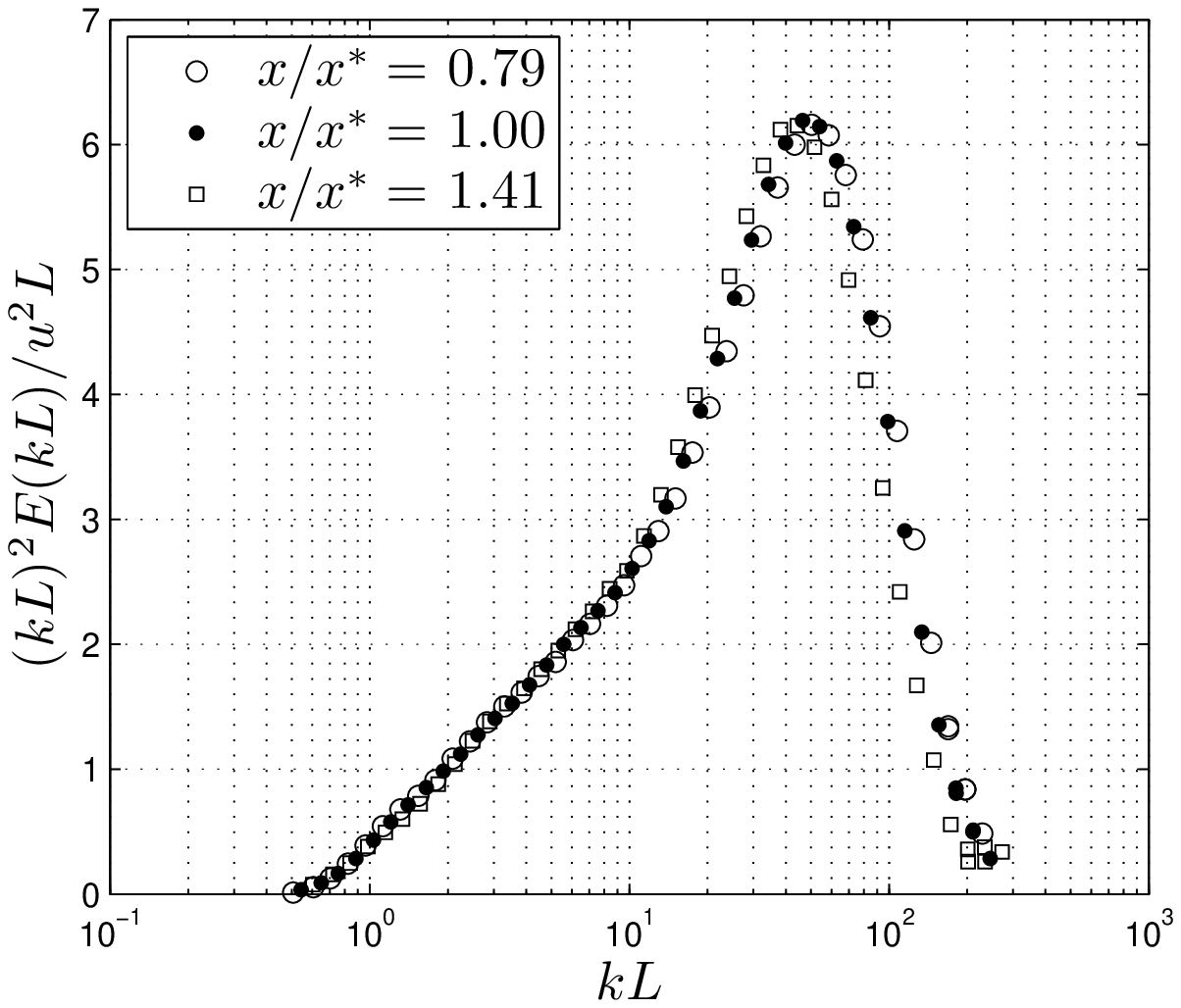}
\includegraphics[trim=10mm 0mm 5mm 2mm, clip=true,width=65mm]
{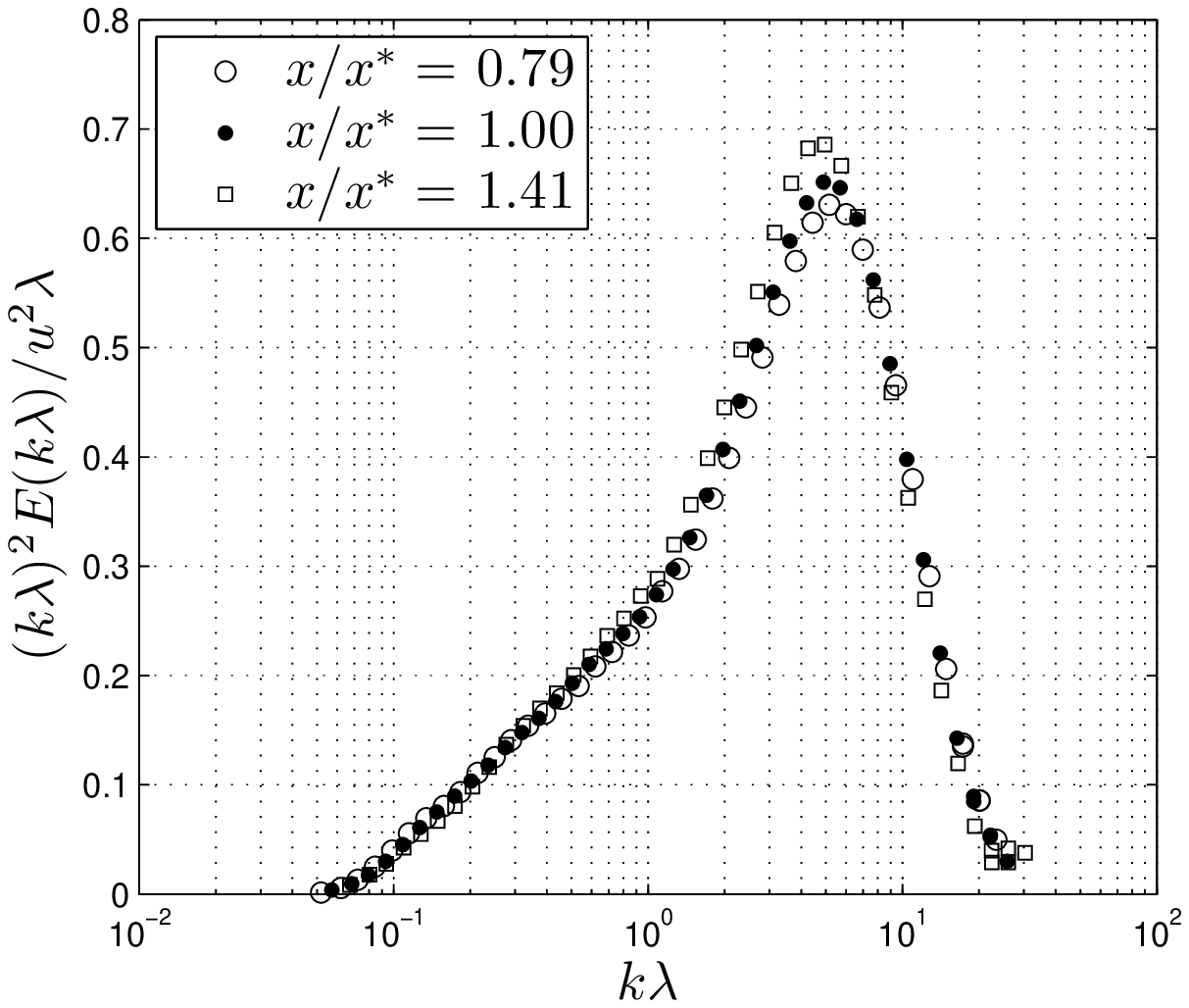}
\caption{Enstrophy spectra of turbulence generated by the fractal square grid at three streamwise downstream locations corresponding to $Re_{\lambda}=300,\,238,\,210$ at $U_{\infty}=15ms^{-1}$ and normalised by (a) $\overline{u^{2}}=2/3\overline{q^{2}}$ and $L$ (b) $\overline{u^{2}}$ and $\lambda$.} 
\label{Fig:CollapseIV} 
\end{figure}

The normalised compensated spectra are shown in figure \ref{Fig:CollapseIII}, while the normalised enstrophy spectra are shown in figure \ref{Fig:CollapseIV}. It is rewarding to see that the collapse of the 3D energy spectrum presents less scatter than the 1D spectrum thus offering support to the self-preserving single-length behaviour of turbulence generated by the fractal square grid. Hence, some of the deviation from single-scale self-similarity collapse of the 1D spectra in figures \ref{Fig:CollapseI} \& \ref{Fig:CollapseII} is due to the moderate level of anisotropy present in the turbulence.

\subsubsection{Second-order structure functions}
\begin{figure} 
\centering
\includegraphics[width=65mm]{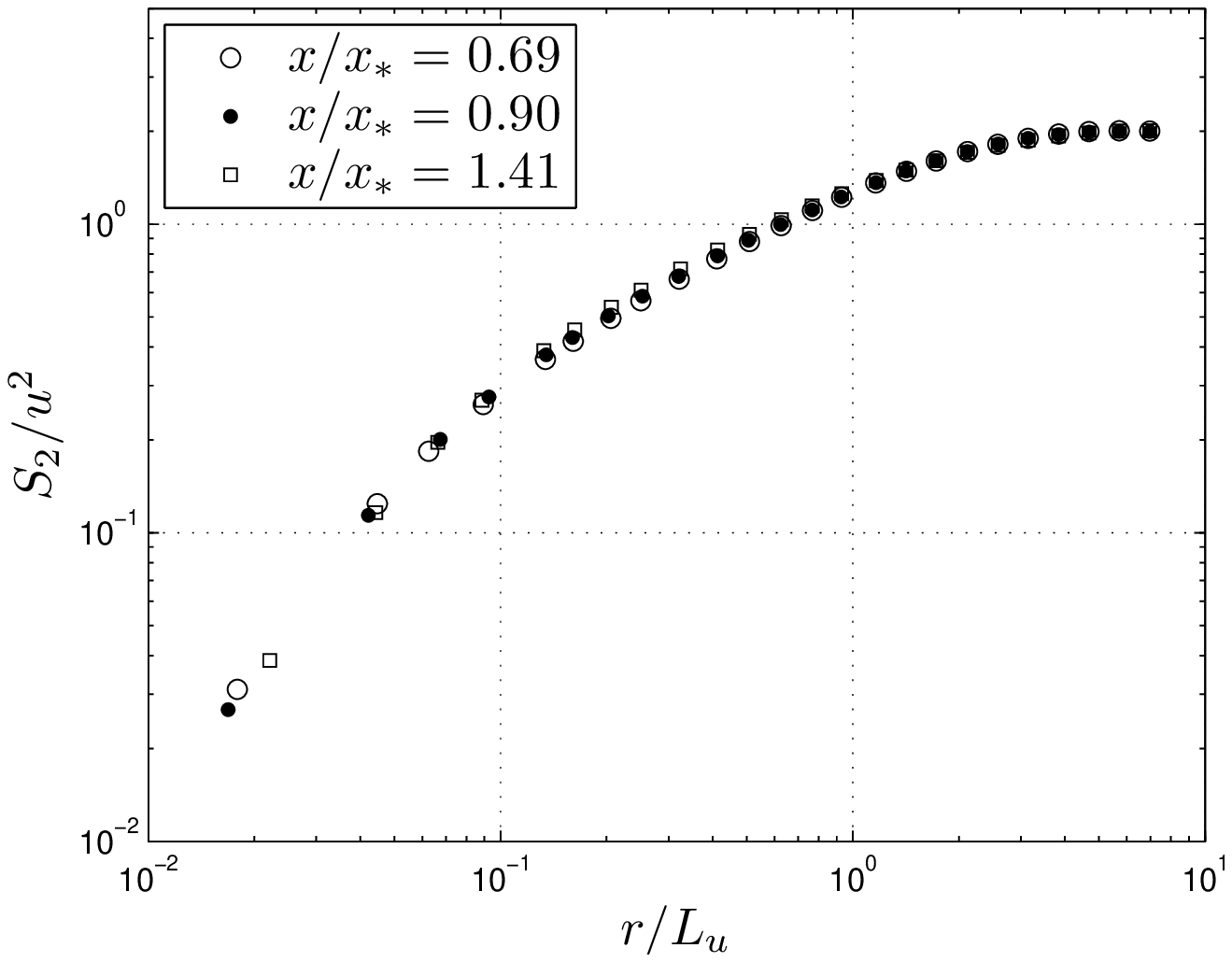}
\includegraphics[width=67mm]{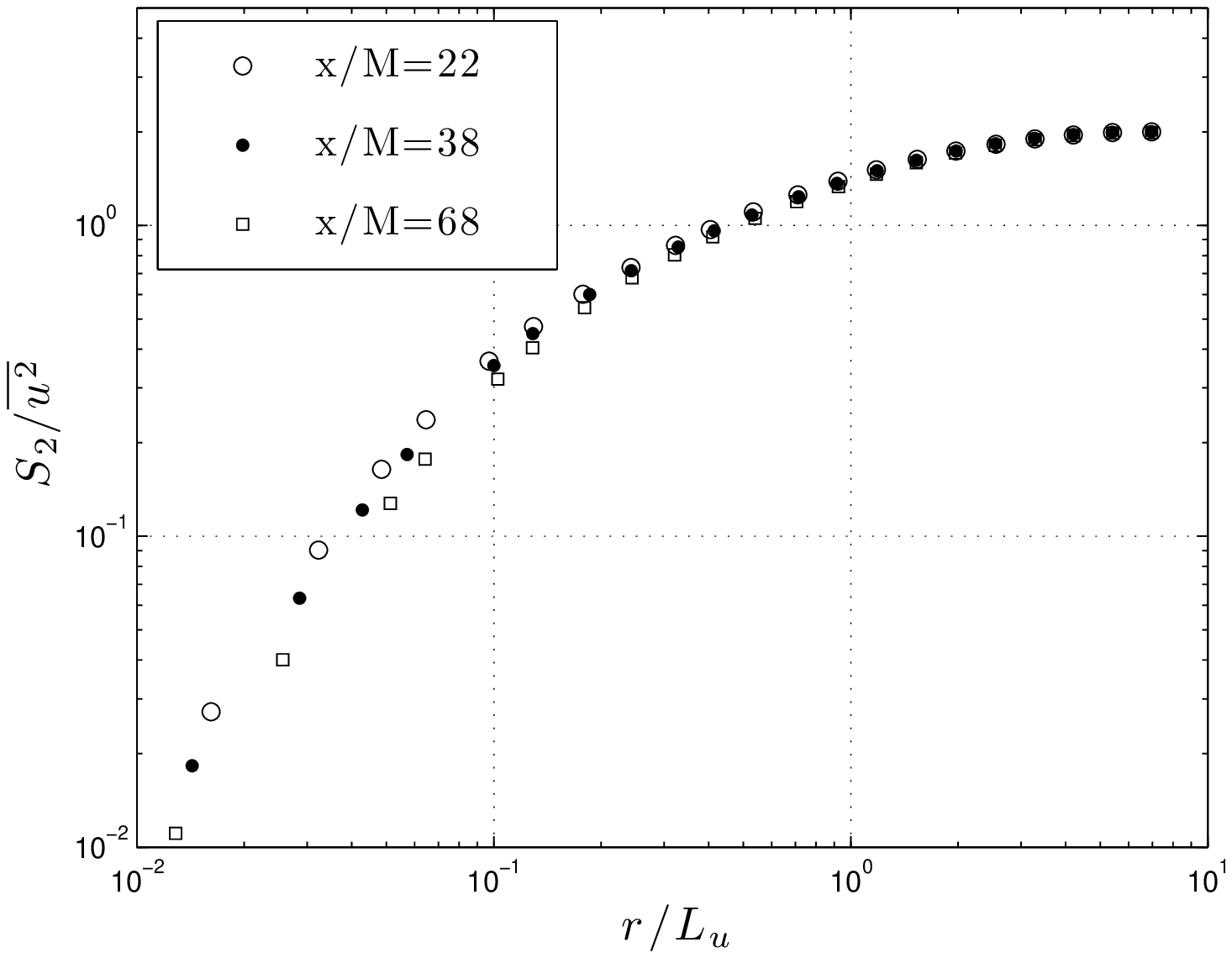}
\caption{Second-order structure function at three streamwise downstream locations normalised by $\overline{u^{2}}$ and $L_{u}$: (a) SFG recorded at $U_{\infty}=15ms^{-1}$, $Re_{\lambda}=323,\,273,\,210$ for $x/x_*=0.69, 0.90, 1.41$  (b) RG  recorded at $U_{\infty}=20ms^{-1}$,  $Re_{\lambda}=156,\,137,\,120$ for $x/M=22, 38, 68$.} 
\label{Fig:CollapseV} 
\end{figure}

The collapse of the second order structure functions using $u'^{2}$ and $L_{u}$ is shown in figure \ref{Fig:CollapseV}a. Similarly to what has already been discussed for the spectra, this structure function collapses well at both low and high separations in the case of our fractal-generated turbulence. However, this is  clearly not the case for the turbulence generated by the regular grid (see figure \ref{Fig:CollapseV}b).

\section{Conclusions and issues raised}\label{sec:concl}

The decay of regular grid- and  fractal square grid-generated turbulence have been experimentally investigated using constant temperature hot-wire anemometry. The main contribution of the present work is to complement previous research on the decay of fractal grid-generated turbulence \cite[\eg][]{H&V2007,S&V2007,M&V2010} by doubling the extent of the assessed decay region with the aim of investigating the persistence (or lack thereof) of the reported high decay exponents and the suppressed Richardson-Kolmogorov cascade. The present experimental investigation also complements the previous research by studying the effect of the hot-wire spatial resolution, carefully assessing the homogeneity of the flow during decay and taking anisotropy into account in the energy spectra.  

We find that for streamwise downstream positions beyond $x/x_{*} \approx 0.6$ the turbulence is close to homogeneous except for a persistence of pressure transport and transverse energy transport and decays such that $L_{u}/\lambda\approx Const$ whilst $Re_{\lambda}$ sharply decreases, at least up to the furthermost downstream position investigated. However $L_{u}/\lambda$  increases with increasing grid Reynolds number, \eg $Re_{0}=U_{\infty} t_0/\nu$. This observation is in direct conflict with the Richardson-Kolmogorov cascade \cite[]{M&V2010}, believed to be dominant at this range of Taylor-based Reynolds numbers $Re_{\lambda}$ in various boundary-free turbulent flows, including regular grid- and active grid-generated turbulence  \cite[]{Burattini2005,Sreeni84,Sreeni98}. It must be noted, however, that the vast majority of existing data is taken at fixed streamwise locations $x$ and varying inlet Reynolds numbers $Re_{0}$ and as we show in section \ref{Sec:Ceps} for fractal grid-generated turbulence, the streamwise downstream Reynolds number dependence $Re_{\lambda}(x)$ isn't necessarily the same. 

We observe that the energy spectra and the 2$^{nd}$ order structure function are much better described in the present fractal square grid-generated turbulence by a single-scale self-similar form than by \cite{kolmogorov1941local} phenomenology. Note that by \cite{kolmogorov1941local} phenomenology we mean, not only the necessity of two dynamically relevant sets of variables, outer and inner, that collapse the low- and the high-frequency part of the spectra respectively, but also that $L_{u}/\lambda\propto Re_{\lambda}$ and $C_{\varepsilon}=Const$, which implicitly dictates the rate of spreading of the high-frequency part of the spectra normalised by outer variables and vice-versa. That turbulence generated by the present fractal square grid does not obey \cite{kolmogorov1941local} phenomenology is clear, for example, from the comparison between figures \ref{Fig:ComparisonRG_SFG}a and \ref{Fig:ComparisonRG_SFG}b.

We also confirm the observations of \cite{H&V2007} and \cite{M&V2010} concerning the abnormally high power-law decay exponents, compared with most boundary-free turbulent flows (see table \ref{Table:OtherFlows}), in particular regular and active grid-generated turbulence ($n_{SFG} >> n_{RG}, n_{AG}$ by a factor between $4/3$ and $2$), and we confirm their persistence further downstream (at least up to $x/x_{*}\approx 1.5$). However, our results do not support the view in \cite{H&V2007} and \cite{M&V2010} that the turbulence decay is exponential or near-exponential. We infer, by comparing our experimental results with the active-grid experiments of \cite{M&W1996}, that the reason for the very unusual turbulence decay properties generated by the fractal square grids cannot be a confinement effect arising from the lateral walls. The two experimental investigations report completely different turbulence properties during decay, even though both experiments were performed on a similar sized wind-tunnel and, in fact, the integral length-scales generated by our fractal square grid are typically less than half the integral length-scales generated by the active-grid. Our fractal-generated turbulence is third-order inhomogeneous in the sense discussed in subsection \ref{Sec:DecayAndDiffusion} but, to our knowledge, no homogeneity studies of active grid-generated turbulence exist to this date which are as thorough as the one presented here, and it is therefore not possible to fully compare homogeneity and isotropy levels of the two types of turbulence. The presence/absence of turbulent transport of pressure and kinetic energy have not been investigated in sufficient detail in either active or regular grid-generated turbulence and it remains unknown to what degree and how far downstream these types of turbulence are third-order homogeneous and isotropic.

Although we find a general agreement with the previous results on the decay of fractal grid-generated turbulence, some new issues are raised by the present experimental results due to the extended wind-tunnel test section. We find that $L_{u}/\lambda$ is in fact not perfectly constant, but slowly decreases with $Re_{\lambda}$, and that the spectral collapse using large scale variables is not perfect at very high wavenumbers as it ought to be for exact single-scale self-preserving turbulence decay. Possible causes for these two observations will be investigated in future work and include: (i) small scale corrections to the single-scale self-preservation, (ii) moderately low Reynolds number limit to the validity of single-scale self-preservation and (iii) excessive thickness of the confining wall boundary layers far downstream interfering with the growth of the largest eddies of the turbulent flow due to insufficient ratio between the wind-tunnel width and the integral length scale. 

As a final remark we note that the study of freely decaying turbulence requires experiments where (i) a wide range of $Re_{0}$  can be achieved by modifying the initial conditions and (ii) a wide range of $Re_{\lambda}$ values must be straddled during decay. 
This is emphasised by the analysis presented in Appendix \ref{ap:appendixB} where we give quantitative criteria for truthful spectral collapse and where we show, in particular, that whereas active grids generate high Reynolds numbers they also generate a narrow logarithmic range of Reynolds numbers during decay thus making it impossible to confirm the Richardson-Kolmogorov cascade via spectral collapse.

So far, only modest ranges of $Re_{\lambda}$ during decay have been achieved with regular grid- and active grid-generated turbulence due to the typically slow decay rates of the turbulence they generate. The fractal square grid-generated turbulence offers the unprecedented possibility of generating high intensity decaying turbulence with a very wide range of $Re_{\lambda}$ values during decay and approximately homogeneous mean flow and turbulence intensity profiles.

\begin{acknowledgments}
We are grateful to Charles Meneveau, Beat L\"{u}thi and one anonymous referee
for numerous suggestions which have helped us to very significantly
improve the paper.
P.C.V. acknowledges the support of the Portuguese Foundation for Science and Technology (FCT) under grant number SFRH/BD/61223/2009. The support from Dantec Dynamics Ltd (UK) in the design of the sine-wave testing hardware is also acknowledged.
\end{acknowledgments}

\appendix
\section{A note on the energy spectra collapse in turbulence generated by active grids}\label{ap:appendixB}

\begin{figure}
\centering
\includegraphics[width=65mm]{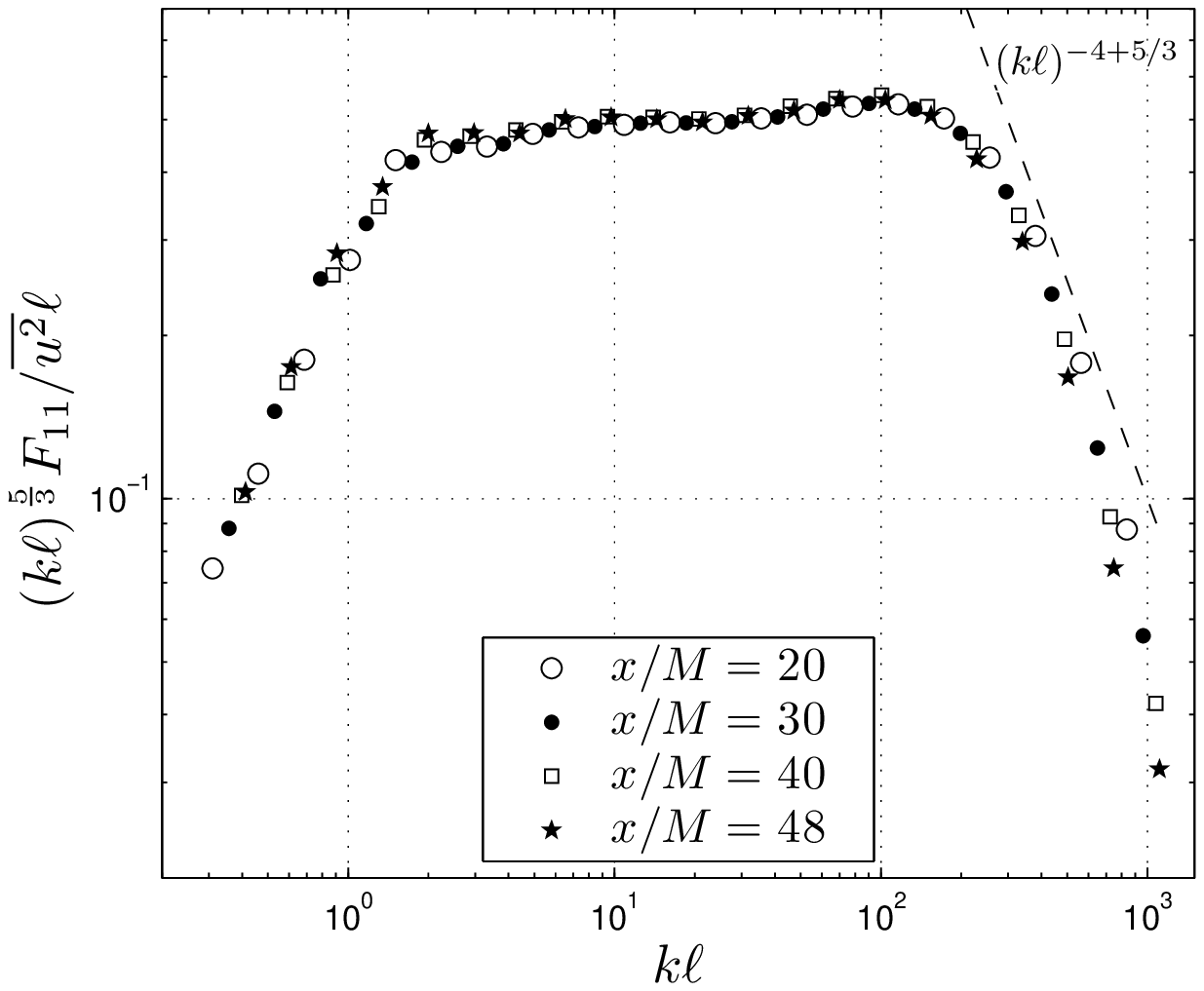}
\includegraphics[width=65mm]{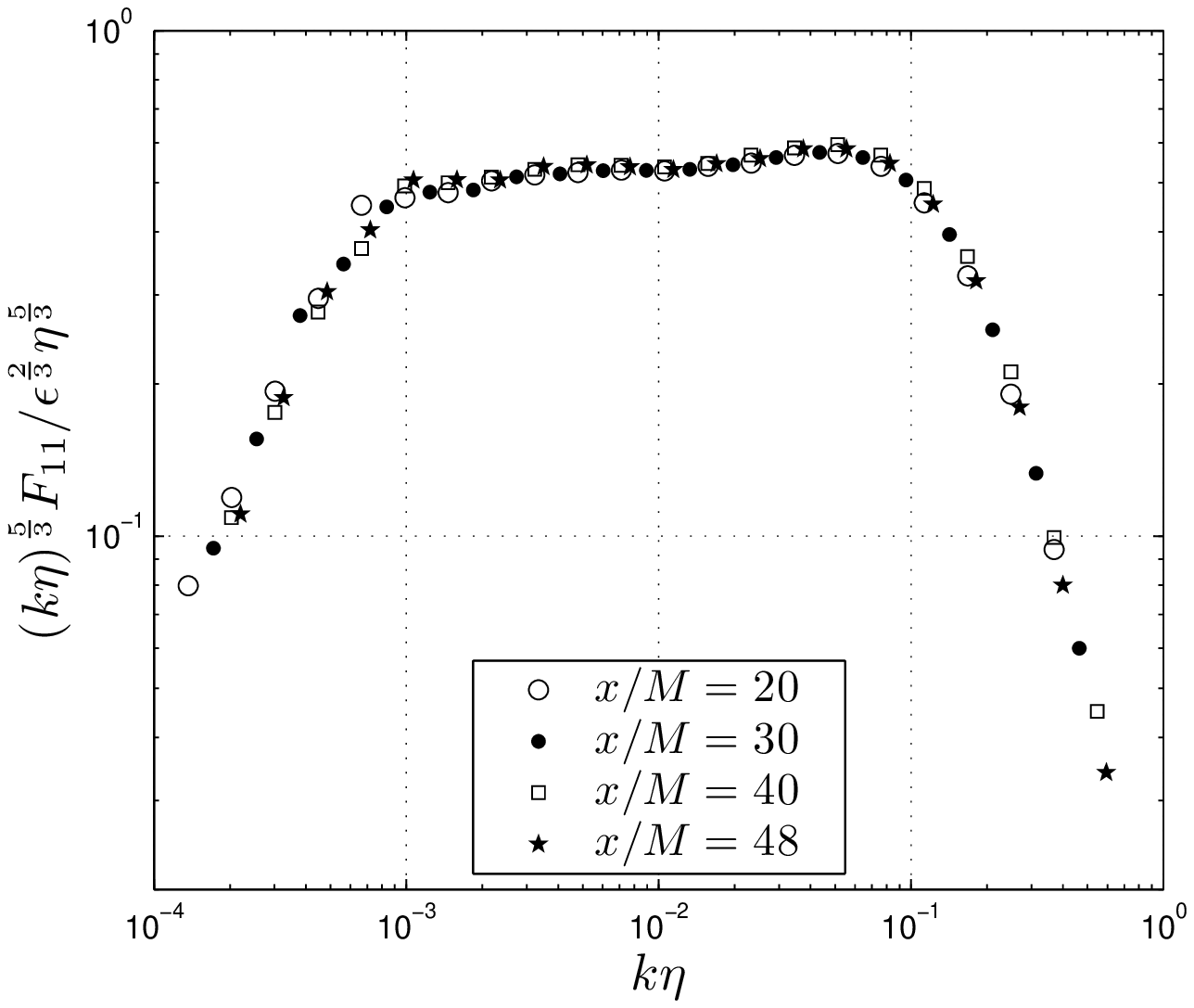}
\caption{Compensated 1D energy spectra at four streamwise locations ($Re_{\lambda}=716,\,676,\,650,\,626$) at $U \approx 11ms^{-1}$, normalized by (a) $\overline{u^{2}}$ and $\ell\equiv 0.9\, u'^{3}/\varepsilon$ (b) $\varepsilon$ and $\eta$. Data from \cite{KCM02}.} 
\label{Fig:CollapseAG} 
\end{figure}

Active-grid experiments can generate relatively high $Re_{\lambda}$ turbulence in a typically sized laboratory wind-tunnel \cite[]{M&W1996} and are thus a potentially good test case to compare the fractal-generated turbulence with. The comparison of the energy decay rate was shown in Sec. \ref{Sec:Decay} and here we focus on the collapse of the energy spectra. The data is taken from the experimental investigation by \cite{KCM02} on the decay of active grid-generated turbulence. In their paper the longitudinal energy spectra for four downstream positions is provided in tabular form and we use this data here to create the plots in figure \ref{Fig:CollapseAG}. 

The pre-multiplied longitudinal energy spectra (figure \ref{Fig:CollapseAG}) are normalised using both outer ($\overline{u^{2}}$ and $\ell$) and inner ($\varepsilon$ and $\eta$) variables. Note that $\ell$ is the pseudo-integral-scale defined as $\ell\equiv 0.9\, u'^{3}/\varepsilon$ which is proportional to the integral scale if and only if the dissipation coefficient $C_{\varepsilon}$ is constant during decay, in this case $C_{\varepsilon}=0.9$. 

At a first glance the results seem striking since both outer and inner variables seem to be collapsing the spectra. Thus one could conjecture that turbulence generated by active grids is self-similar and has only one determining length-scale. This would in fact be the case if the Reynolds number $Re_{\lambda}$ remained constant during the turbulent kinetic energy decay and consequently $L\propto \lambda \propto \eta$, which is the first ever self-preserving decay proposed \cite[]{K&H1938}. Instead $Re_{\lambda}$ decreases during decay. Hence this collapse can only be apparent, but not real.

A simple method of estimating the necessary range of Reynolds numbers $Re_{\lambda}$ for the collapse to be meaningful is now presented where it is shown that the collapse (or spread) of a normalised spectrum at two streamwise locations is only significant if the logarithm of the respective Reynolds numbers' ratio is large, typically $\log\left(Re_{\lambda_1}/Re_{\lambda_2}\right) > 1/4$. The starting point in this methodology is the assumption that a given scaling is correct (e.g. Kolmogorov or single-length scalings) which then allows the quantification of the spread for a given $Re_{\lambda}$ range of any other attempted normalisation.

\begin{figure}
\centering
\includegraphics[trim=10 20 80 110, clip=true, width=100mm]{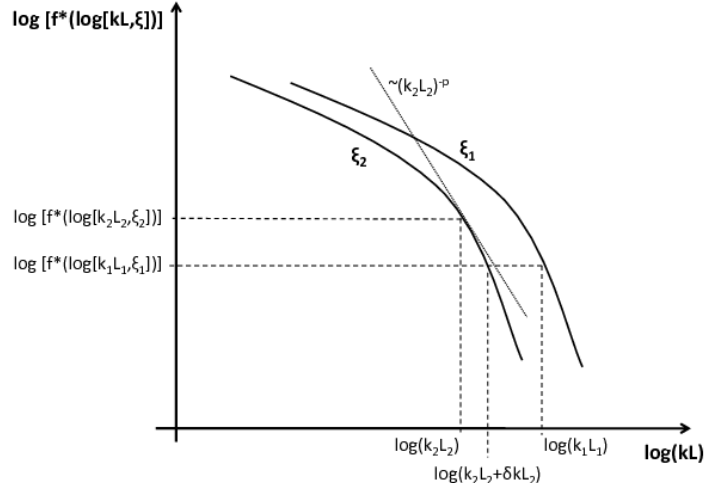}
\caption{Sketch of two spectra at two streamwise positions $x=\xi_1$ and $x=\xi_2$ normalised with outer variables spreading at high frequency.} 
\label{Fig:Sketch} 
\end{figure}

We outline the method by considering the dissipation range of the longitudinal spectrum and assuming the Kolmogorov scaling is correct, i.e. $F_{11}(k,x)=\varepsilon^{2/3}\eta^{5/3}f(k\eta)$, but this methodology is easily extendable to the energy containing range of the spectrum as well as to the case where the single-length-scaling is correct.  

Consider two streamwise distances $x=\xi_1$ and $x=\xi_2$ and write $\eta_1=\eta(\xi_1)$, $\eta_2=\eta(\xi_2)$, $\lambda_1=\lambda(\xi_1)$, $\lambda_2=\lambda(\xi_2)$, $L_1=L_{u}(\xi_1)$, $L_2=L_{u}(\xi_2)$, $u'_1=u'(\xi_1)$, $u'_2=u'(\xi_2)$, $\varepsilon_1=\varepsilon(\xi_1)$, $\varepsilon_2=\varepsilon(\xi_2)$ for the Kolmogorov scales, Taylor micro-scales, integral scales, r.m.s. turbulence velocities and dissipation rates at these two locations. We take $\xi_2 > \xi_1$ so that $Re_{L_1} \equiv u'_1 L_1 / \nu > Re_{L_2} \equiv u'_2 L_2 / \nu$.

Choose two wavenumbers $k_1$ and $k_2$ in the dissipation range such that $k_1 \eta_1=k_2 \eta_2$ and $f(k_1 \eta_1)=f(k_2 \eta_2)$  by assumption. If one would normalise the same spectra in this range using $u'^{2}$ \& $L_{u}$, the dependence of the normalised spectra on $x$ would explicitly resurface, i.e. $F_{11}(k,x)=u'^{2}\, L_{u}\, f^{*}(kL_{u},x)$ (see figure \ref{Fig:Sketch}). Since $\varepsilon = C_{\varepsilon} u'^3 / L_{u}$ with $C_{\varepsilon}$ independent of $x$ in the Richardson-Kolmogorov phenomenology, it follows that $L_{u}/\eta=C_{\varepsilon}^{1/4}Re_{L}^{3/4}$ and it is possible to show that
\begin{align}
f^{*}(k_1 L_1, \xi_1)=f^{*}(k_2 L_2, \xi_2)\left(\frac{\eta_1}{L_1}\frac{L_2}{\eta_2}\right)^{5/3}=f^{*}(k_2 L_2, \xi_2)\left(\frac{Re_{L_2}}{Re_{L_1}}\right)^{5/4}
\label{B1}
\end{align}
and
\begin{align}
k_1 L_1 = k_2 L_2 \left( \frac{\eta_2}{L_2}\frac{L_1}{\eta_1}\right)=k_2 L_2 \left(\frac{Re_{L_1}}{Re_{L_2}}\right)^{3/4},
\label{B2}
\end{align}
so that $f^{*}(k_1 L_1, \xi_1)\neq f^{*}(k_2 L_2, \xi_2)$ and $k_1 L_1 \neq k_2 L_2$.  

We define the spectral spread which characterises the degree of non-collapse by the form $F_{11}=u'^{2} L_{u} f^{*}(kL_{u})$ as 
\begin{equation}
\Psi = \log(k_1 L_1)-\log(k_2 L_2 + \delta kL_2),
\end{equation}
where $f^{*}(k_1 L_1, \xi_1)=f^{*}(k_2 L_2+ \delta kL_2, \xi_2)$, see figure \ref{Fig:Sketch}. There are two contributions to the spectral spread, one from the rescaling of the abscissas, $k_1 L_1 \neq k_2 L_2$, and another from the rescaling of the ordinates. From $Re_{L_1} > Re_{L_2} $ and equations \eqref{B1},  \eqref{B2} we know that $k_1 L_1 > k_2 L_2$ and $f^{*}(k_1 L_1, \xi_1)< f^{*}(k_2 L_2, \xi_2)$ so that the two contributions to the spectral spread can actually in principle, cancel each other. 
However, the second contribution depends on the functional form of $f^{*}(kL_{u})$ and therefore it is not possible to quantify its spectral spread contribution without an analytical expression for $f^{*}(kL_{u},\xi)$. Nonetheless, as is shown below, we can estimate a bound for this contribution, so that in the end we can estimate a upper and lower bound for the expected spectral spread $\Psi$ characterising the degree of non-collapse by the alternative scaling.

The contribution to the spread $\Psi$ from the abscissa's rescaling alone (which is the upper bound) is given by (using \eqref{B2})
\begin{align}
\Psi_{max}=\log(k_1 L_1)-\log(k_2 L_2)=\frac{3}{4} \log\left(\frac{Re_{L_1}}{Re_{L_2}}\right)=\frac{3}{2} \log\left(\frac{Re_{\lambda_1}}{Re_{\lambda_2}}\right).
\label{B5}
\end{align}
(for the last equality, \eqref{Eq:LOverLambda} was used to relate the integral scale to the Taylor micro-scale with $C_{\varepsilon}=Const$ from Richardson-Kolmogorov phenomenology).

The contribution to the spread $\Psi$ from the ordinate's rescaling is measured as a fraction of the abscissa's rescaling
\begin{align}
\Phi \equiv \frac{\log(k_2 L_2+ \delta kL_2)-\log(k_2 L_2)}{\log(k_1 L_1)-\log(k_2 L_2)},
 \label{eq:Phi}
\end{align}
so that $\Phi=0$ for $\delta kL_2=0$ (ordinate rescaling has no effect) and $\Phi=1$ for $\delta kL_2=k_1 L_1-k_2 L_2$ (ordinate rescaling cancels the abscissas rescaling). It is possible to show using a first order Taylor expansion in logarithmic coordinates  that we can re-write the function $\Phi$ to leading order as 
\begin{align}
\Phi = -\frac{5}{3}\left(\left.\frac{\partial\, \log(f^{*}(\log(kL_{u}), \xi))}{\partial\, \log(kL_{u})}\right|_{kL_{u}=k_2 L_2}\right)^{-1}.
\label{B4}
\end{align}

Since the spectra in the dissipation range roll-off faster than any power law we can always find a high enough wavenumber $k_t (p)L_{u}$ for which the tangent of the spectrum (in logarithmic coordinates) is steeper than $(kL)^{-p}$ given an exponent $p$ (see figure \ref{Fig:Sketch}). Consequently, for a given choice of $p$, we get an upper bound for $\Phi$ for wavenumbers above $k_t (p) L_{u}$ which is $\Phi_{max}=5/(3p)$. Therefore we can estimate a lower bound for the spectral spread as $\Psi_{min} = \Psi_{max}-\Phi_{max}$ and thus 
\begin{align}
 \frac{3}{2} log\left(\frac{Re_{\lambda_1}}{Re_{\lambda_2}}\right) - \frac{5}{3p} < \Psi <  \frac{3}{2} log\left(\frac{Re_{\lambda_1}}{Re_{\lambda_2}}\right).
 \label{eq:Psi}
\end{align}

In figure \ref{Fig:CollapseAG}a we plot the $(kL_{u})^{-p}$ function with $p=4$ and it can be seen that for wavenumbers higher than $k_t L_{u}\approx 400$, the tangent of the spectra (in logarithmic coordinates) are steeper. Hence, according to \eqref{eq:Psi}, for $k_t L_{u} > 400$ (taking into account the $Re_{\lambda}$ range of the data presented in figure \ref{Fig:CollapseAG}a) the spectral spread will be around $5\% < \Psi < 9\% $ of a decade, which can easily be confounded with scatter. Therefore, the apparent collapse observed in the spectra from the active-grid experiments of \cite{KCM02} (see figure \ref{Fig:CollapseAG}a) may be misleading as it is simply the result of a small range of $Re_{\lambda}$ variation during decay (from 716 to 637). Note that the same misleading collapse occurs in the low wavenumber range of the spectra plotted in figure \ref{Fig:CollapseAG}b where the Kolmogorov inner variables were used for the normalisation. \\

We can repeat the exact same analysis for the case where the Richardson-Kolmogorov cascade is suppressed 
and assume the validity of $F_{11} (k,x)= u'^2 L_{u} f^{*}(kL_{u})$ and $L_{u}/\lambda \approx Const$. We can then carry out the same calculation as above to obtain the spectral spread $\Psi = \log(k_1 \eta_1)-\log(k_2 \eta_2 + \delta k\eta_2)$ for $k_1 L_1=k_2 L_2$ when attempting to collapse the data with Kolmogorov variables. We would then obtain the same expression to quantify the spread contribution caused by the rescaling of the ordinates relative to the total spread, \eqref{eq:Phi} with a suitably redefined $\Phi$ where the outer scales $L_{u}$ have been replaced by inner scales $\eta$. Note that spread of the high frequency spectra normalised by outer variables assuming that $F_{11}(k,x)=\varepsilon^{2/3}\eta^{5/3}f(k\eta)$ holds is the same as the spread of the high frequency spectra normalised by Kolmogorov inner variables assuming that $F_{11} (k,x)= u'^2 L_{u} f^{*}(kL_{u})$ holds. 
Hence, using $L_{u}\sim \lambda$, the spectral spread resulting from an attempt to collapse with Kolmogorov inner variables spectra which obey complete self-similarity is
\begin{align}
\Psi = \log(k_1 \eta_1)-\log(k_2 \eta_2)=\frac{1}{2} \log\left(\frac{Re_{\lambda_1}}{Re_{\lambda_2}}\right)\hspace*{1mm} .
\label{B7}
\end{align}

It is interesting to observe that the rate of spread in this case is three times slower than the rate of spread when the Richardson-Kolmogorov cascade dominates and one tries to collapse with outer variables. Hence, the spectral spread observed in figure \ref{Fig:ComparisonRG_SFG}a is in agreement with the view that regular grid turbulence at the Reynolds numbers of this figure obeys Richardson-Kolmogorov interscale dynamics. However, the high-frequency behaviours in figures \ref{Fig:ComparisonRG_SFG}b, \ref{Fig:CollapseI}b and \ref{Fig:CollapseI}c fall within the uncertainty defined by \eqref{B7} and we are therefore unable to conclude whether our fractal grid-generated turbulence obeys complete or incomplete self-similarity even though it is clear that $L_{u}\propto \lambda$ is a good approximation. By complete self-similarity we refer to the property that $F_{11} (k,x)= u'^2 L_{u} f^{*}(kL_{u})$ is exact at all frequencies and by incomplete self-similarity we refer to the property that 
deviations to $F_{11} (k,x)= u'^2 L_{u} f^{*}(kL_{u})$ can appear at the very highest frequencies. We stress that this does not imply Kolmogorov scaling even if these high frequencies may be collapsed by Kolmogorov inner variables for the simple reason that $L_{u}\propto \lambda$ and therefore $\varepsilon$ is not proportional to $u'^{3}/L_{u}$.

\bibliographystyle{jfm}
\bibliography{mybib}

\end{document}